\documentclass[traditabstract]{aa}  
\usepackage{graphicx}
\usepackage{textcomp}
\usepackage{lscape}
\usepackage{multicol}
\usepackage[varg]{txfonts}
\usepackage{mathrsfs}
\usepackage{amssymb}
\usepackage{natbib}
\usepackage[colorlinks, citecolor={blue}]{hyperref}
\usepackage{color}
\usepackage{float}

\authorrunning{S.~Duarte~Puertas et al.,}
\titlerunning{Searching for star formation regions in Stephan's Quintet: II. }

\begin{document} 

   \title{Searching for intergalactic star forming regions in Stephan's Quintet with SITELLE: II. Physical properties and metallicity}

\author{
S.~Duarte~Puertas\inst{1}
\and 
J.~M.~Vilchez\inst{1}
\and 
J.~Iglesias-P\'{a}ramo\inst{1}
\and
L. Drissen\inst{2,3}
\and
C.~Kehrig\inst{1}
\and
T. Martin\inst{2,3}
\and
E. P\'erez-Montero\inst{1}
\and
A. Arroyo-Polonio\inst{1}
}
\institute{
Instituto de Astrof\'{\i}sica de Andaluc\'{\i}a - CSIC, Glorieta de la Astronom\'{\i}a s.n., 18008 Granada, Spain\label{inst1} \\ \email{salvini@iaa.es}
\and
D\'epartement de physique, de g\'enie physique et d'optique, Universit\'e Laval, Qu\'ebec (QC), G1V 0A6, Canada\label{inst3}
\and
Centre de recherche en astrophysique du Qu\'ebec\label{inst4}
}

   \date{Received \today; accepted \today}

 \abstract 
{Based on SITELLE spectroscopy data, we studied the ionised gas emission for the 175 H$\alpha$ emission regions in the Stephan's Quintet (SQ). In this paper we perform a detailed analysis of the star formation rate (SFR), oxygen abundance, and nitrogen-to-oxygen abundance ratio (N/O) of the SQ regions, with the intention of exploring the provenance and evolution of this complex structure. According to the BPT diagram, we found 91 HII, 17 composite, and 7 active galactic nucleus-like regions in SQ. Several regions are compatible with fast shocks models without a precursor for solar metallicity and low density (n = 0.1 cm$^{-3}$), with velocities in the range of 175 -- 300 km s$^{-1}$. We derived the total SFR in SQ (log(SFR/M$_\odot\,yr^{-1}$=0.496). Twenty-eight percent of the total SFR in SQ comes from starburst A, while 9\% is in starburst B, and 45\% comes from the regions with a radial velocity lower than 6160 km s$^{-1}$. For this reason, we assume that the material prior to the collision with the new intruder does not show a high SFR, and therefore SQ was apparently quenched. When considering the integrated SFR for the whole SQ and the new intruder, we found that both zones have a SFR consistent with those obtained in the SDSS star-forming galaxies. At least two chemically different gas components cohabit in SQ where, on average, the regions with high radial velocities (v$>$6160 km s$^{-1}$) have lower values of oxygen abundance and N/O than those with low radial velocities (v$\leq$6160 km s$^{-1}$). The values found for the line ratios considered in this study, as well as in the oxygen abundance and N/O for the southern debris region and the northernmost tidal tail, are compatible with regions belonging to the outer part of the galaxies. We highlight the presence of inner-outer variation for metallicity and some emission line ratios along the new intruder strands and the young tidal tail south strand. Finally, the SQ H$\alpha$ regions are outside the galaxies because the interactions have dispersed the gas to the peripheral zones.}


   \maketitle

\section{Introduction}
The Stephan's Quintet (SQ) is a compact group of galaxies discovered by \cite{1877MNRAS..37..334S}. SQ is formed by two elliptical (NGC7317 and NGC7318A), three spiral (NGC7318B, NGC7319, NGC7320c), and one new dwarf galaxy described recently in \citet[][hereafter paper I]{2019A&A...629A.102D}. Also, in the field of view (FoV) of SQ we can see a foreground galaxy (NGC7320) with a discordant distance. In SQ we found diverse structures (e.g. tidal tails), but probably the most remarkable one is the large-scale shock region \citep[LSSR, e.g.][]{1972Natur.239..324A,1998ApJ...492L..25O,2012A&A...539A.127I} produced by the collision between NGC7318B and both NGC7319 and debris material produced from previous interactions. In paper I we found five velocity systems in SQ: i) v=[5600-5900] $km\, s^{-1}$; ii) v=[5900-6100] $km\, s^{-1}$; iii) v=[6100-6600] $km\, s^{-1}$; iv) v=[6600-6800] $km\, s^{-1}$; and v) v=[6800-7000] $km\, s^{-1}$ (see paper I for more details).

Previous works studied the physical and chemical properties of star-forming objects in SQ \citep[e.g.][]{2004ApJ...605L..17M,2014ApJ...784....1K}. The work by \cite{2014ApJ...784....1K} studied the properties of 40 H$\alpha$ emitting regions in the LSSR, the new intruder (NI), and the southern debris region (SDR) and did not find shocked gas associated with the HII regions or the intrusive galaxy. Massive star formation in regions of the shock zone appears substantially suppressed while the starburst A (SQA) region appears to be the most active star-forming region in the SQ \citep{2014ApJ...784....1K}. The shocked gas presents two velocity components: low velocity ($\leq$ 6160 $km\,s^{-1}$ and nearly solar metallicities\footnote{Solar metallicity: 12 + log(O/H) = 8.69 \citep{2009ARA&A..47..481A}.}) and high velocity ($>$ 6160 $km\,s^{-1}$ and low metallicity) \citep{2012A&A...539A.127I,2014MNRAS.442..495R}. The SDR presents typical star formation rate (SFR) values of nearby galaxies \citep{2014ApJ...784....1K} and its HII regions have subsolar metallicities, presenting a metallicity gradient along the spiral arm of NGC7318B \citep{2012A&A...539A.127I}. 

In this work we performed a detailed analysis of SQ based on the spectroscopic data from SITELLE \citep{2012SPIE.8446E..0UG}, an imaging Fourier transform spectrometer, attached to the Canada-France-Hawaii Telescope. SITELLE has a large FoV \citep[11$^\prime$x11$^\prime$; 0.32$^{\prime\prime}$ per pixel,][]{2019MNRAS.485.3930D}, and covers the optical range from 3500 $\AA$ to 7500 $\AA$. It allows us to obtain a more complete view of the ionisation structure of the SQ and its physical and chemical properties. Previous spectroscopic studies are based on long-slit observations or integral field units with a smaller FoV, which do not cover the whole area of the SQ \citep[e.g.][]{2012A&A...539A.127I,2014ApJ...784....1K}. Thanks to the large FoV of SITELLE, we can cover almost the entire SQ with a reasonable spectroscopic and spatial resolution.

In this work the determination of the gas metallicity, using oxygen abundance (O/H) as a proxy, allows us to distinguish between evolved metal-rich gas from the SQ galaxies and tidal tails, and identify unevolved metal-poor gas from the intergalactic medium. We know that while representative dwarf galaxies present typical values of 12+log(O/H) between 7.1 and 8.4 \citep{2006ApJ...636..214V,2009ApJ...705..723C,2012ApJ...754...98B,2018MNRAS.480.2719L}, tidal dwarf galaxies present values higher than 8.4 \citep{2012MNRAS.426.2441D,2014MNRAS.440.1458D,2015A&A...584A.113L}. The combination of the results obtained in the three-dimensional study from paper I with the SQ properties derived here (e.g. SFR, O/H, N/O, extinction, and diagnostic line ratios) leaves us in an excellent position to try to answer fundamental questions about the SQ, such as: is star formation inhibited for this group of galaxies? Where is the star formation in SQ located? Does the star formation occur outside the galaxies of SQ or not? Does the new intruder chemically enrich the environment? Is extinction associated with the highest star formation zones? Are we seeing different gas components cohabiting in SQ?

The structure of this paper is organised as follows: in Sect.~\ref{sec:results} we detail our main results. The discussion and summary of our work are presented in Sect.~\ref{sec:conclu}. Throughout the paper, we assume a Friedman-Robertson-Walker cosmology with $\Omega_{\Lambda 0}=0.7$, $\Omega_{\rm m 0}=0.3$, and $\rm H_0=70\,km\,s^{-1}\,Mpc^{-1}$.

\section{Results}
\label{sec:results}

\begin{figure}
    \centering
    \includegraphics[width=\columnwidth]{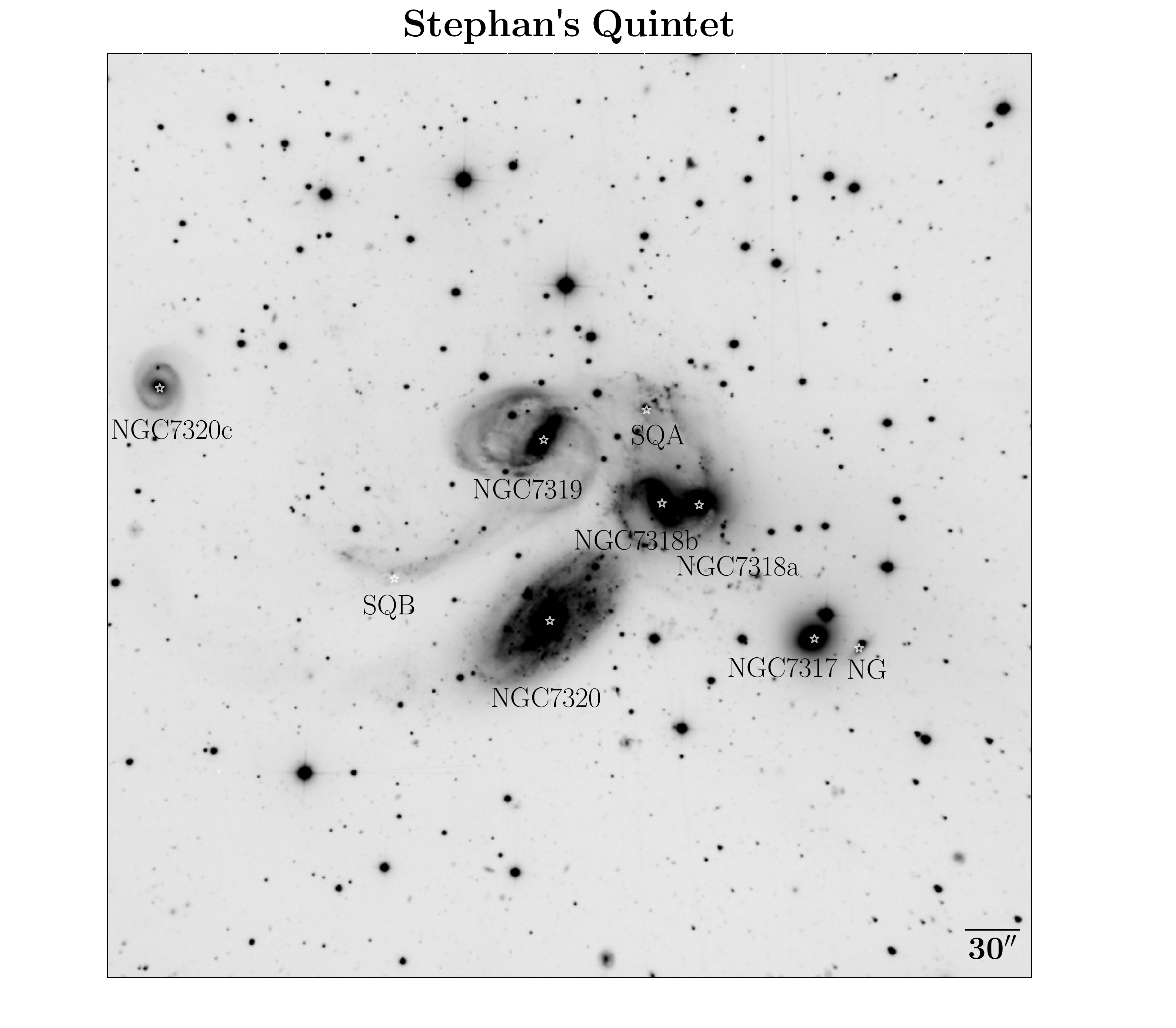}
    \caption{SITELLE deep-greyscale image of SQ composed using SN1, SN2, and SN3 data cubes. North is top and East is left. The distance considered for SQ in this paper is d = 88.6 Mpc (from the NASA/IPAC Extragalactic Database known as NED). At the distance of SQ, 30$^{\prime\prime}$ corresponds to $\sim$13.04 kpc.}
    \label{fig:SQ}
\end{figure}

\subsection{SITELLE spectroscopy: Line fluxes}
\label{subsec:line}
Figure~\ref{fig:SQ} shows a SITELLE deep-greyscale image of the SQ field. We considered the sample of 175 SQ H$\alpha$ emission regions defined in paper I. For each H$\alpha$ emission region we fitted the emission lines that we found in the SN1, SN2, and SN3 data cubes: e.g. [\ion{O}{ii}]$\lambda$3727, H$\beta$, [\ion{O}{iii}]$\lambda\lambda$4959,5007, H$\alpha$, [\ion{N}{ii}]$\lambda\lambda$6548,6584, and [\ion{S}{ii}]$\lambda\lambda$6717,6731. The data analysis followed is described in detail in paper I. It should be noted that the emission lines were fitted with sincgaussian functions (the convolution of a Gaussian with a sinc function) using the Python-based package \textsc{ORCS} \citep{2015ASPC..495..327M}. The fit output parameters are the radial velocity, broadening, intensity peak, total flux, and the corresponding uncertainties. All SQ H$\alpha$ regions were defined according to the following criteria: the radial velocity of the region is within the radial velocity range of SQ (between $\sim$5600 and $\sim$7000 $km\,s^{-1}$) and at least one additional emission line besides H$\alpha$ was detected in the data cubes. From the sample of 175 SQ H$\alpha$ regions, 127 (73\%) present signal-to-noise ratio (S/N), S/N([\ion{O}{ii}]$\lambda$3727) $\geq$ 3, 169 (96\%) S/N(H$\beta$) $\geq$ 3, 131 (75\%) S/N([\ion{O}{iii}]$\lambda$5007) $\geq$ 3, and 146 (83\%) S/N([\ion{N}{ii}]$\lambda$6584) $\geq$ 3. We defined the S/N as the ratio of the flux to the statistical error flux, calculated with the pipeline \textsc{ORCS} \citep{2015ASPC..495..327M}. [\ion{O}{iii}]$\lambda$5007 and [\ion{O}{ii}]$\lambda$3727 are not detected in some regions of the north lobe, NGC7319 `arm', and shock strands (Shs). In Fig.~\ref{fig:fluxes} we show the emission line maps for [\ion{O}{ii}]$\lambda$3727, H$\beta$, [\ion{O}{iii}]$\lambda$5007, H$\alpha$, and [\ion{N}{ii}]$\lambda$6584 for SQ when the S/N in each of them is greater than or equal to 3. For the regions with two velocity components, we consider the sum of the flux value of both components. As noted in paper I, we separated the SQ emission regions into two sub-samples: lower radial velocity sub-sample (LV) for those regions where the radial velocity is lower than or equal to 6160 $km\,s^{-1}$; and the higher radial velocity sub-sample (HV) for all regions with a radial velocity higher than 6160 $km\,s^{-1}$.

We corrected the emission line fluxes for reddening using the theoretical case B recombination (theoretical Balmer decrement, $I_{H\alpha/H\beta}=2.86$; electron temperature $T = 10^{4}\,K$, and low-density limit $n_e \sim 10^2\,cm^{-3}$; \citealt{1989agna.book.....O,1995MNRAS.272...41S}) together with the \cite{1989ApJ...345..245C} extinction curve with $R_v=A_v/E(B-V)=3.1$ \citep[][]{1994ApJ...422..158O,1998ApJ...500..525S}, where A$_v=2.5\,c(H\beta)$. When the reddening coefficient, c(H$\beta$), is negative we set it as zero. The Galactic extinction is very small in the direction of SQ, E(B - V) = 0.07 according to NASA/IPAC Extragalactic Database (NED)\footnote{\texttt{http://ned.ipac.caltech.edu/}}, and its contribution to A$_v$ derived for each region is not relevant.

In Table~\ref{table:table3} we present emission line fluxes divided by H$\beta$ and corrected for reddening for the SQ H$\alpha$ emission regions. Only line fluxes with S/N greater than or equal to 3 are listed in Table~\ref{table:table3}. In Column 1 the region name is presented. Columns 2, 3, 4, 5, 6, and 7 show the [\ion{O}{ii}]$\lambda$3727, [\ion{O}{iii}]$\lambda$5007, H$\alpha$, [\ion{N}{ii}]$\lambda$6584, [\ion{S}{ii}]$\lambda$6716, and [\ion{S}{ii}]$\lambda$6731 emission line fluxes, respectively. Column 8 shows A$_v$, and Column 9 tells us whether the velocity of the region belongs to the LV (0) or HV (1).

\begin{figure*}
    \centering
    \includegraphics[width=.7\textwidth]{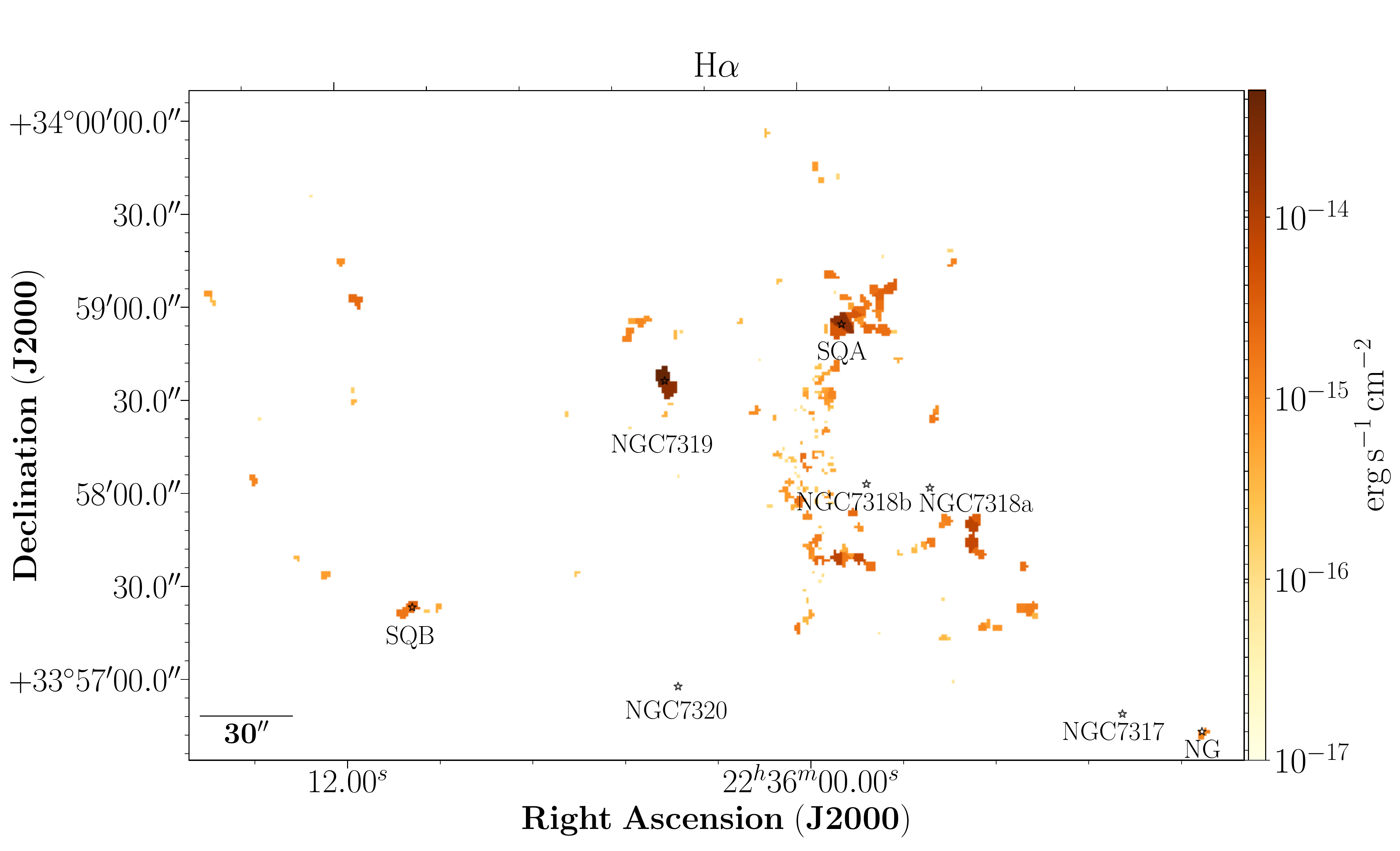}\\
    \includegraphics[width=.7\textwidth]{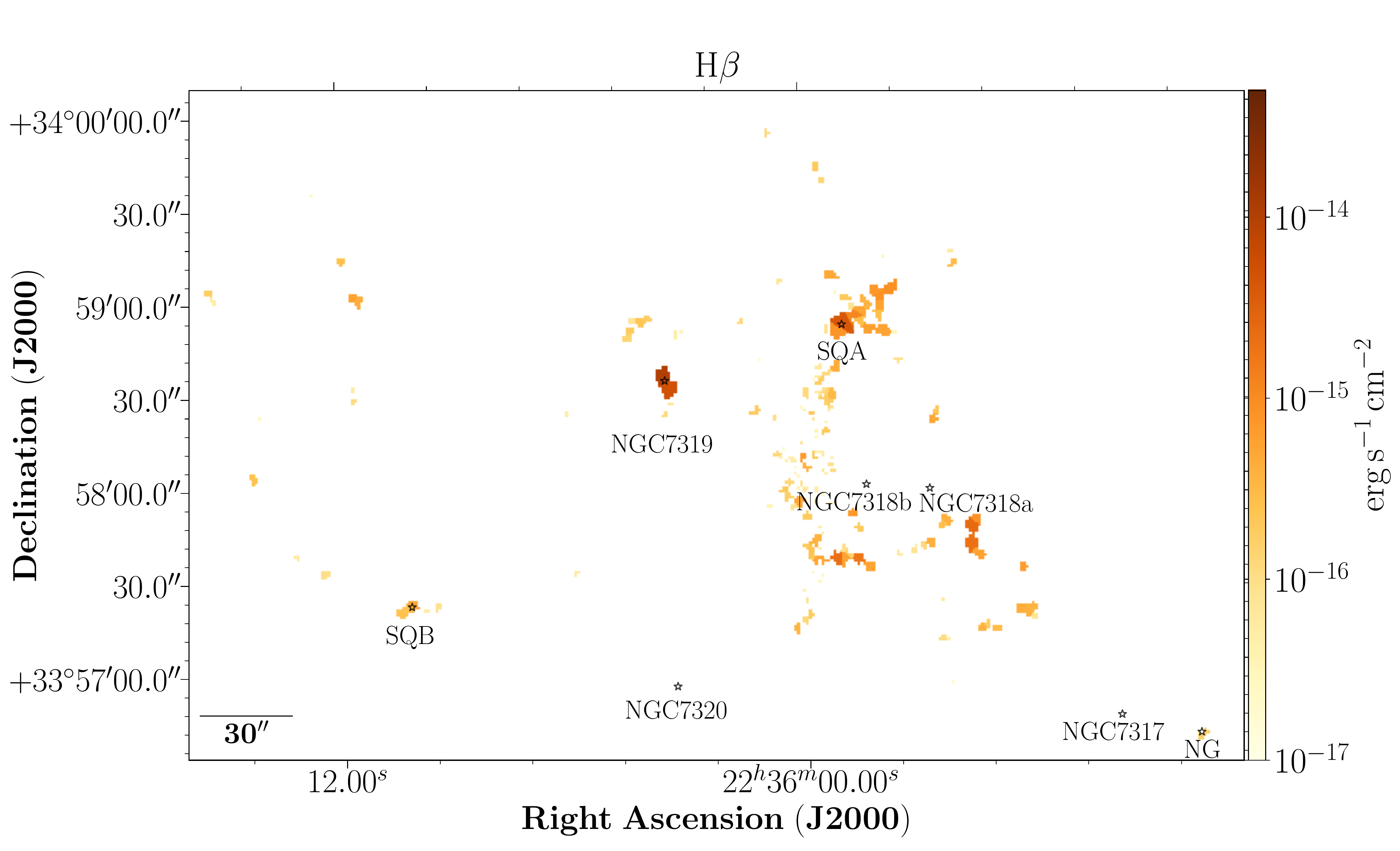}\\
    \includegraphics[width=.7\textwidth]{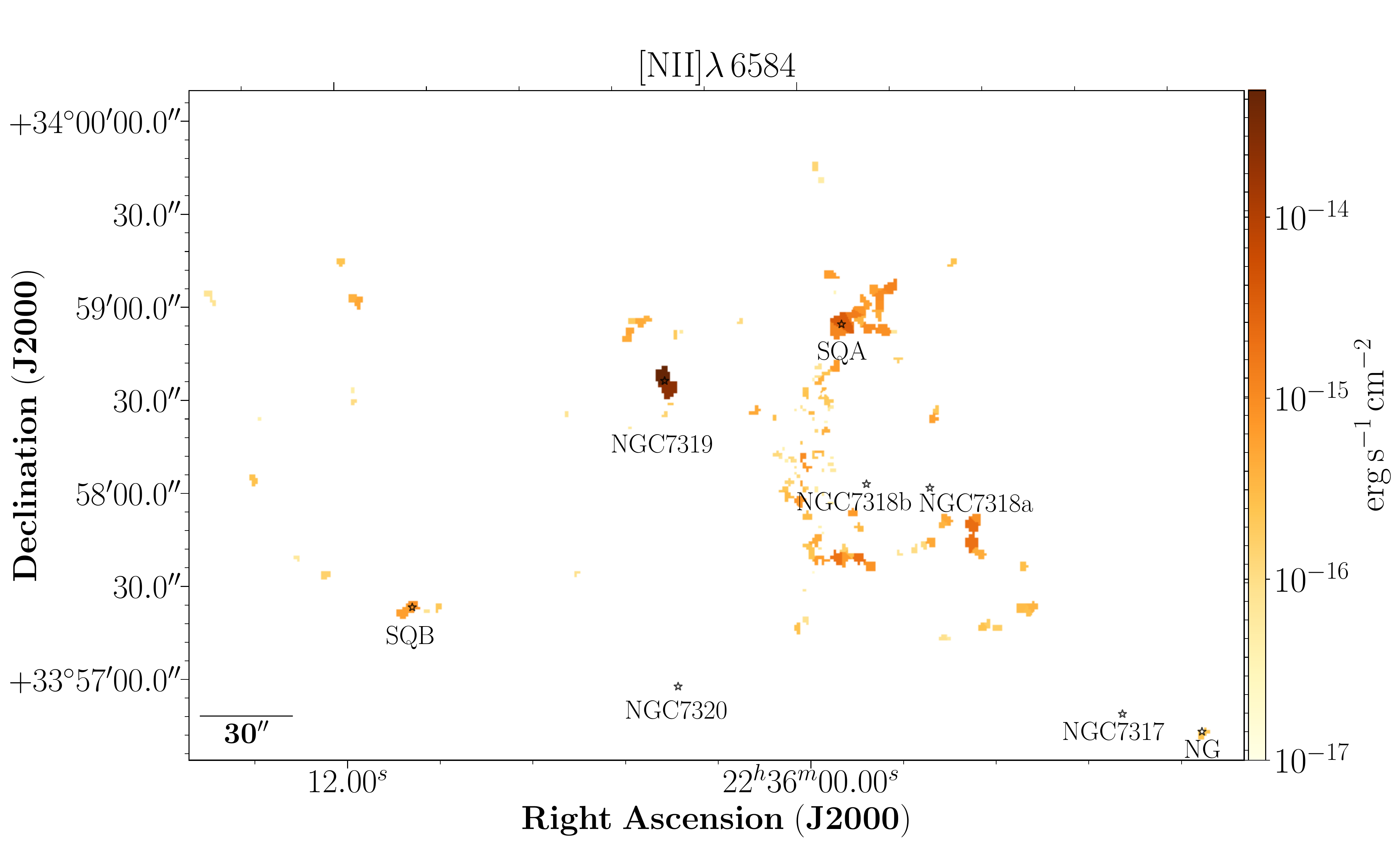}
\caption{Emission-line flux maps of SQ. From top to bottom: H$\alpha$, H$\beta$, [\ion{N}{ii}], [\ion{O}{iii}], and [\ion{O}{ii}].}
\label{fig:fluxes}
\end{figure*}
\addtocounter{figure}{-1}

\begin{figure*}
    \centering
    \includegraphics[width=.8\textwidth]{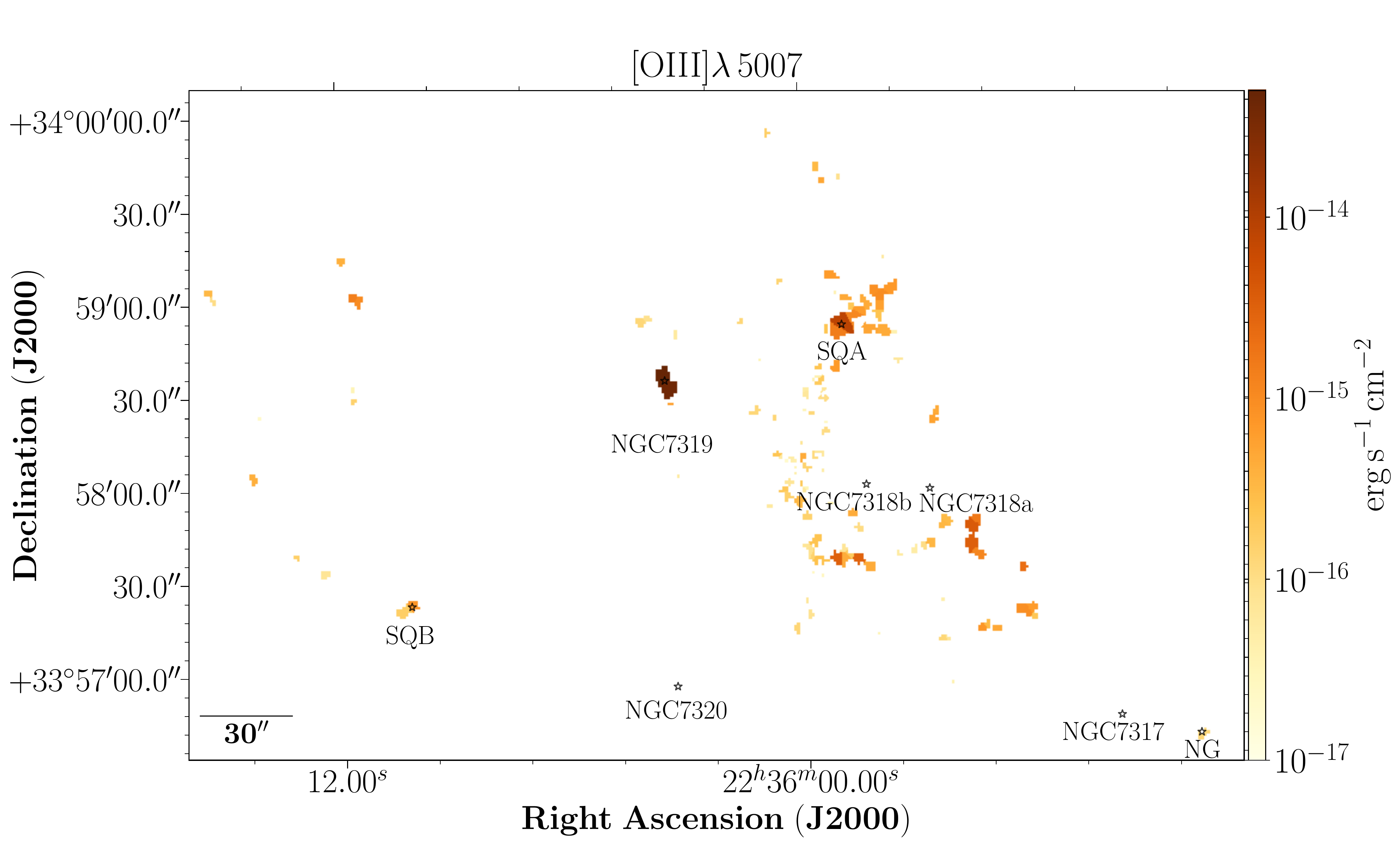}\\
    \includegraphics[width=.8\textwidth]{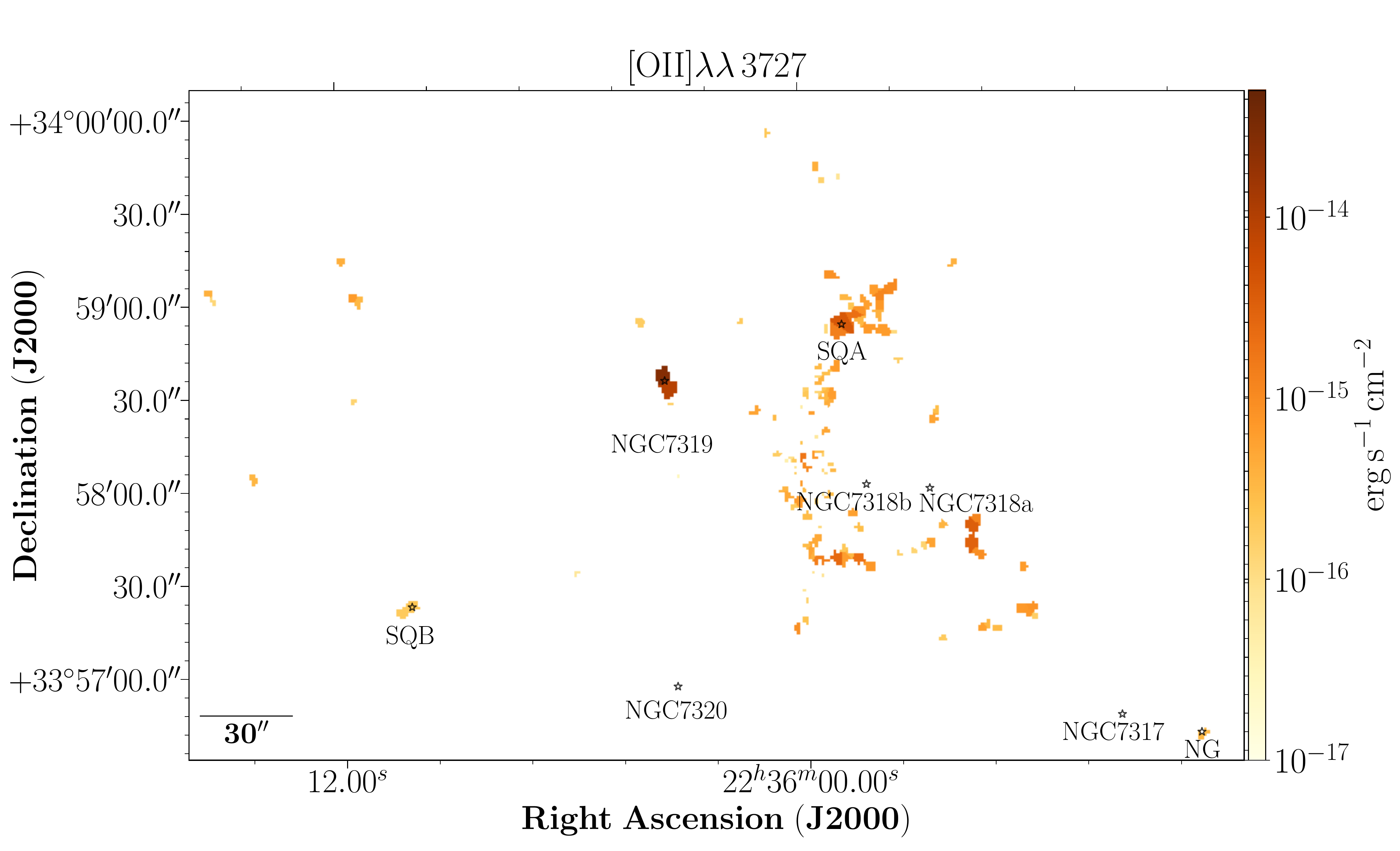}\\
\caption{(continued)}
\end{figure*}

\begin{landscape}
\begin{figure}
    \centering
    \includegraphics[width=\columnwidth]{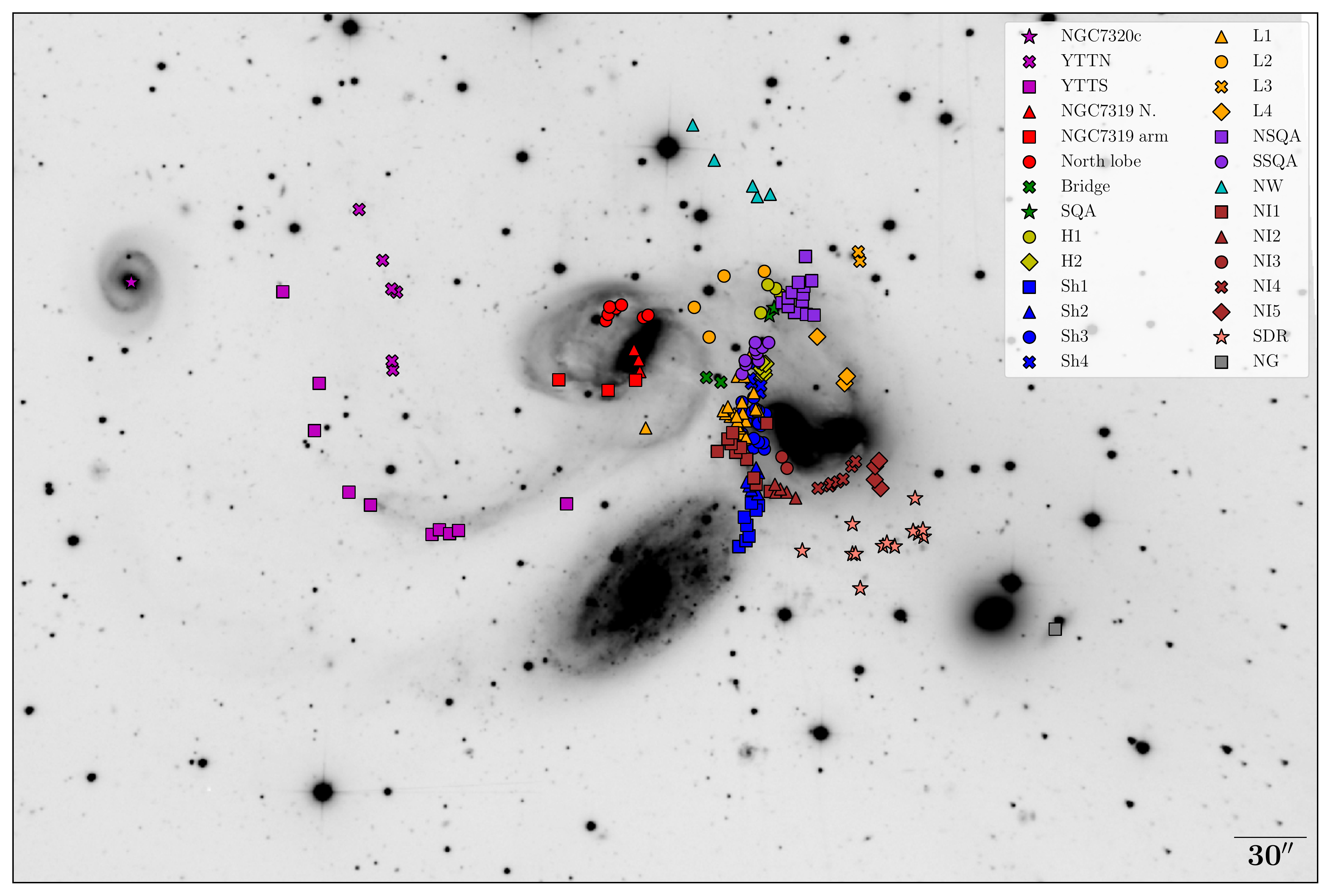}
    \caption{Different systems of emission line objects defined here and indicated on SITELLE deep-greyscale image of SQ. The figure shows are the positions of YTTN and YTTS (magenta crosses and squares, respectively); NGC7319 nucleus, `arm', and north lobe (red triangles, squares, and circles, respectively); bridge (green crosses); SQA (green stars); Hs (H1: yellow circles; H2: yellow diamonds); Shs (Sh1: blue squares; Sh2: blue triangles; Sh3: blue circles; Sh4: blue crosses); Ls (L1: orange triangles; L2: orange circles; L3: orange crosses; L4: orange diamonds); NSQA and SSQA (violet squares and circles); NW (cyan triangles); NIs (NI1: brown squares; NI2: brown triangles; NI3: brown circles; NI4: brown crosses; NI5: brown diamonds); SDR (salmon stars); and NG (grey squares).}
    \label{fig:regions}
\end{figure}
\end{landscape}

In Fig.~\ref{fig:regions} we show the 12 zones and 28 sub-zones defined in paper I for SQ as follows: i) young tidal tail (YTT) \citep[e.g.][]{2002A&A...394..823L}, so that north and south strands are respectively YTTN and YTTS, and NGC7320c is the old intruder (OI); ii) NGC7319 (NGC7319 nucleus, NGC7319 `arm', north lobe); iii) H$\alpha$ `bridge'; iv) high radial velocity strands, Hs (H1 and H2); v) SQA \citep[e.g.][]{1999ApJ...512..178X}; vi) low radial velocity strands, Ls (L1, L2, L3, and L4); vii) shock strands, Shs (Sh1, Sh2, Sh3, and Sh4); viii) north and south of SQA (NSQA and SSQA, respectively); ix) tidal tail at north of NSQA \citep[NW, e.g.][]{2010ApJ...724...80R}; x) NI \citep[e.g.][]{1997ApJ...485L..69M} strands, NIs (NI1, NI2, NI3, NI4, and NI5); xi) southern debris region \citep[SDR, e.g.][]{2011AJ....142...42F}; and xii) the new dwarf galaxy (NG). In Appendix~\ref{append:app1} we detail all the zones defined here.


\subsection{Distribution of emission line ratios}
\label{subsec:linerat}
In this section we study the spatial distribution in the SQ of several emission-line ratios considered in the literature as proxies of oxygen abundance indicators (e.g. [\ion{N}{ii}]/H$\alpha$, R$_{23}$, O3N2) and of the ionisation degree of the regions (e.g. [\ion{O}{iii}]/[\ion{O}{ii}]). Some of them were not corrected for reddening given the proximity in wavelength of the emission lines involved (e.g. [\ion{N}{ii}]/H$\alpha$ or [\ion{O}{iii}]/H$\beta$). On the contrary, R$_{23}$, [\ion{O}{iii}]/[\ion{O}{ii}], and [\ion{N}{ii}]/[\ion{O}{ii}] need to be corrected for reddening according to Sect.~\ref{subsec:line}. We calculated the values of R$_{23}={{[\ion{O}{ii}]\lambda3727+[\ion{O}{iii}]\lambda\lambda 4959,5007}\over{H\beta}}$ \citep{1979MNRAS.189...95P}, [\ion{O}{iii}]/[\ion{O}{ii}]=${{[\ion{O}{iii}]\lambda\lambda 4959,5007}\over{[\ion{O}{ii}]\lambda 3727}}$ \citep{2000MNRAS.318..462D}, and O3N2=${{[\ion{O}{iii}]\lambda 5007/H\beta}\over{[\ion{N}{ii}]\lambda 6584/H\alpha}}$ \citep{1979A&A....78..200A} for the H$\alpha$ regions.

\begin{figure}
    \centering
    \includegraphics[width=\columnwidth]{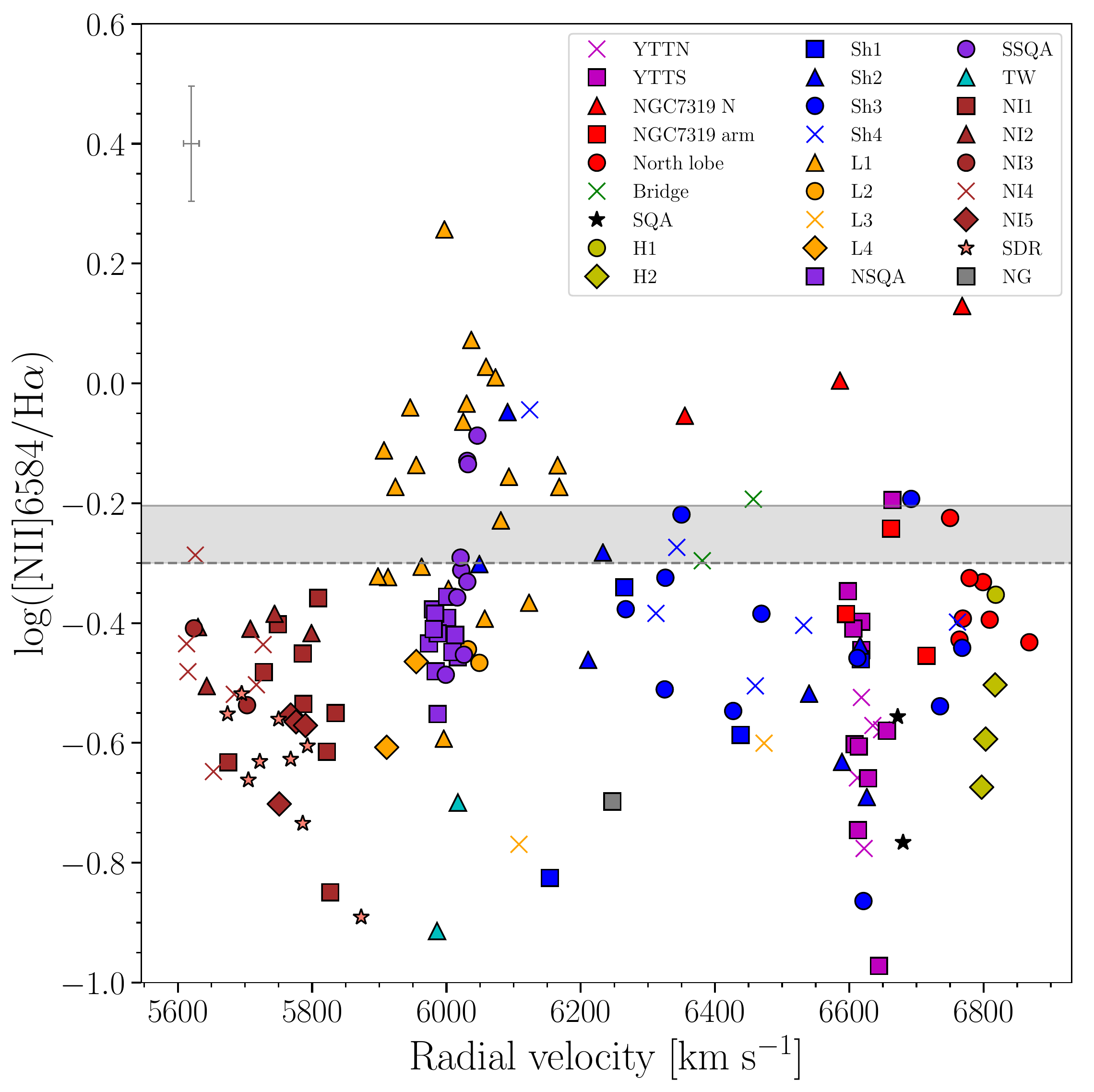}
    \caption{Log([\ion{N}{ii}]/H$\alpha$) versus radial velocity diagram. All the points in the figure have the same colours and markers as Fig.~\ref{fig:regions}. The horizontal dashed grey line and the grey band correspond to the reference value at log([\ion{N}{ii}]/H$\alpha$)=-0.3 and the uncertainties for the log([\ion{N}{ii}]/H$\alpha$) ratio, respectively. The upper left cross indicates the typical error of both parameters.}
    \label{fig:NIIHa}
\end{figure}

Figure~\ref{fig:NIIHa} shows the relation between log([\ion{N}{ii}]/H$\alpha$) and radial velocity for all the regions in SQ. In the figure, the horizontal dashed grey line and the grey band correspond to the reference value at log([\ion{N}{ii}]/H$\alpha$)=-0.3 and the uncertainties for the log([\ion{N}{ii}]/H$\alpha$) ratio, respectively. Regions with log([\ion{N}{ii}]/H$\alpha$) $\leq$ -0.2 are assumed to be HII-like regions, while those with log([\ion{N}{ii}]/H$\alpha$) $>$ -0.2 are related to shocked gas or active galactic nuclei (AGN, see Sect.~\ref{subsec:extinction}). For the LV sub-sample, practically all the regions in NIs, SDR, NSQA, L2, L3, L4, and NW have values lower than -0.2 for log([\ion{N}{ii}]/H$\alpha$). In contrast, L1 and SSQA show values higher than -0.2 for log([\ion{N}{ii}]/H$\alpha$). In the HV sub-sample, most of the regions have values lower than -0.2 for log([\ion{N}{ii}]/H$\alpha$), except the NGC7319 nucleus, H$\alpha$ `bridge', and several regions from Shs, north lobe, and the NGC7319 `arm'.

Figure~\ref{fig:ratio} shows the line ratio maps in SQ considering the emission line maps from Fig.~\ref{fig:fluxes} (from top to bottom: log([\ion{N}{ii}]/[\ion{O}{ii}]), log([\ion{N}{ii}]/H$\alpha$), log([\ion{O}{iii}]/H$\beta$), log(O3N2), log([\ion{O}{iii}]/[\ion{O}{ii}]), and log(R$_{23}$)). The values for the line ratios for SDR and NW are compatible with those of regions detected in the outer discs of galaxies since they have low [\ion{N}{ii}]/H$\alpha$ values, and high [\ion{O}{iii}]/H$\beta$, O3N2, and [\ion{O}{iii}]/[\ion{O}{ii}] values \citep[see][for more information about inner and outer HII regions over the discs of spiral galaxies]{2018A&A...609A.102R}. The values derived for SQA are compatible with those of the outer regions of the galaxies, and could also match typical values for a tidal dwarf galaxy. YYTS has an inner-outer gradient presenting lower [\ion{N}{ii}]/H$\alpha$ and higher values for [\ion{O}{iii}]/H$\beta$ and O3N2 in the west than in the east. The LSSR presents high [\ion{N}{ii}]/[\ion{O}{ii}] values, and low [\ion{O}{iii}]/H$\beta$, O3N2, and [\ion{O}{iii}]/[\ion{O}{ii}] values. With these results we can see how the regions from the LSSR have values of emission-line ratios that are completely different from the rest of the SQ zones studied.

\begin{figure*}[h!]
    \centering
    \includegraphics[width=.65\textwidth]{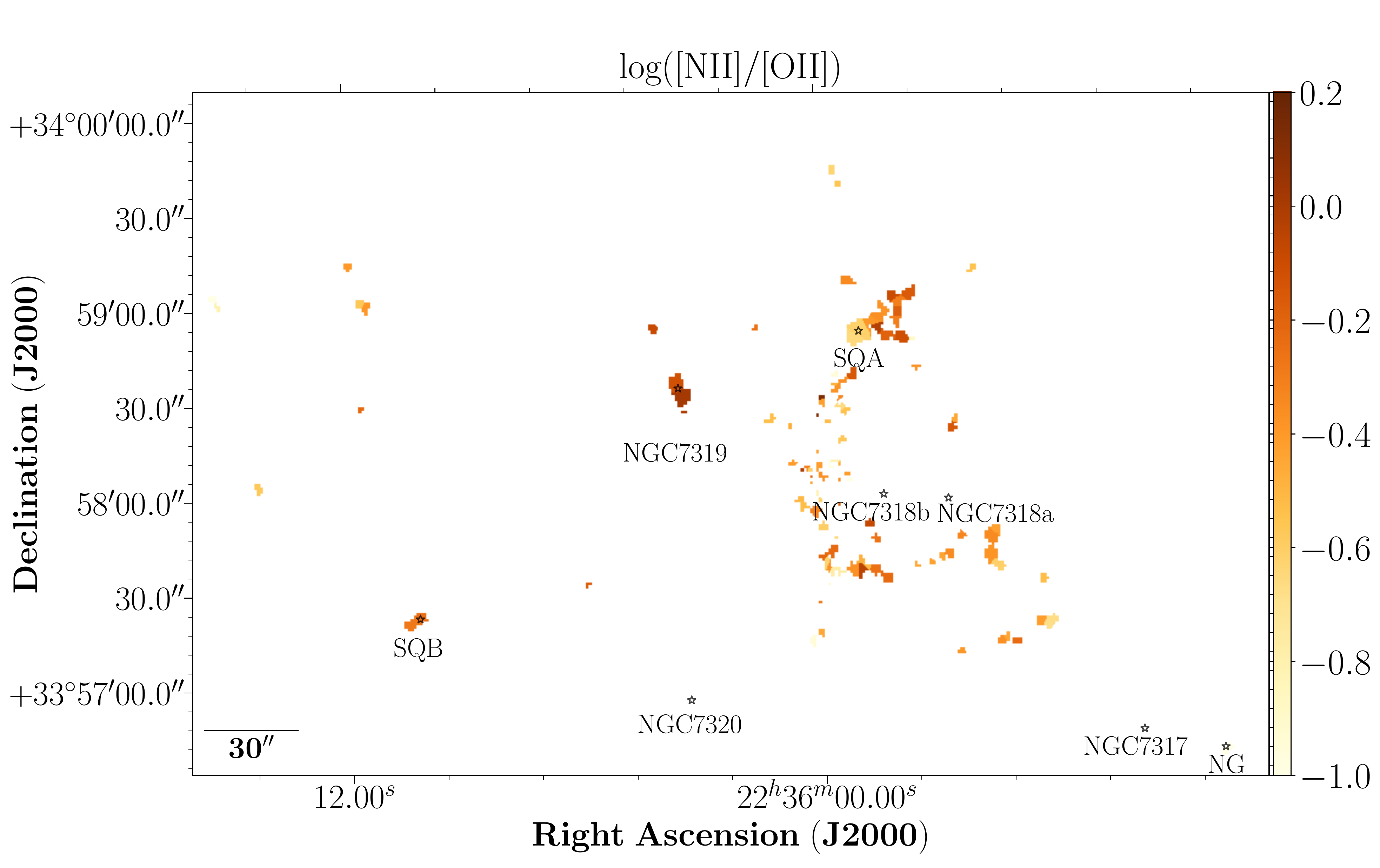}\\
    \includegraphics[width=.65\textwidth]{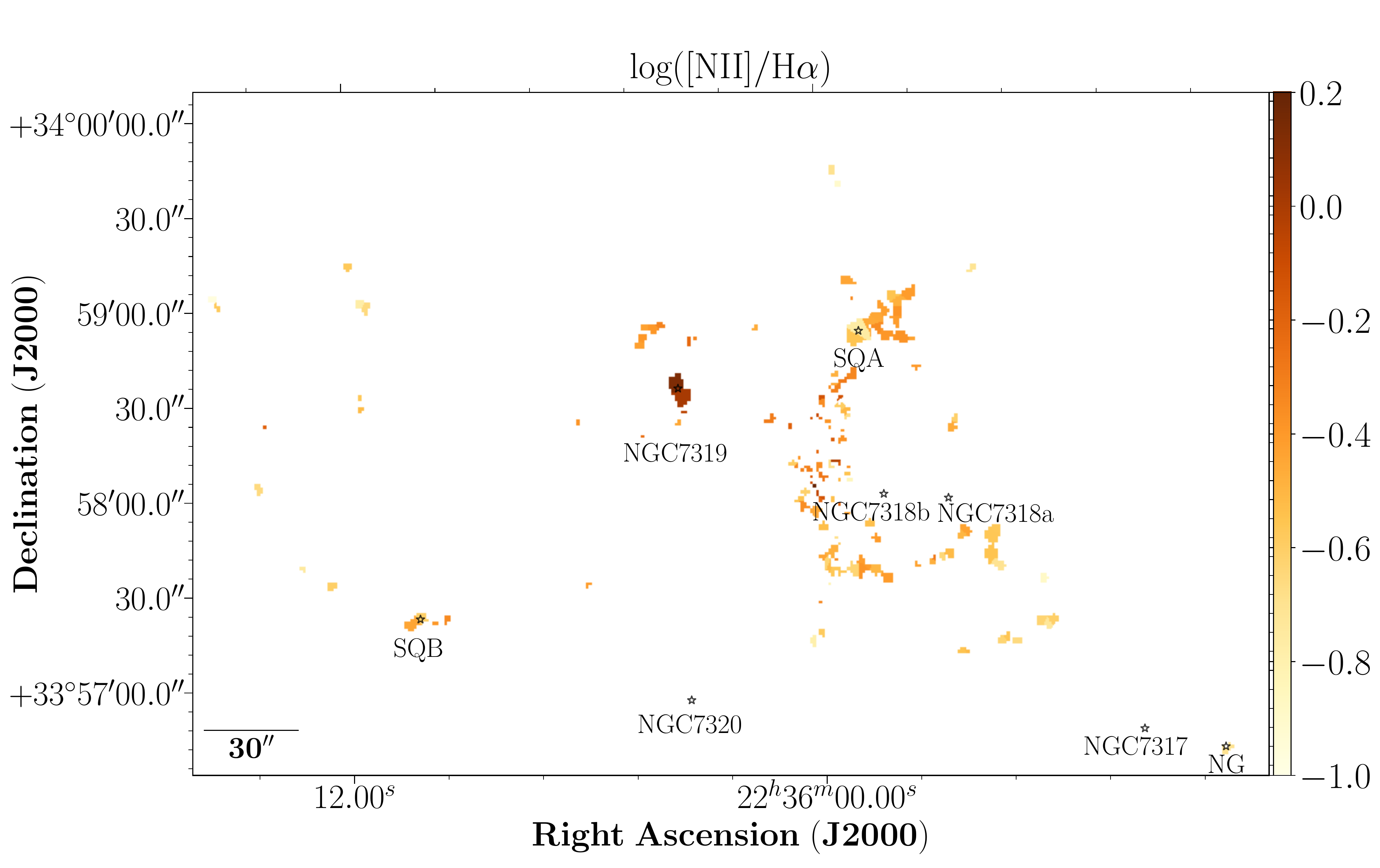}\\ 
    \includegraphics[width=.65\textwidth]{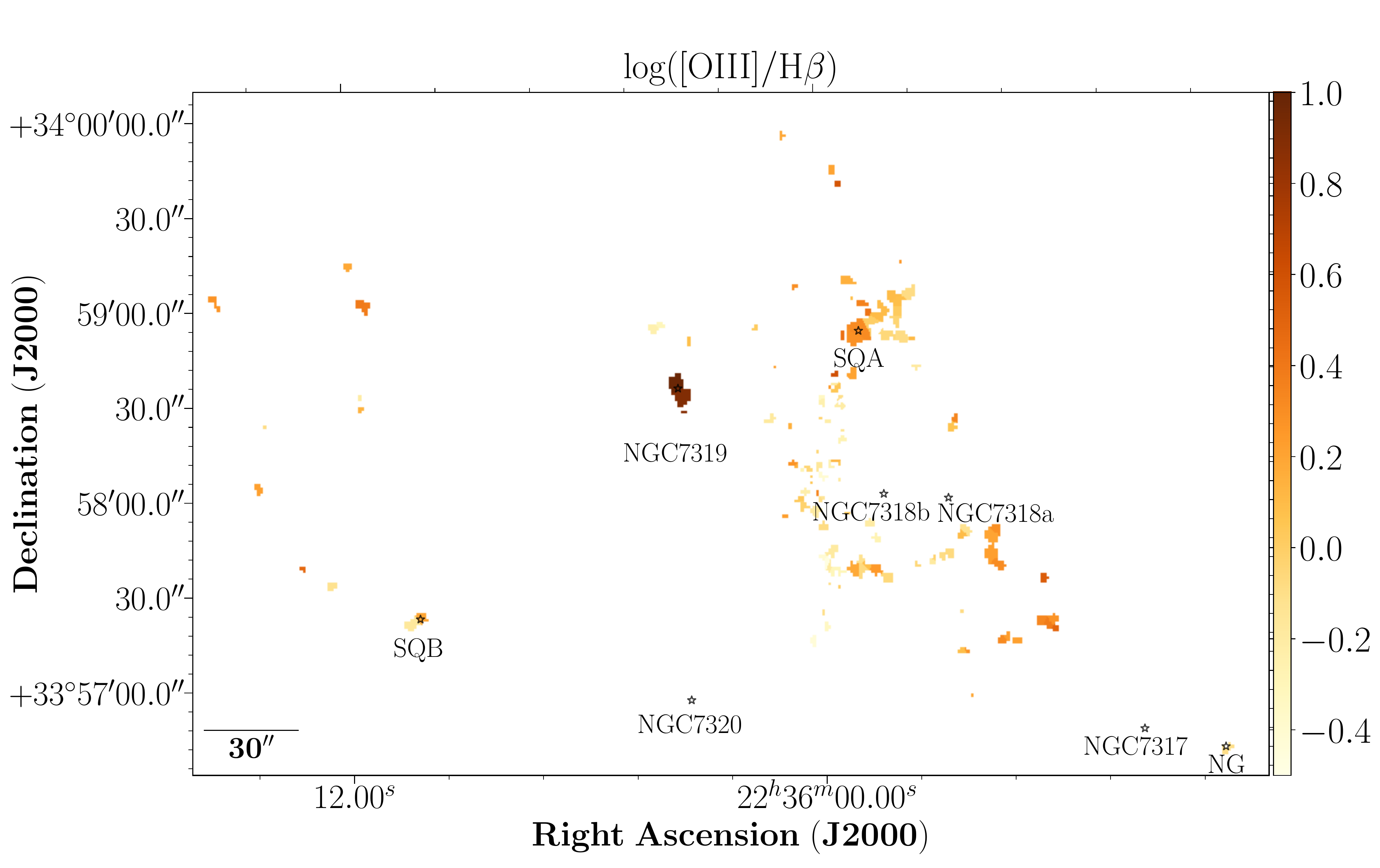}\\ 
\caption{Line ratios maps. From top to bottom: [\ion{N}{ii}]$\lambda$6584/[\ion{O}{ii}]$\lambda$3727, [\ion{N}{ii}]$\lambda$6584/H$\alpha$, [\ion{O}{iii}]$\lambda$5007/H$\beta$, O3N2, [\ion{O}{iii}]$\lambda$5007/[\ion{O}{ii}]$\lambda$3727, R$_{23}$.}
\label{fig:ratio}
\end{figure*}
\addtocounter{figure}{-1}

\begin{figure*}
    \centering
    \includegraphics[width=.65\textwidth]{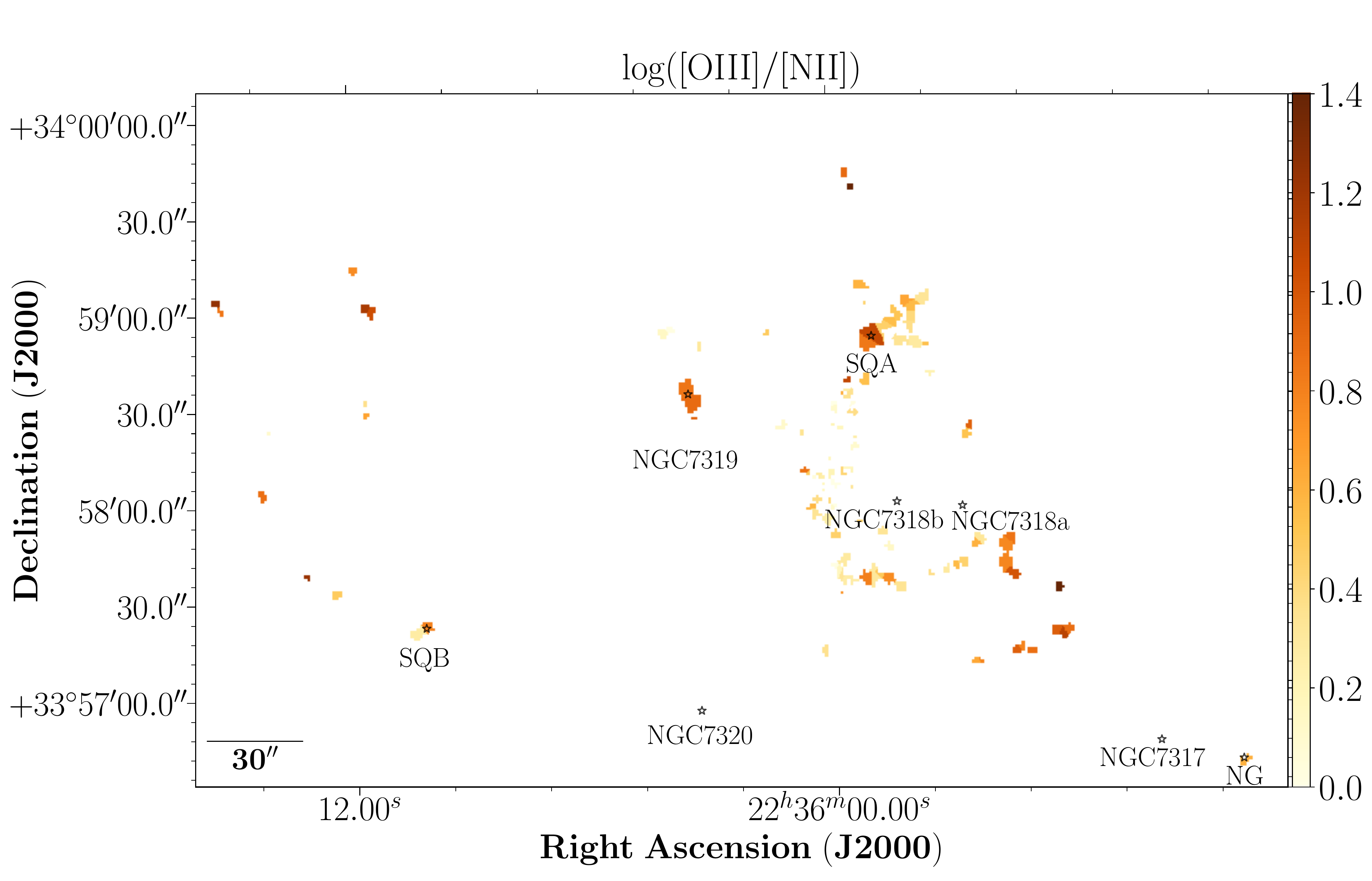}\\
    \includegraphics[width=.65\textwidth]{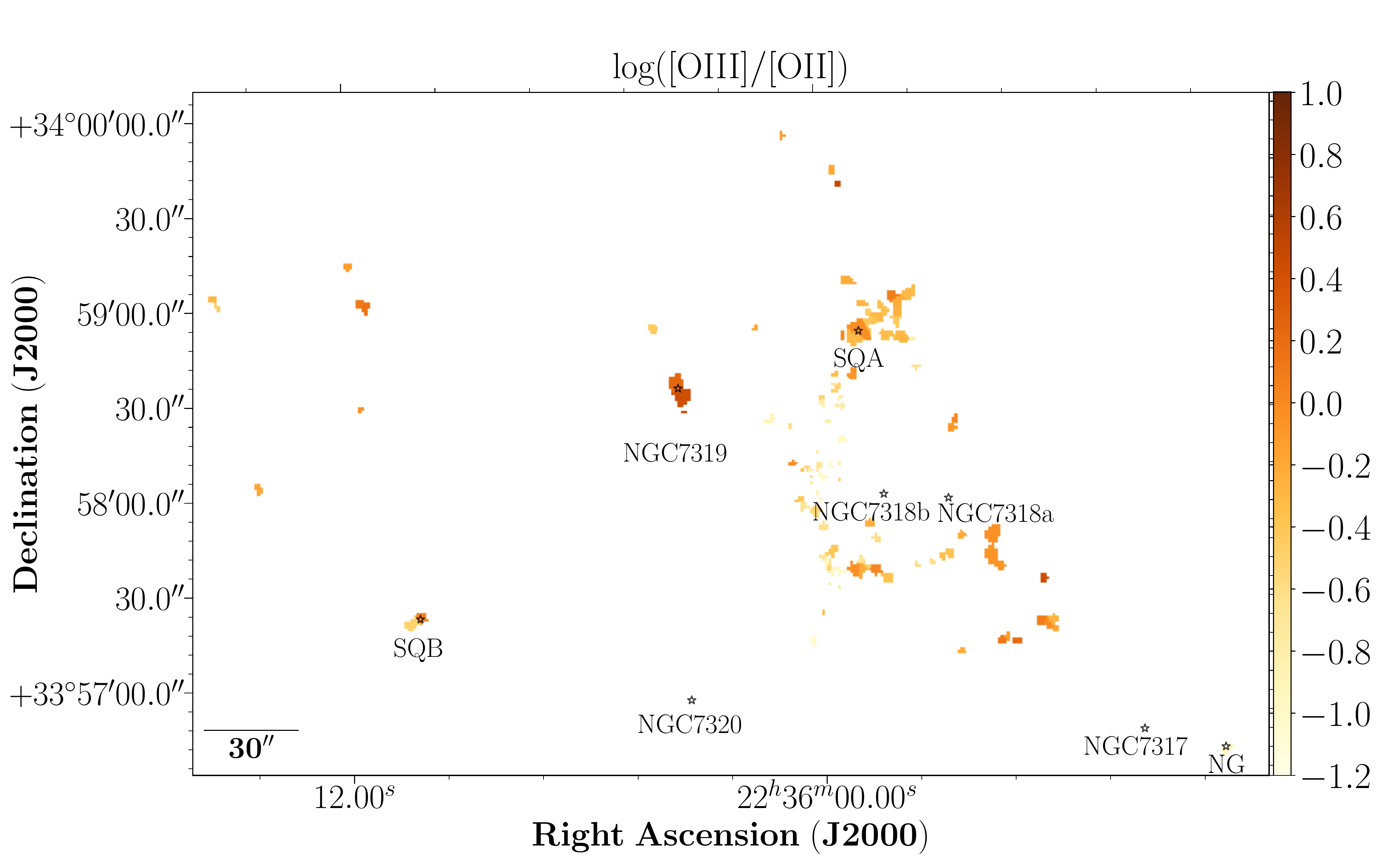}\\
    \includegraphics[width=.65\textwidth]{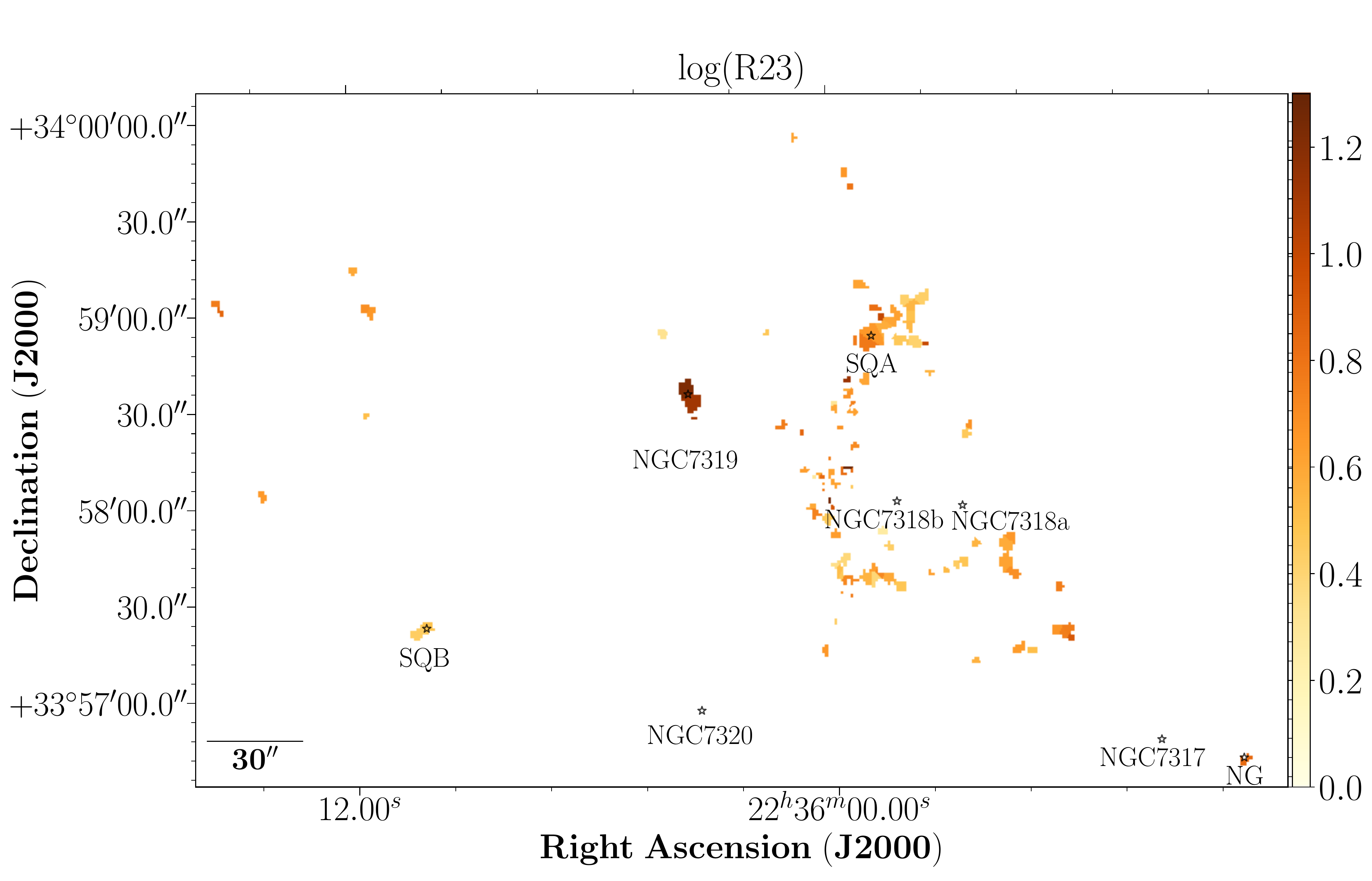}\\ 
\caption{(continued)}
\end{figure*}

\begin{figure*}[h!]
    \centering
    \includegraphics[width=.8\textwidth]{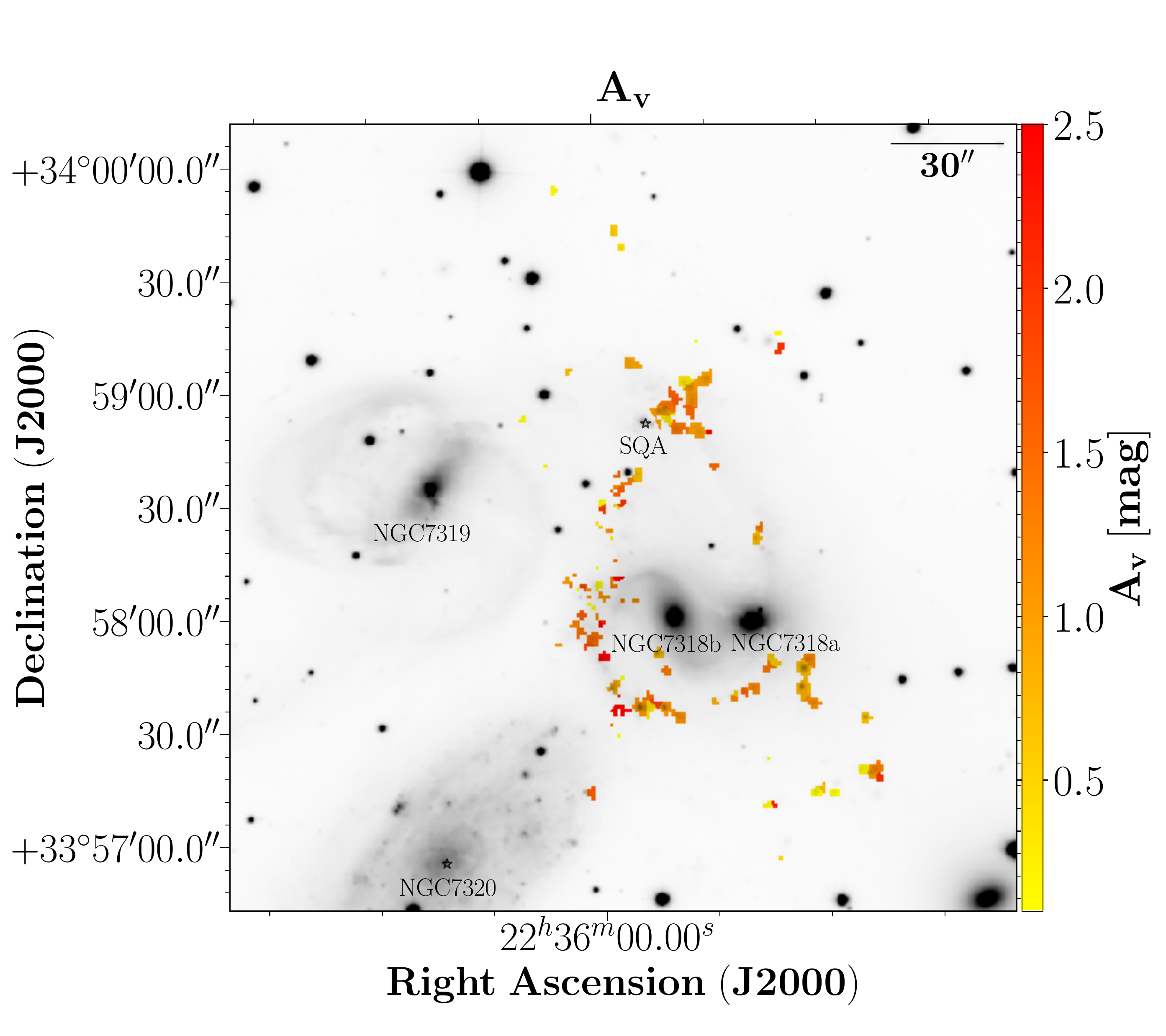}\\ \vspace{-0.5cm}
    \includegraphics[width=.9\textwidth]{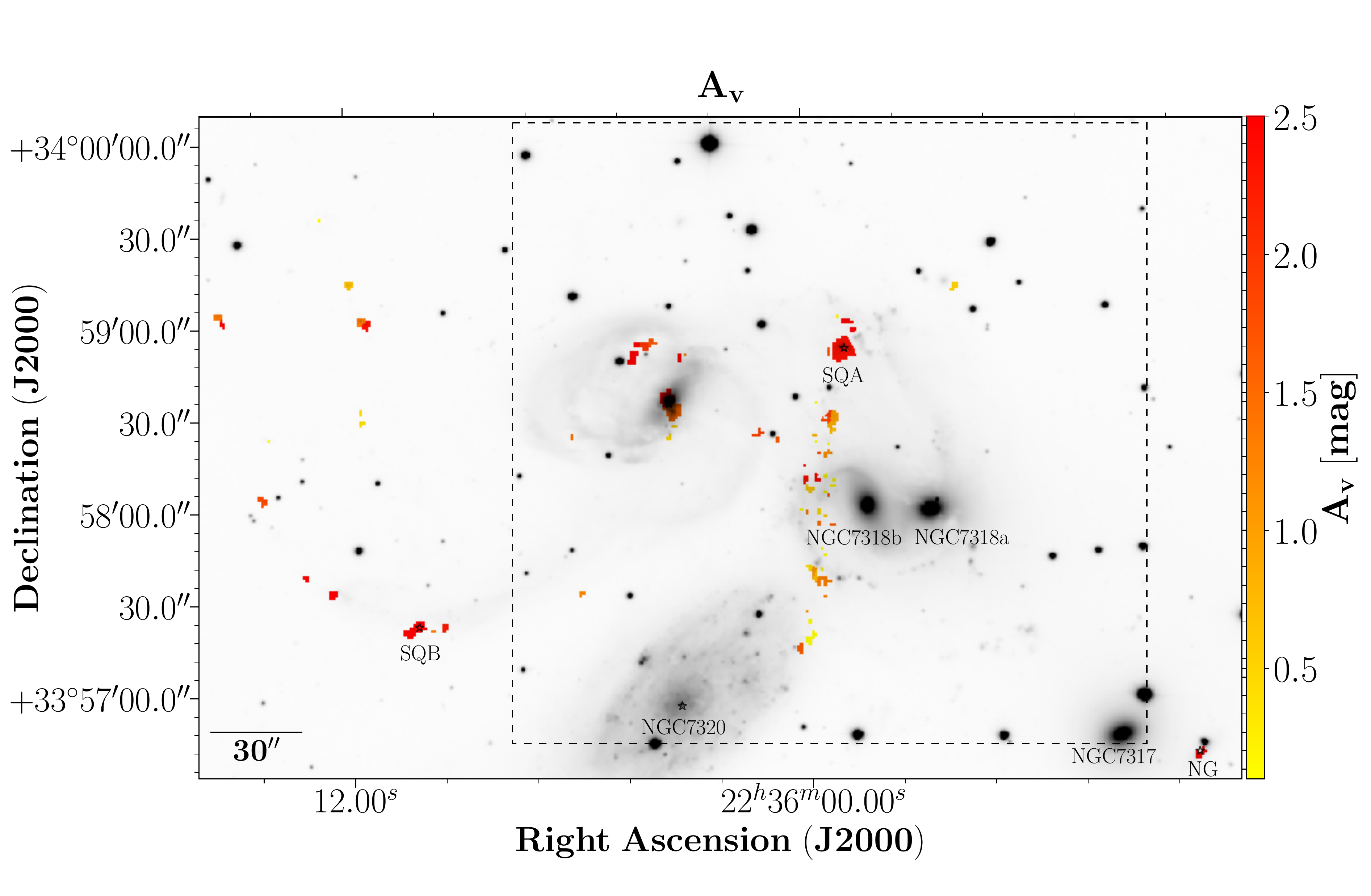}
    \caption{Stephan's Quintet spatial map colour coded according to their A$_v$ extinction for the LV sub-sample (upper panel) and the HV sub-sample (lower panel).}
    \label{fig:Av} 
\end{figure*}
\clearpage

\subsection{Maps of extinction and excitation}
\label{subsec:extinction}

In Fig.~\ref{fig:Av} we show the extinction map in magnitudes in the V band, A$_v$, for the emission regions found in SQ. We separate the maps into LV (upper left panel) and HV (upper right panel) sub-samples according to the definition given in Sect.~\ref{subsec:line}. The extinction A$_v$ derived in this work goes from 0 to 4 mag. The median is A$_v$ = 1.1 mag, in agreement with the value obtained in \cite{2018A&A...609A.102R} for a sample of star-forming regions over the discs of spiral galaxies. On average, the regions found in LV show lower values of A$_v$ ($\sim$1 mag) than those located in HV ($\sim$1.3 mag), in line with previous works \citep[e.g.][]{2014ApJ...784....1K}. In the LV extinction map we can see several interesting distributions. SDR presents A$_v$ values $\sim$0.5 mag. Moreover, in the north and south of the strong HII region SQA (NSQA and SSQA, respectively) a mixture of A$_v$ values are found. Generally, SSQA has higher A$_v$ values than NSQA (SSQA: $\sim$1.3; NSQA: $\sim$1). Region 66 has a A$_v\sim$2.9 (from L1), being the region with the greatest value in the LV map. On the other hand, SQA has an extinction value of A$_v\sim$2.3 mag. In the AGN north lobe, we can see A$_v>$1.2 mag, where the greatest value is seen in the region 20 (A$_v\sim$4.2 mag, see Table~\ref{table:table3}). NG also has high extinction (on average A$_v\sim$3.5 mag). YTTS presents higher extinction values, on average, than in the north regions, where the extinction values are lower than 2.5 mag. Conversely, in the LSSR zone, a mixture of extinction values are found (with a median value of A$_v\sim$1 mag). It is important to note that at the centre of the LSSR zone, several regions with A$_v>$2 mag (e.g. region 83 and 60) are present in both the LV and HV maps.

\begin{figure}
    \centering
    \includegraphics[width=\columnwidth]{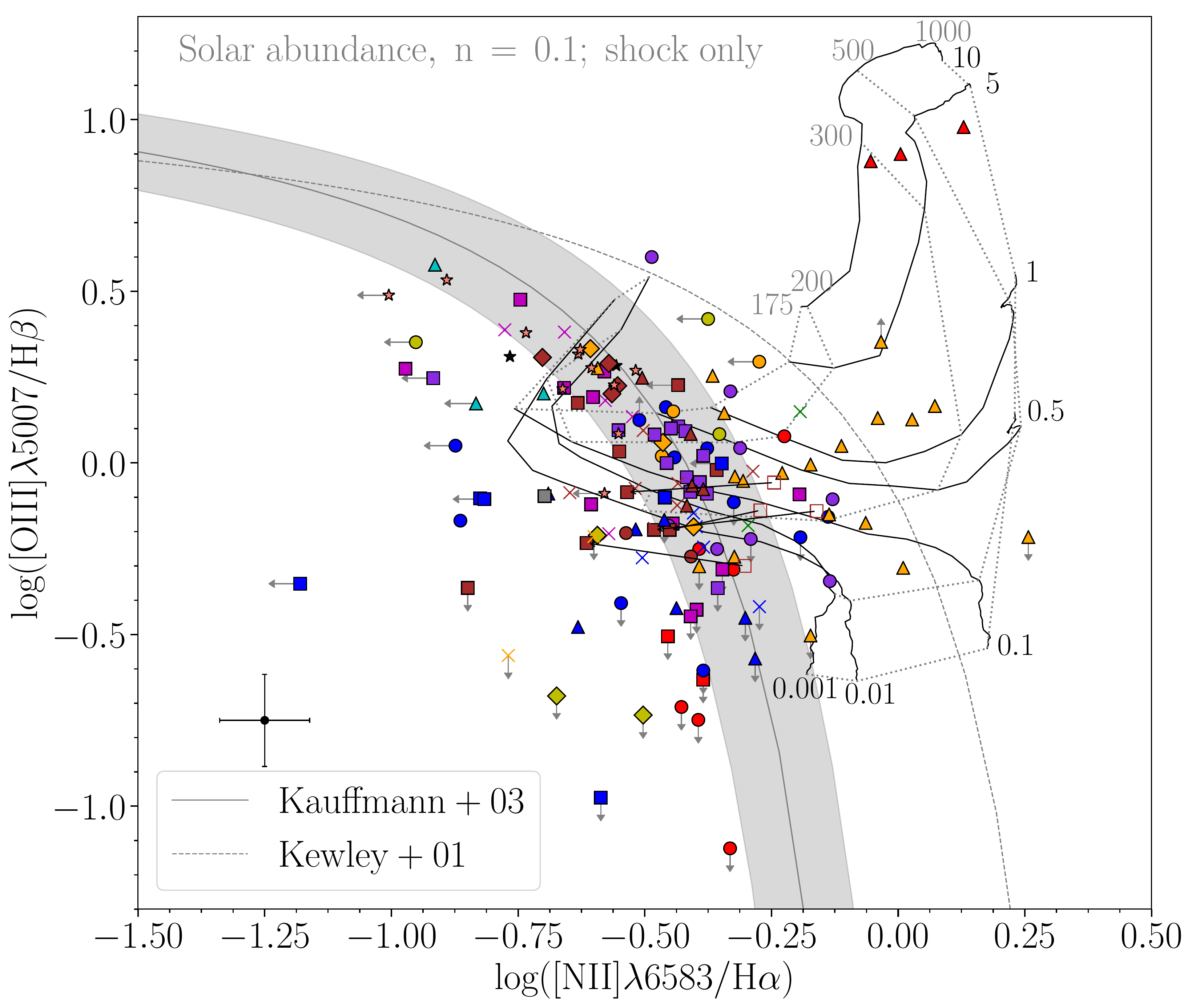}
    \caption{Diagnostic diagram (BPT) of [\ion{O}{iii}]$\lambda$5007/H$\beta$ versus [\ion{N}{ii}]$\lambda$6584/H$\alpha$ for the SQ H$\alpha$ emission regions. Regions without an arrow have a S/N higher than 3 for the fluxes in the strong emission lines H$\alpha$, H$\beta$, [\ion{O}{iii}]$\lambda$5007, and [\ion{N}{ii}]$\lambda$6584. The typical error in each axis is represent with a black cross. Regions with an ascending grey arrow have a S/N(H$\beta$)$<$3, regions with a decreasing grey arrow have a S/N([\ion{O}{iii}]$\lambda$5007)$<$3, while regions with a grey arrow pointing left have a S/N([\ion{N}{ii}]$\lambda$6584)$<$3. All the points in the figure have the same colours and markers as Fig.~\ref{fig:regions}. The grey dashed line shows the \cite{2001ApJ...556..121K} demarcation and the grey continuous line shows the \cite{2003MNRAS.346.1055K} curve. The grey band shows the uncertainties for the BPT line ratios (i.e. [\ion{O}{iii}]$\lambda$5007/H$\beta$ and [\ion{N}{ii}]$\lambda$6584/H$\alpha$) to the Kauffmann demarcation. The black lines correspond to the shock only models of \cite{2008ApJS..178...20A} for solar metallicity and low density (n = 0.1 cm$^{-3}$), while the dotted lines correspond to the shock velocity.}
    \label{fig:bpt_sq}
\end{figure}

In this work, we are interested in studying the properties of the SQ HII regions. In order to produce a classification of the emitting regions, we study the diagnostic diagram \citep[][hereafter BPT]{1981PASP...93....5B,2001ApJ...556..121K,2003MNRAS.346.1055K}. In Fig.~\ref{fig:bpt_sq} we show the BPT diagram (log([\ion{O}{iii}]$\lambda$5007/H$\beta$) versus log([\ion{N}{ii}]$\lambda$6584/H$\alpha$)) for the SQ regions. The predicted models from \cite{2008ApJS..178...20A} are shown for the ionisation of gas considering fast shocks without a precursor for solar metallicity and low density (n = 0.1 cm$^{-3}$), with velocities between 175 and 1000 $km\, s^{-1}$. The grey band in Fig.~\ref{fig:bpt_sq} corresponds to the mean uncertainties measured for the [\ion{O}{iii}]$\lambda$5007/H$\beta$ and [\ion{N}{ii}]$\lambda$6584/H$\alpha$ line ratios. As expected, the regions from the NGC7319 nucleus appear in the AGN zone in the BPT diagram. The remaining SQ regions are displayed within the star-forming and composite (C) zones. We found 91 star-forming regions, 17 C regions, and 7 AGN-like regions. \cite{2014MNRAS.442..495R} found a fraction of pixels showing emission typical of the composite zone of the BPT diagram and claiming that HII regions can be found in the C zone \citep[see also][]{2014A&A...563A..49S}. We believe that these C regions are contaminated by the emission of the shocked regions located in its vicinity.\footnote{We note that the [\ion{N}{ii}]$\lambda$6584 emission line is more sensitive to shocks and non-thermal processes than the H$\alpha$ emission line \citep{2001AJ....122.2993S}.} Here, the SQ regions are classified as HII regions whenever they fulfil the two following conditions: i) they are located in the star-forming (SF) zone according to the Kauffmann demarcation or in the plotted grey band from Fig.~\ref{fig:bpt_sq}; and ii) where the S/N for the H$\alpha$, H$\beta$, [\ion{O}{iii}]$\lambda$5007, and [\ion{N}{ii}]$\lambda$6584 emission fluxes are higher than or equal to 3. AGN-like and composite regions are shown in several diagrams and maps. No values of SFR, nitrogen-to-oxygen abundance ratio (N/O) or oxygen abundance have been derived for composite or AGN-like regions. 

\begin{figure*}
    \centering
    \includegraphics[width=.88\textwidth]{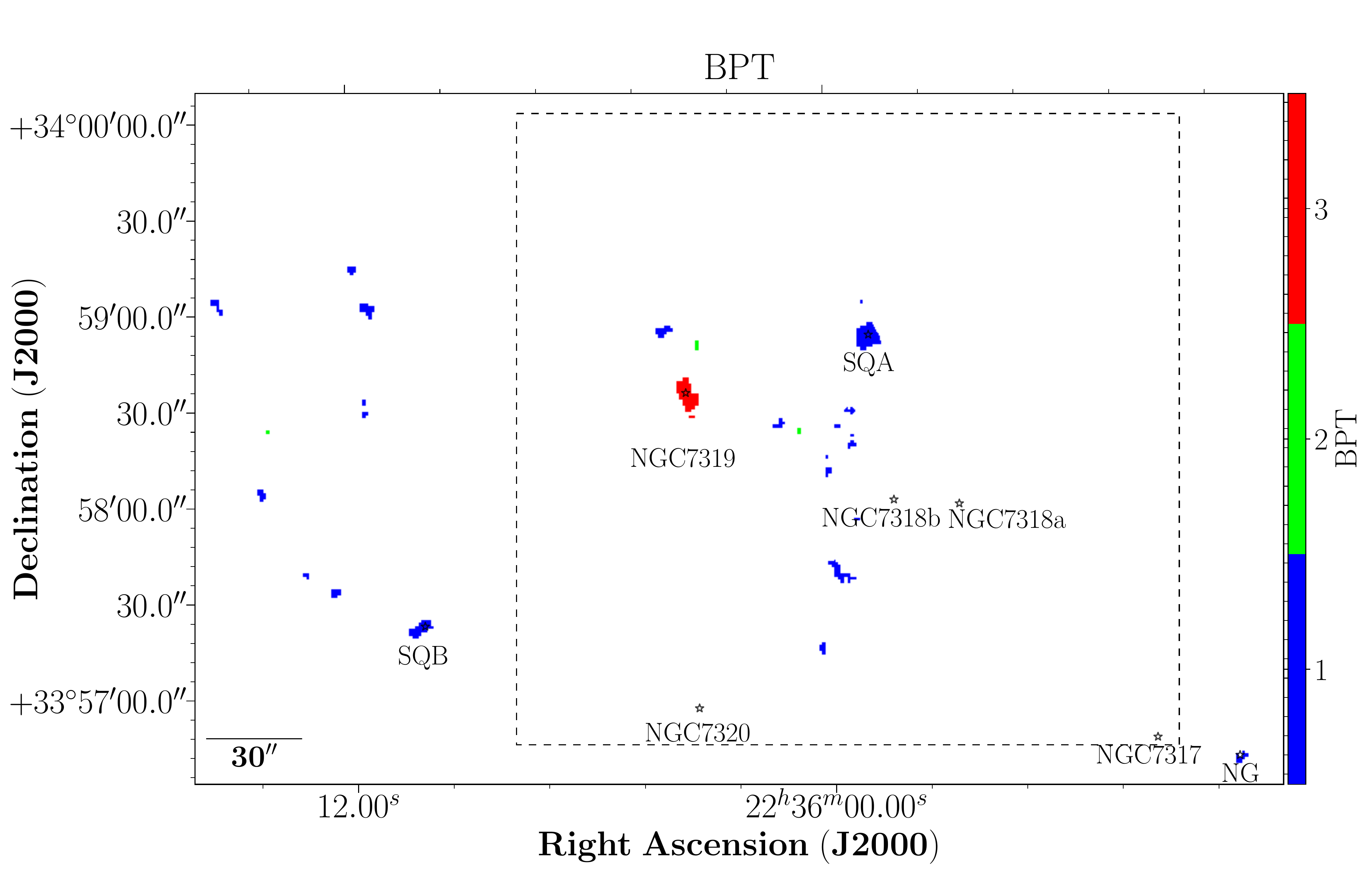}\
    \includegraphics[width=.49\textwidth]{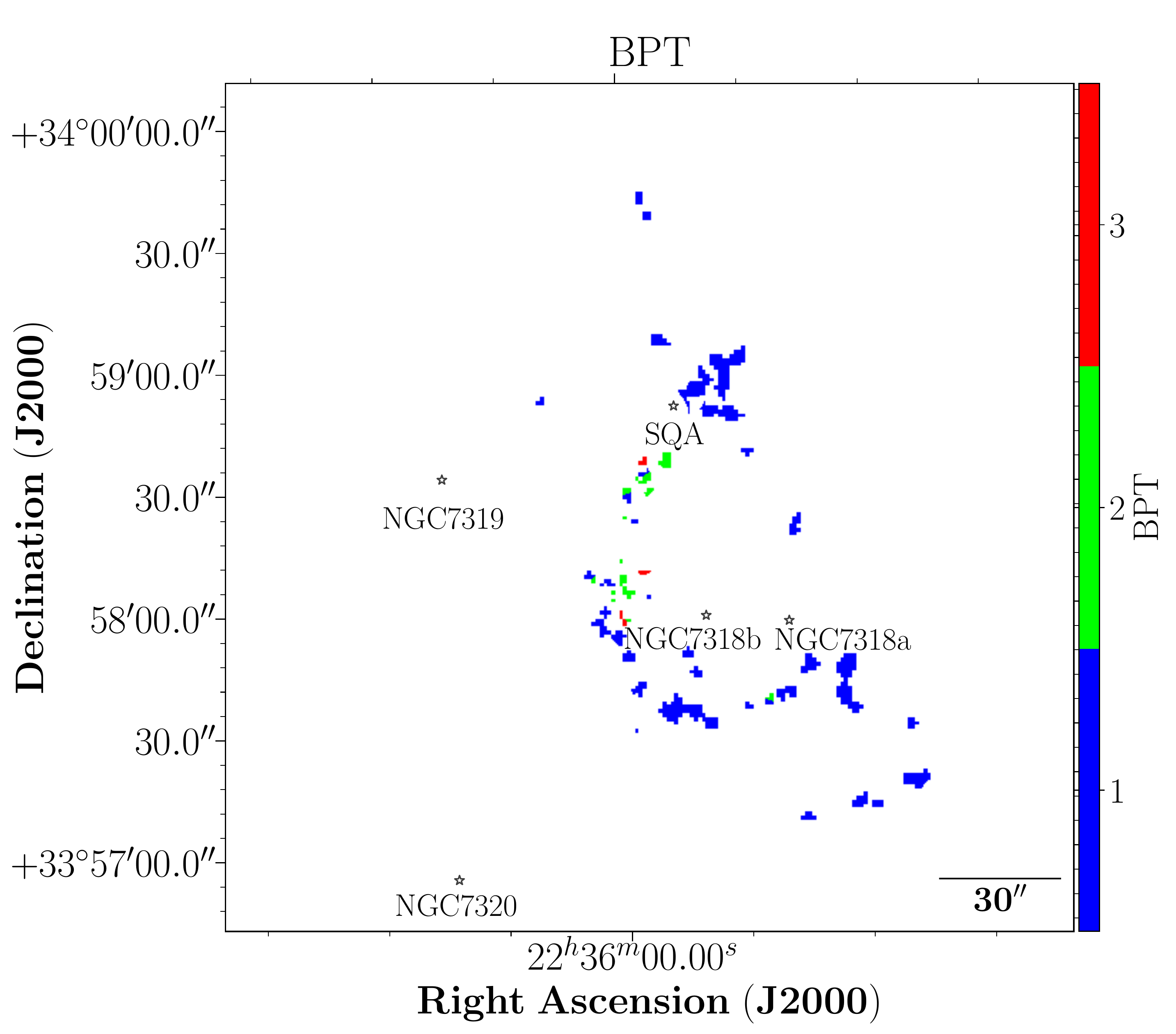}
    \includegraphics[width=.49\textwidth]{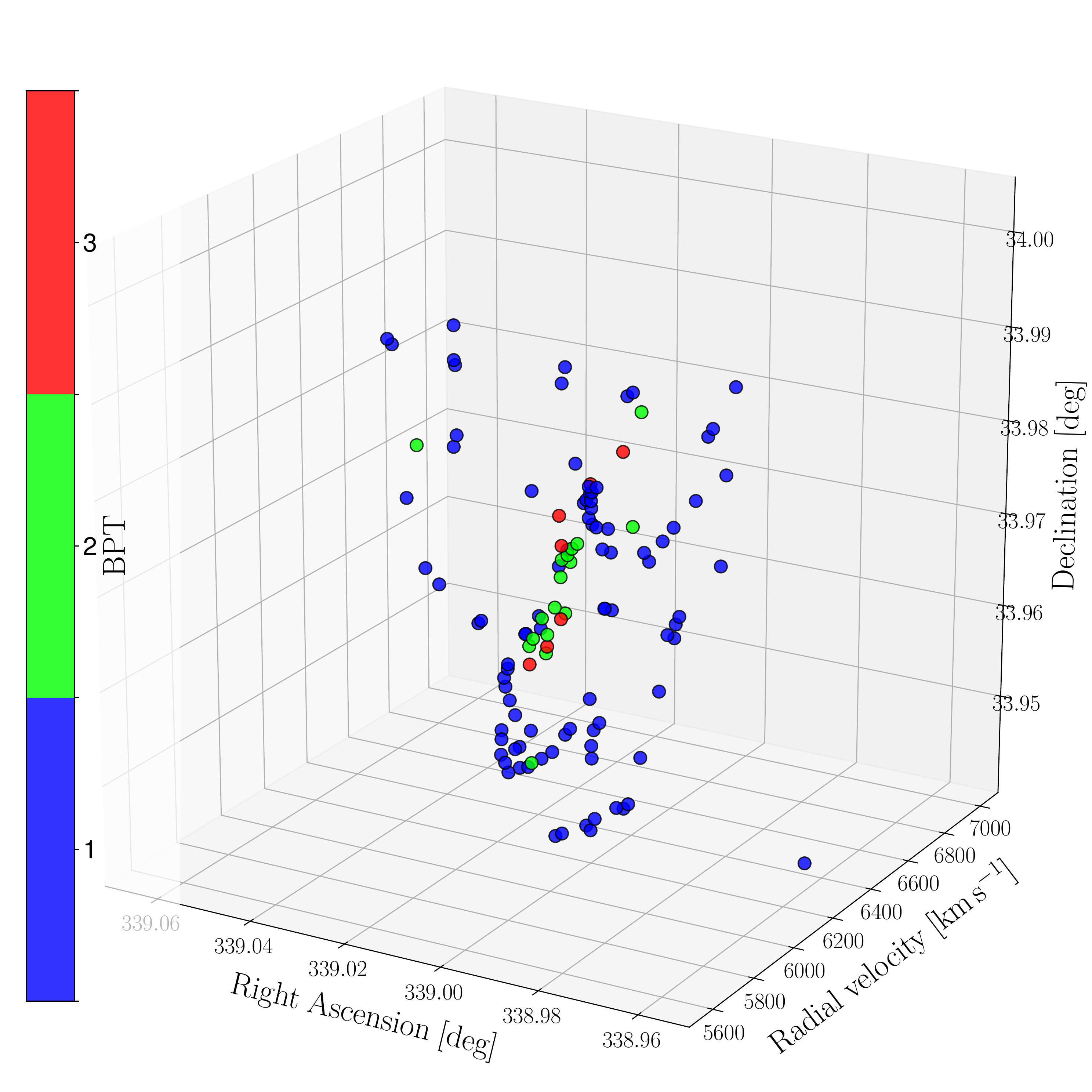}
    \caption{Stephan's Quintet spatial map colour coded according to their position in the BPT diagnostic diagram for the HV sub-sample (upper panel), the LV sub-sample (lower left panel), and the three-dimensional distribution of the BPT class ($\alpha$, $\delta$, radial velocity; lower right panel). Blue, green, and red pixels represent star-forming (1), composite (2), and AGN-like (3) regions, respectively.}
    \label{fig:BPT_map_flux_Ha_zoom}
\end{figure*}

In Fig.~\ref{fig:BPT_map_flux_Ha_zoom} (upper panel) we show the BPT map for the HV sub-sample. In the lower left panel, we present the BPT map for the LV sub-sample and in the lower right panel we display the three-dimensional distribution of the BPT class ($\alpha$, $\delta$, radial velocity) for all the SQ regions. We colour coded the SQ regions according to their position in the BPT diagram (see Fig.~\ref{fig:bpt_sq}): i) star-forming regions (BPT colour bar equal to 1, blue regions); ii) C regions (BPT colour bar equal to 2, green regions); iii) AGN-like regions (BPT colour bar equal to 3, red regions). The LV map shows that all regions found in the SDR and NIs are star forming. It is important to note that LSSR contaminates the NI1 spectra, where [\ion{N}{ii}]$\lambda$6584 from NI1 coincides in the same wavelength with H$\alpha$ from the LSSR, and must be take into account when fitting the H$\alpha$ and [\ion{N}{ii}] emission lines.\footnote{This fact tell us that care should be exercised when analysing narrow-band photometric observations without spectroscopic information.} Also, all regions from NSQA, SSQA, L2, L4, and the northernmost regions (i.e. NW tidal) are star-forming regions. According to Fig.~\ref{fig:bpt_sq}, most regions from L1 are compatible with the shock models adopted in this work. The HV map shows that all regions in the YTT and in SQA are star-forming regions. NG is a star-forming galaxy. Also, the regions found in the H$\alpha$ `bridge' are C and star-forming regions. As expected, NGC7319 presents AGN-like regions, but the north lobe presents star-forming and composite regions.

\subsection{Derivation of the SFR and chemical abundances}
\label{subsec:sfrohno_der}

Once we defined the star-forming objects among the sample of H$\alpha$ emission regions in SQ, we derived the SFR, as well as the oxygen abundance and N/O. In order to derive the SFR, we use the equation

\begin{equation}
  SFR(M_\odot\,yr^{-1})=7.9\times\,10^{42} L_{H\alpha}(erg\,s^{-1})
\end{equation}
\noindent where L$_{H\alpha}$=4$\pi$D$^{2}\,$F(H$\alpha$) is the H$\alpha$ luminosity corrected for reddening (see Sect.~\ref{subsec:line}), D is the distance in Mpc \citep{1998ARA&A..36..189K}, and F(H$\alpha$) is the H$\alpha$ flux. We assume that the average distance of SQ is 88.6 Mpc \citep{1982ApJ...255..382H,2000ApJ...529..786M,2015MNRAS.449.2937F}. We computed the SFR for the 91 HII regions found in SQ according to the classification in Fig.~\ref{fig:bpt_sq}. In Table~\ref{table:table4} we present the SFR values for each HII region. The HII regions from SQ span log(SFR/$M_\odot yr^{-1}$) values between -3.26 (region 109) and -0.13 (region 122).

To derive the oxygen abundance, we use three empirical calibrators\footnote{These are calibrated against direct derivations of abundances calculated with electron temperature measurements.}: N2 (Eq.~\ref{eq:EPM_n2}) and O3N2 (Eq.~\ref{eq:EPM_o3n2}) from \cite{2009MNRAS.398..949P}, and R from \cite{2016MNRAS.457.3678P} (see Eqs.~\ref{eq:Pi16_h} and \ref{eq:Pi16_l}). The R calibrator uses the line ratios N2, R2, and R3.\footnote{We are aware that the derivation of oxygen abundance using bright-lines calibrators can be dependent on the particular indicator used. For the sake of a comparison with previous work, and taking into account the typical errors obtained, we selected these three calibrators for O/H.}

\begin{figure}[h!]
    \centering
    \includegraphics[width=\columnwidth]{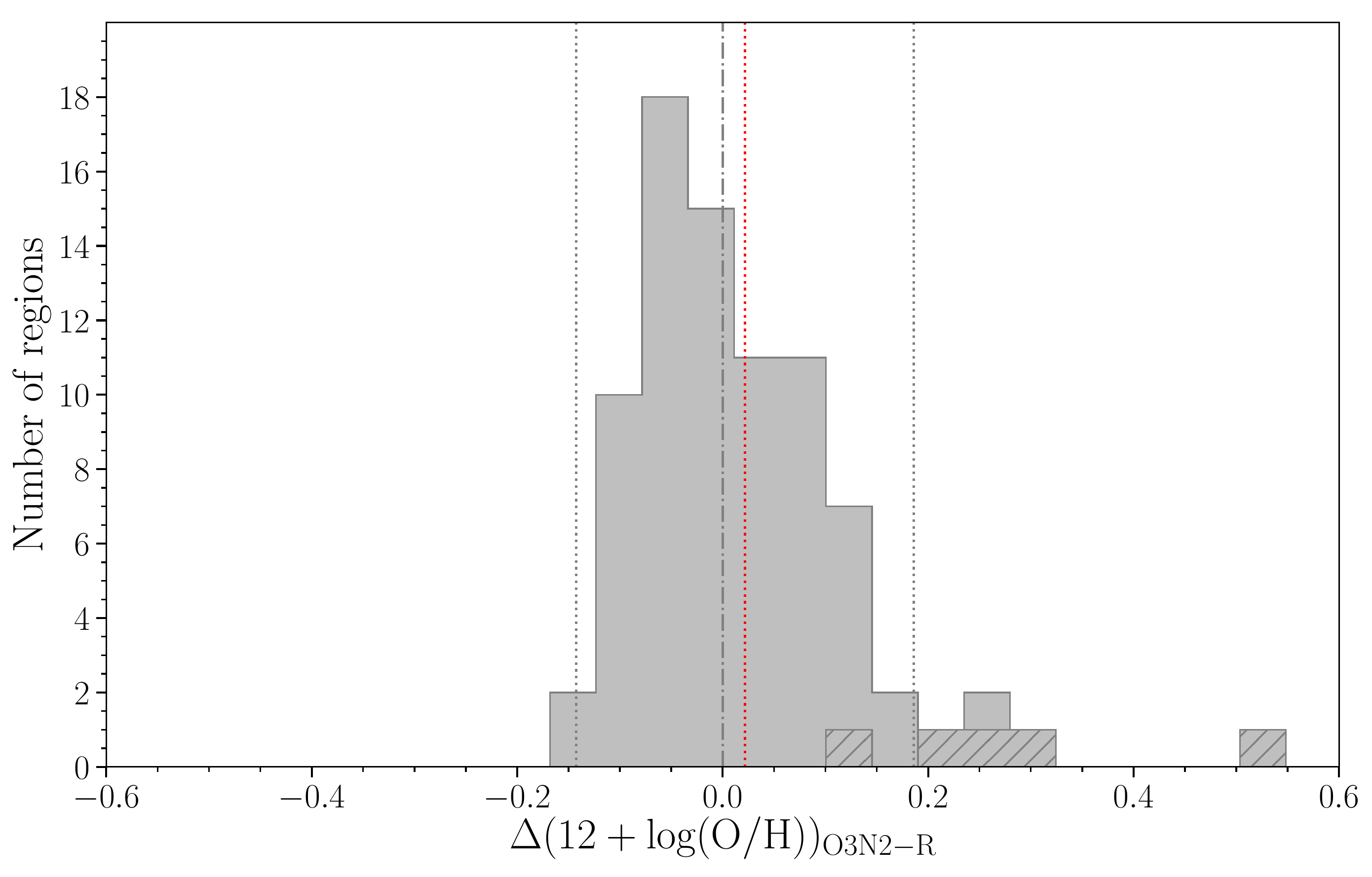}
    \caption{Distribution of difference of the oxygen abundance derived with O3N2 and R calibrations for SQ HII regions. The grey dashed line indicates a difference equal to zero, and the red dashed line corresponds to the mean oxygen abundance difference between O3N2 and R calibrators (0.02 dex) for SQ HII regions; grey dotted lines indicate the 3$\sigma$ rms. The dashed histogram represents the sample of SQ HII regions when 12+log(O/H) $<$ 8.4, for which the R calibrator was adopted.}
    \label{fig:diff_OH}
\end{figure}

\noindent The \cite{2009MNRAS.398..949P} equations are:
\begin{equation}\label{eq:EPM_n2}
12 + log(O/H) = 0.79\times\,N2 + 9.07
\end{equation}
\begin{equation}\label{eq:EPM_o3n2}
12 + log(O/H) = 8.74 - 0.31\times\,O3N2
\end{equation}

\noindent where N2 = log([\ion{N}{ii}]$\lambda$6584/H$\alpha$) and

\noindent O3N2 = log([\ion{O}{iii}]$\lambda$5007/H$\beta$ $\times$ H$\alpha$/[\ion{N}{ii}]$\lambda$6584), 

\noindent where O3N2 $<$ 2.

\noindent The \cite{2016MNRAS.457.3678P} equations are:
if log(N2) $>$ -0.6:
\begin{eqnarray}\label{eq:Pi16_h}
12 + \log(O/H) = 8.589 + 0.022 \log\left(\frac{R3}{R2}\right) + 0.399 \log\left(N2\right) + \nonumber\\
\left(- 0.137 + 0.164 \log\left(\frac{R3}{R2}\right) + 0.589 \log\left(N2\right)\right) \times \log\left(R2\right)
\end{eqnarray}

\noindent on the contrary, if log(N2) $\leq$ -0.6:
\begin{eqnarray}\label{eq:Pi16_l}
12 + \log(O/H) = 7.932 + 0.944 \log\left(\frac{R3}{R2}\right) + 0.695 \log\left(N2\right) + \nonumber\\
\left(0.970 - 0.291 \log\left(\frac{R3}{R2}\right) - 0.019 \log\left(N2\right)\right) \times \log\left(R2\right)
\end{eqnarray}

Figure~\ref{fig:diff_OH} shows the distribution of difference between the O3N2 and R calibrators used to derive the oxygen abundance for the sample of HII SQ regions.\footnote{For ten HII regions for which no [\ion{O}{ii}]$\lambda$3727 was measured, the R calibrator cannot be used.} The resulting oxygen abundances derived using both calibrators are consistent within errors when 12+log(O/H) $\geq$ 8.4, but not when 12+log(O/H) $<$ 8.4, a likely consequence of the range of validity adopted for the O3N2 parameter.\footnote{A similar situation holds for N2.} When 12+log(O/H) $<$ 8.4, we adopted the results from the R calibration \citep{2016MNRAS.457.3678P}. For the sake of consistency, we used this last work to derive N/O (Eq.~\ref{eq:nopi16}) as

\begin{eqnarray}\label{eq:nopi16}
\log(N/O) = - 0.657 - 0.201 \log\left(N2\right) +\nonumber\\
\left(0.742 - 0.075 \log\left(N2\right)\right) \times \log\left(\frac{N2}{R2}\right)
\end{eqnarray}

\noindent where:
\begin{equation}
R2 = I([\ion{O}{ii}])\lambda\lambda 3727/I(H\beta),\nonumber\\
\end{equation}
\begin{eqnarray}
N2 = I([\ion{N}{ii}])\lambda 6548+\lambda 6584/I(H\beta)=\nonumber\\1.333 I([\ion{N}{ii}])\lambda 6584/I(H\beta),\\
R3 = I([\ion{O}{iii}])\lambda 4959+\lambda 5007/I(H\beta)=\nonumber\\1.333 I([\ion{O}{iii}])\lambda 5007/I(H\beta).\nonumber
\end{eqnarray}

In Table~\ref{table:table4} we show several properties derived in this section. In Column 1 the region name is presented. Columns 2 and 3 present the SFR and H$\alpha$ luminosity. Column 4 shows the classification according to the BPT diagram. Column 5 tells us whether the velocity of the region belongs to the LV (0 in the column) or HV (1 in the column).

In Table~\ref{table:table5} we show the O/H and N/O values for the SQ HII regions derived in this section. In Column 1 the region name is presented. Columns 2, 3, and 4 show the oxygen abundances using N2, O3N2, and R calibrators, respectively. Column 5 displays the N/O using \cite{2016MNRAS.457.3678P}. Column 6 tells us whether the velocity of the region belongs to the LV (0 in the column) or HV (1 in the column) sub-samples.

\subsection{Spatially resolved analysis of SFR, O/H and N/O}
\label{subsec:SFR}
In order to study the provenance of the gas it is necessary to carry out a comprehensive study of the spectroscopic information presented in the previous sections. Consequently, in Sect.~\ref{subsec:sfrohno_der} we derived the SFR, as well as the oxygen abundance and N/O for our sample of HII SQ regions.

\begin{figure*}[h!]
    \centering
    \includegraphics[width=.9\textwidth]{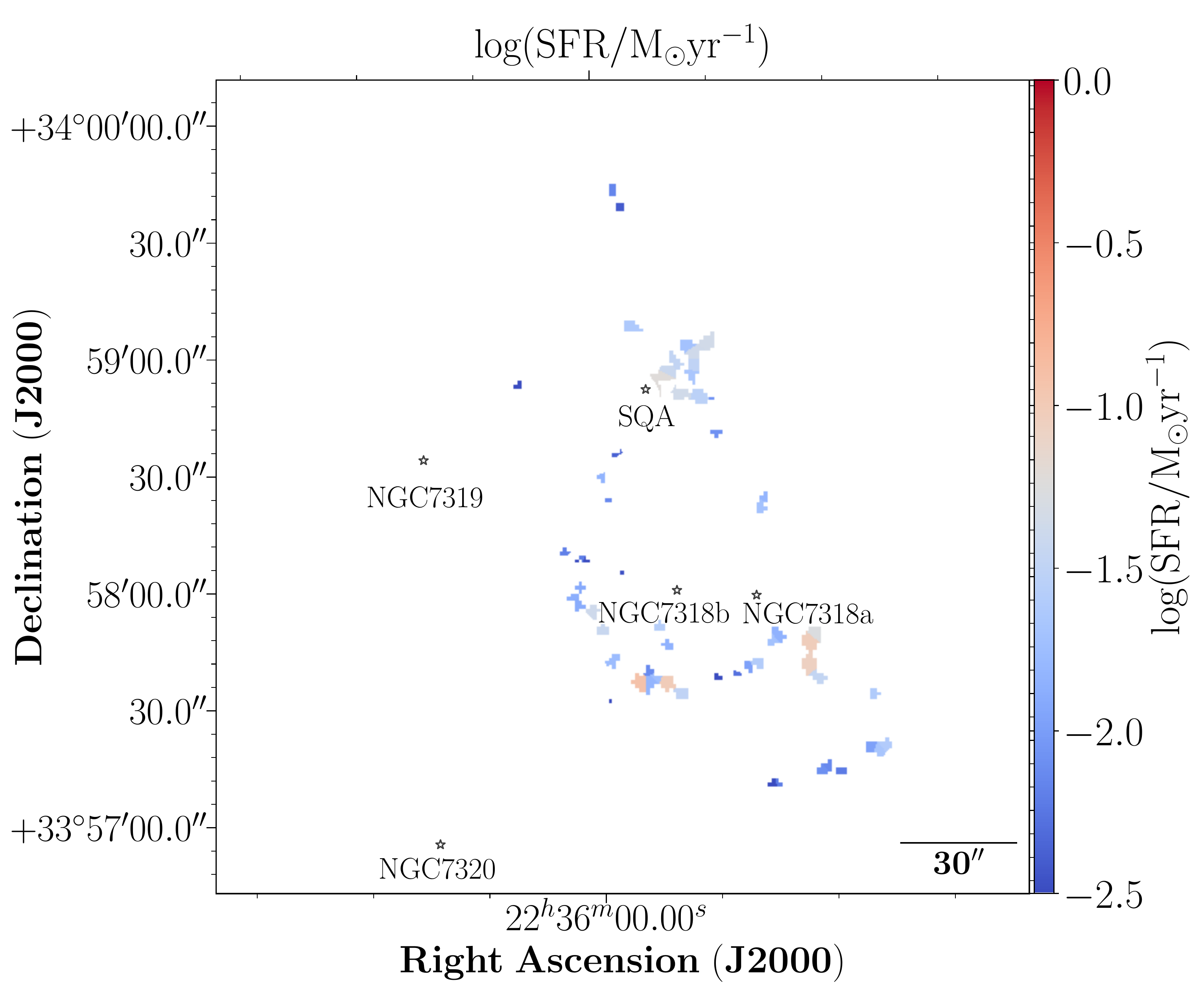}\\
    \includegraphics[width=.9\textwidth]{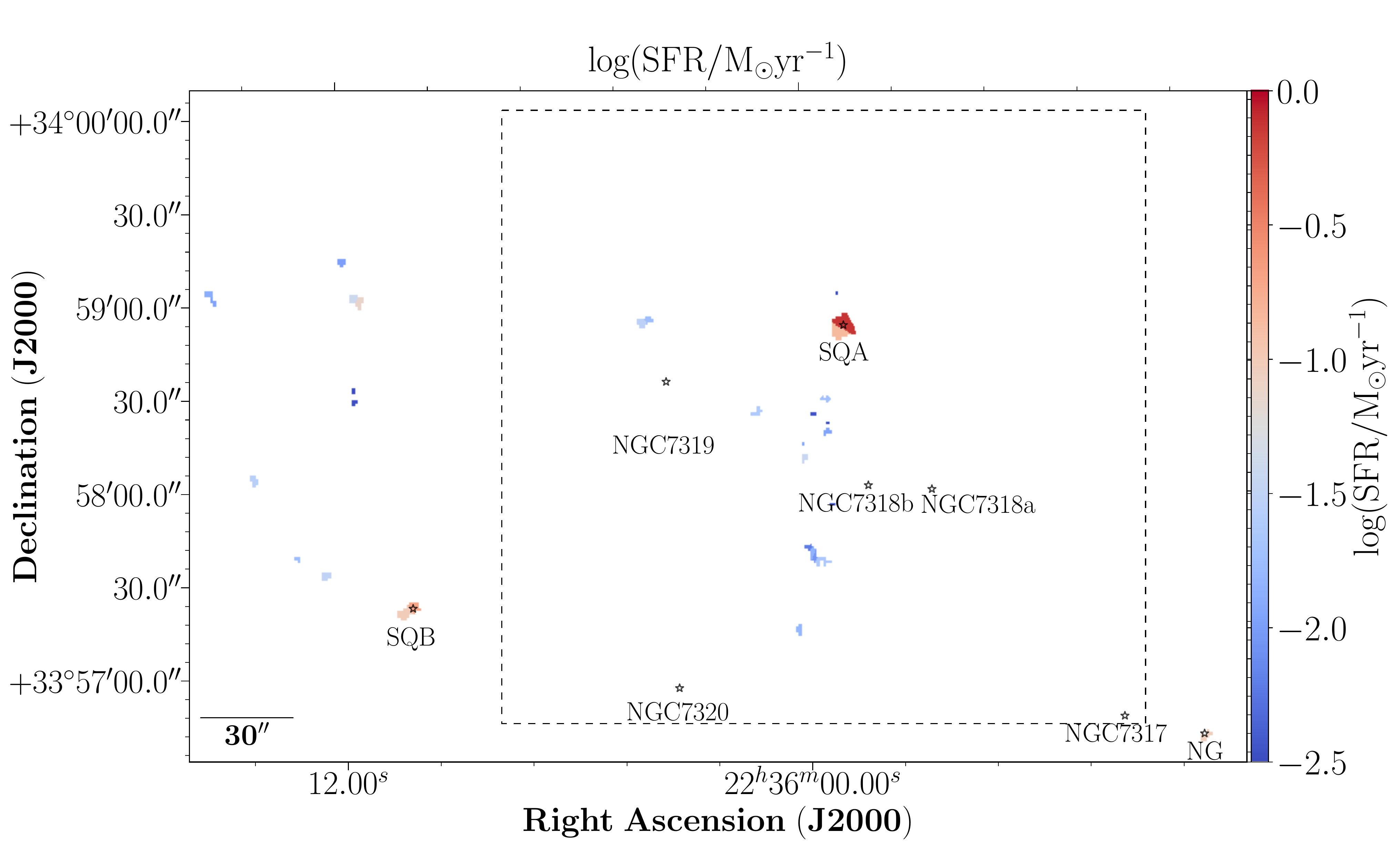}
    \caption{Stephan's Quintet spatial map colour coded according to their SFR for the LV sub-sample (upper panel) and the HV sub-sample (lower panel).}
    \label{fig:SFR}
\end{figure*}

Figure \ref{fig:SFR} shows the SFR map for the SQ. The upper panel presents HII regions belonging to the LV sub-sample, while the lower panel shows regions from the HV sub-sample. The HII regions from SQ, except SQA and SQB, have log(SFR/M$_\odot yr^{-1}$) lower than -1. We see that the regions from Shs, NW tidal tail, L1, L2, and L4 have quite a low SFR (log(SFR/M$_\odot yr^{-1}$)$<$-2). Moreover, NIs, SDR, Hs, NSQA, SSQA, and the north lobe have a low SFR (log(SFR/M$_\odot yr^{-1}$)$<$-1). Conversely, SQA has the greatest SFR value found in the SQ (log(SFR/M$_\odot yr^{-1}$)$\sim$-0.06). SQB also has a SFR higher than the average SFR found in SQ (region 15, log(SFR/M$_\odot yr^{-1}$)$\sim$-0.7). On average, the HII regions from YTTS have higher SFR values than YTTN (log(SFR/M$_\odot yr^{-1}$)=-1.5 versus -2, respectively). NG has a value of log(SFR/M$_\odot yr^{-1}$)=-1 and the north lobe has log(SFR/M$_\odot yr^{-1}$)$\sim$-1.5.

\begin{figure*}[h!]
    \centering
    \includegraphics[width=.8\textwidth]{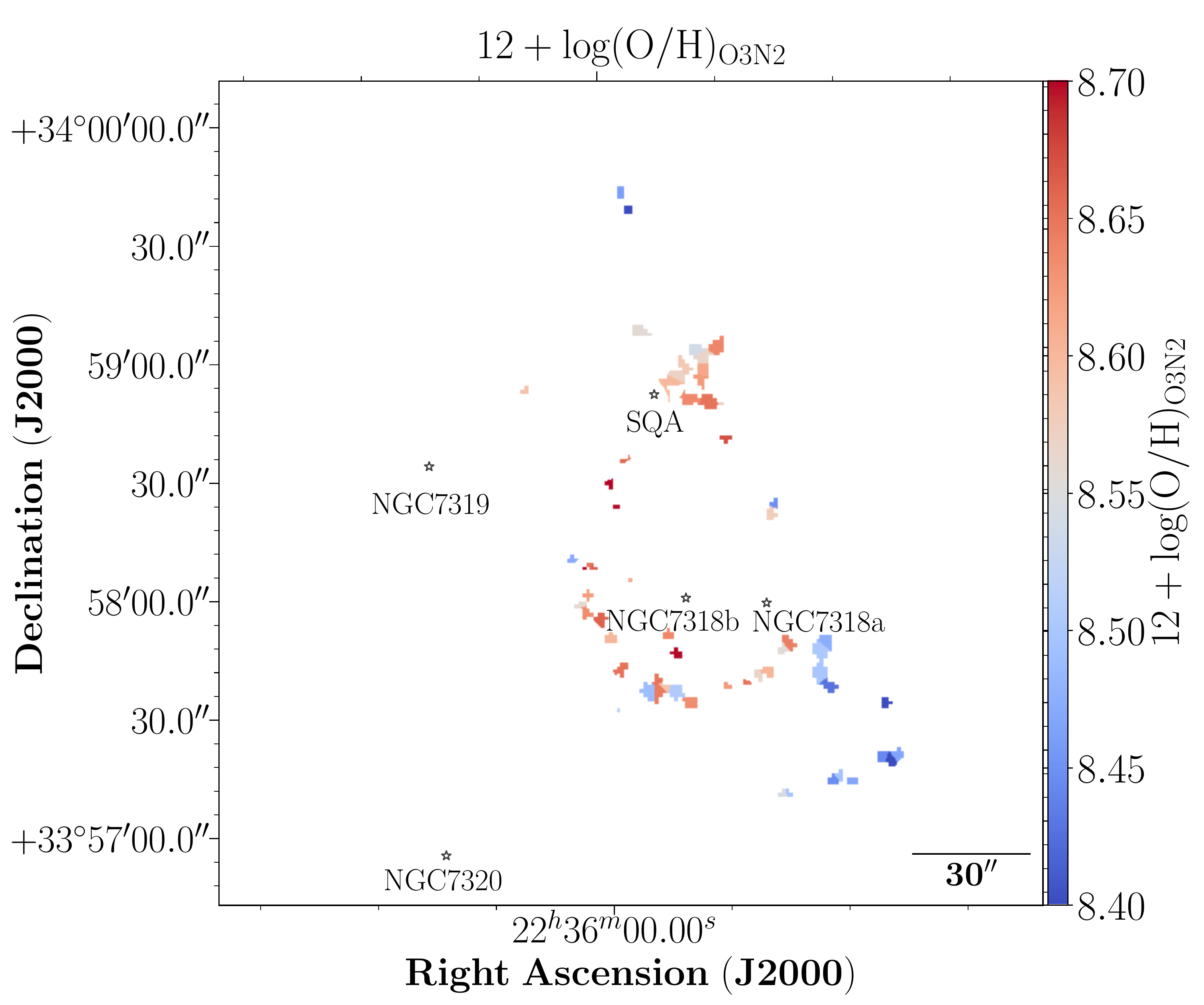}\\ 
    \includegraphics[width=.9\textwidth]{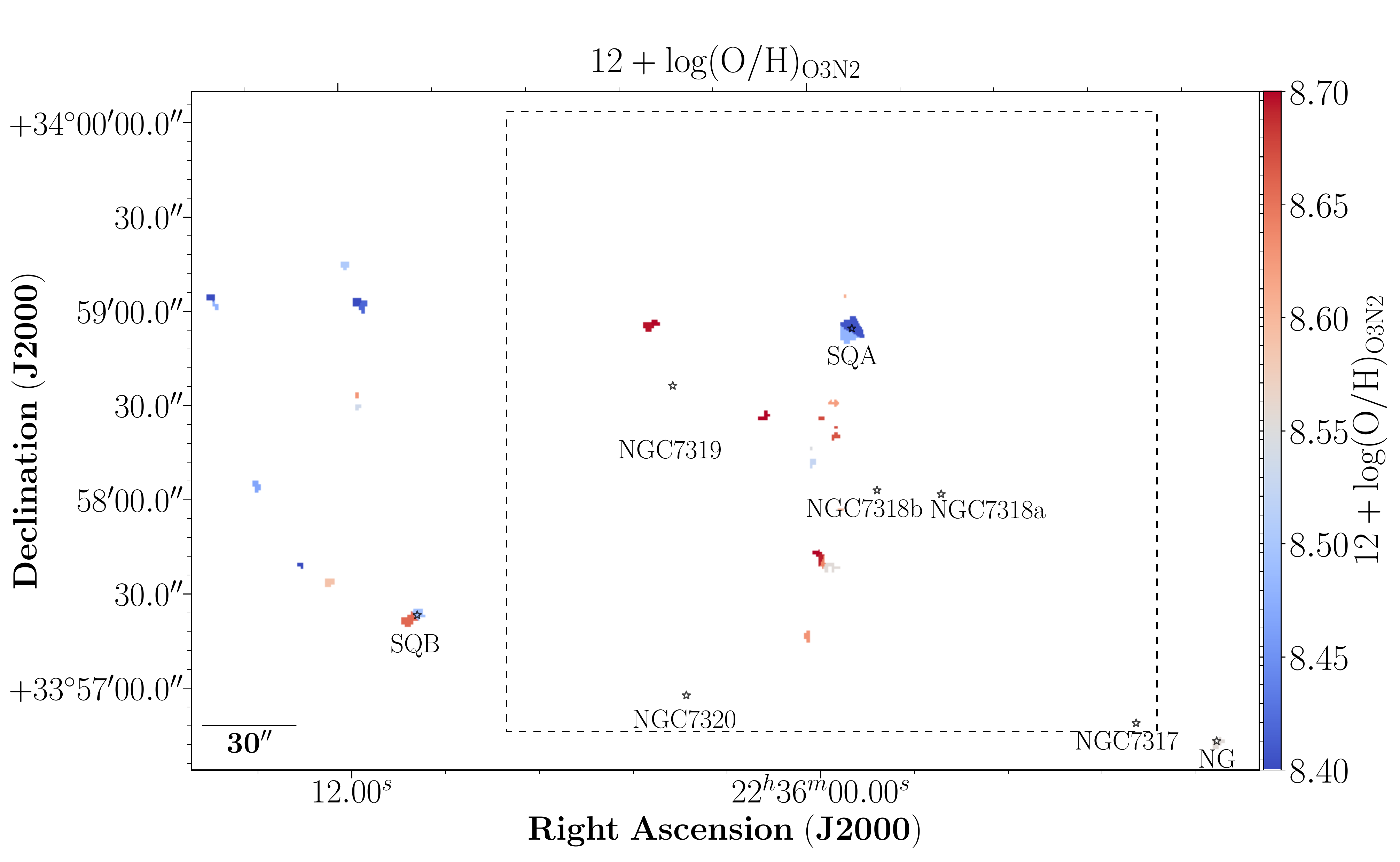}\\
    \caption{Stephan's Quintet spatial map for the LV sub-sample (upper panel) and the HV sub-sample (lower panel) colour coded according to their 12+log(O/H) derived using the O3N2 calibrator from \cite{2009MNRAS.398..949P}.}
    \label{fig:metall_max}
\end{figure*}

In Fig.~\ref{fig:metall_max} we show the oxygen abundance maps for the LV (upper panel) and HV (lower panel) sub-samples using the O3N2 calibrator from \cite{2009MNRAS.398..949P}. In Appendix~\ref{append:app3} we show the oxygen abundance maps using the R calibrator from \cite{2016MNRAS.457.3678P}. As we explained in Sect.~\ref{subsec:sfrohno_der}, the oxygen abundance calibrators (O3N2 and R) are consistent when 12+log(O/H)$\geq$8.4, and either can be adopted; and when 12+log(O/H)$<$8.4, we used the results from the R calibrator. In general, the two metallicity maps for the LV sub-sample show similar trends regardless of the calibrator used. On average, NIs span metallicity values from solar (12+log(O/H)=8.69) to half-solar (12+log(O/H) = 8.4). The regions from the NI1 and NI2 zones, which coincide with the pointings in \cite{2012A&A...539A.127I}, show oxygen abundance values in the range of 8.5 to 8.66, in accordance with \cite{2012A&A...539A.127I}. Also, SDR has an average metallicity of 12+log(O/H)$\sim$8.45 but the minimum value calculated in this zone is 8.3 (region 172); moreover, regions from the tidal tail NW present metallicity values close to 12+log(O/H)=8.45 (regions 97 and 82, respectively), in agreement with \cite{2012MNRAS.426.2441D}. As we can see in Fig.~\ref{fig:metall_max} and Table~\ref{table:table5}, NSQA has slightly higher metallicity values than SSQA (8.69 compared to 8.6), on average. The two HII regions from L2 (we recall that this zone connects NSQA with SSQA) present a metallicity of 12+log(O/H)$\sim$8.57, as well as the L4 zone. On the other hand, when we focus on the HV sub-sample (lower panel in Fig.~\ref{fig:metall_max}) we can see that SQA, region 10 (in YTTN), and region 2 (in YTTS) show lower values of metallicity (12+log(O/H)=8.4 on average), where SQB has 12+log(O/H)=8.5. For NG we found 12+log(O/H)=8.27. The HII regions from Shs span metallicity values between 8 $<$ 12+log(O/H) $<$ 8.7. Moreover, the regions from YTT present metallicity values from 12+log(O/H)=8.2 (region 2) to solar metallicity (region 14), in agreement with \cite{2004ApJ...605L..17M}. We found oversolar metallicity values in the north lobe and in region 35 (H$\alpha$ `bridge'). The average metallicity derived in SQ is $\langle 12+log(O/H)\rangle$=8.6, and the minimum value found is 12+log(O/H)=8.0 (region 59).

\begin{figure*}[h!]
    \centering
    \includegraphics[width=.9\textwidth]{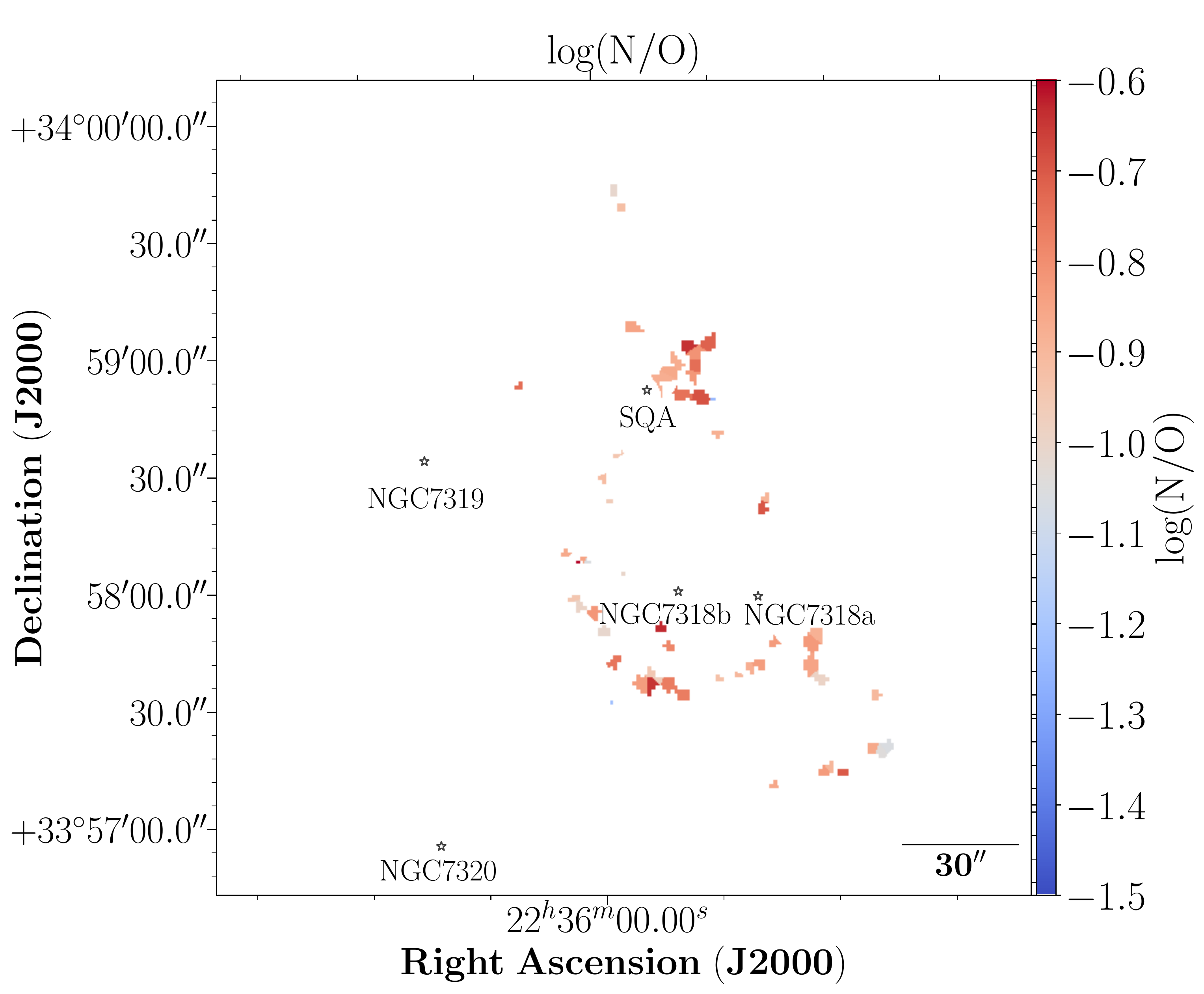}\\
    \includegraphics[width=.9\textwidth]{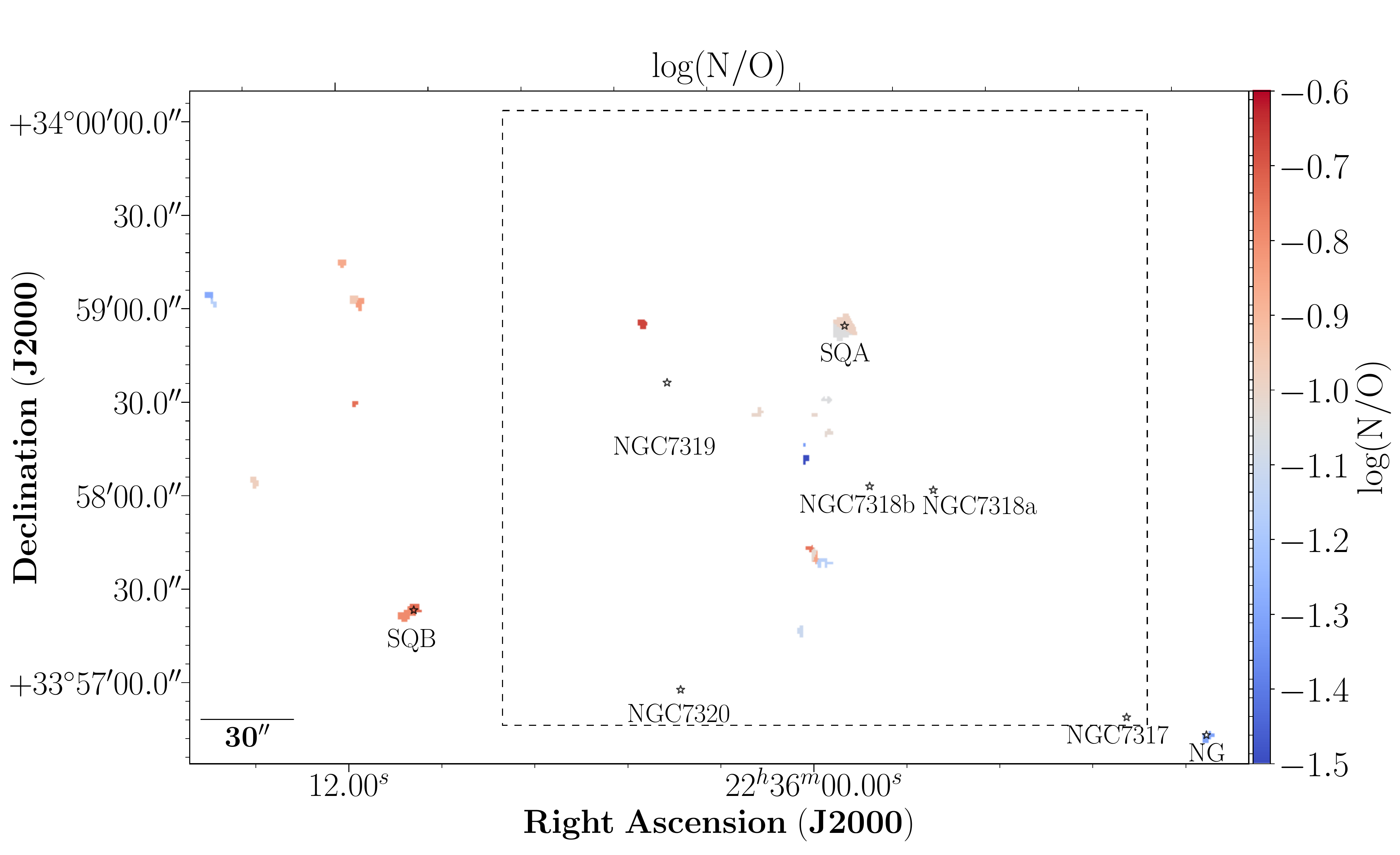}
    \caption{Stephan's Quintet spatial map for the LV sub-sample (upper panel) and the HV sub-sample (lower panel) colour coded according to their log(N/O) derived from the \cite{2016MNRAS.457.3678P} calibrator.}
    \label{fig:NO_Pi}
\end{figure*}

In Fig.~\ref{fig:NO_Pi} we show the N/O maps for LV (upper panel) and HV (lower panel) sub-samples of SQ HII regions. In this work, we used the calibrator from \cite{2016MNRAS.457.3678P} as the representative N/O for each HII region. It is important to note that the regions found in the HV sub-sample have lower values of N/O than those located in LV (log(N/O)=-1 and -0.86, respectively). The average value of N/O derived in SQ is -0.88. Except in the north lobe, on average the remaining HII regions in the HV sub-sample have subsolar N/O values.\footnote{$log(N/O)_\odot=-0.86$ \citep{2009ARA&A..47..481A,2011MNRAS.412.1367T}.} However, in the LV sub-sample, we found two regions with subsolar N/O values (SSQA and NW with mean values of log(N/O)=-0.93 and -0.97, respectively). This may indicate that both parts have different behaviour coming from diverse galaxies. LV could come from NGC7318B, whilst HV could come from NGC7319 and the debris produced in any past interaction between NGC7319 and other galaxies (e.g. NGC7317, NGC7318A, or NGC7320c). Only 6 out of 54 HII regions with N/O in the LV sub-sample have values log(N/O) $<$ -1 (regions 50, 65, 82, 147, 173, and 174). No evidence of a N/O gradient is found in NIs ($\sim$solar value, see Sect~\ref{sec:conclu}). NSQA presents higher values than SSQA on average (log(N/O)=-0.8 and -1, respectively). We can divide NSQA into two parts, east and west. The eastern part has lower N/O values than the western part (log(N/O)=-0.9 versus $\gtrsim$-0.75). The SDR zone on average presents the same N/O as NI5. However, except the north lobe (log(N/O)=-0.64), all zones from the HV sub-sample have log(N/O)$\lesssim$-1. Only 7 out of 33 HII regions with N/O in the HV sub-sample have log(N/O) $>$ -0.86 (regions 11, 13, 14, 15, 24, 64, and 77). Additionally, NG has a low value of log(N/O)=-1.3. The minimum value for N/O in SQ is log(N/O)=-1.6 (region 59).

\clearpage

\clearpage

\section{Discussion and summary}
\label{sec:conclu}
A sample of 175 SQ H$\alpha$ emitting regions, 22 of them presenting two velocity components, was defined according to the following criteria: i) the velocity of the region is within the velocity range of SQ (between $\sim$5500 and $\sim$7000 $km\, s^{-1}$); ii) the detection of at least one additional emission line beside H$\alpha$. We selected the SF regions from the BPT diagram. Taking this into account, we found 91 HII regions, 17 composite, and 7 AGN-like regions. We found all composite and AGN regions in the LV sub-sample, located in the L1, SSQA, Shs, and north lobe zones. This is also confirmed by the log([\ion{N}{ii}]$\lambda$6584/H$\alpha$) versus radial velocity diagram. We need to keep in mind that the spectra from NI1 are contaminated by the shock zone, since the [\ion{N}{ii}]$\lambda$6584 from NI1 coincides in wavelength with the H$\alpha$ from the shock. It is important to correct for this in order not to find HII regions in the composite zone \citep[e.g.][]{2014MNRAS.442..495R} to avoid misclassifications.

We found three AGN-like spectra in the L1 zone and another one in NSQA. As expected, the NGC7319 nucleus has an AGN classification, but the north lobe also presents star-forming and composite regions. The composite and AGN regions are in L1, SSQA, and Shs, being consistent with fast shock ionisation without a precursor for solar metallicity and low density (n = 0.1 cm$^{-3}$), with velocities between 175 and 300 $km\,s^{-1}$ using the models from \cite{2008ApJS..178...20A}. The star-forming regions are located in the north and south zone of the shock \citep[NW-LV, NW-HV, and SW zones in ][]{2002AJ....123.2417W}, where there is HI. Conversely, the regions in the YTT, the new intruder (NI, NGC7318B) and SDR are star forming. 

We derived the total SFR for the sample of 91 HII regions in SQ. The total SFR found for SQ is log(SFR/$M_\odot\,yr^{-1}$)=0.496. Fifty-five percent of the SFR (log(SFR/$M_\odot\,yr^{-1}$)=0.24) comes from the HV sub-sample, while the LV sub-sample is 45\% (log(SFR/$M_\odot\,yr^{-1}$) = 0.15). Twenty-eight percent of the total SFR in SQ comes from SQA (log(SFR/$M_\odot\,yr^{-1}$)=-0.06), while 9\% (log(SFR/$M_\odot\,yr^{-1}$)=-0.54) is in SQB. So, except for SQA and SQB, the material prior to the collision with NI does not show a high SFR, therefore SQ was apparently quenched. There are differences between the SFR derived for SQA from \cite{2005ApJ...619L..95X} and the one derived here. The difference is likely due to the fact that \citet[][see region VI]{2005ApJ...619L..95X} derive the SFR for a larger area that includes SQA, NSQA, L2, and H1. When considering all these zones we obtained log(SFR/$M_\odot\,yr^{-1}$)=0.12, which is in agreement with \cite{2005ApJ...619L..95X}, who found log(SFR/$M_\odot\,yr^{-1}$)=0.125.

\begin{figure}[h!]
    \centering
    \includegraphics[width=\columnwidth]{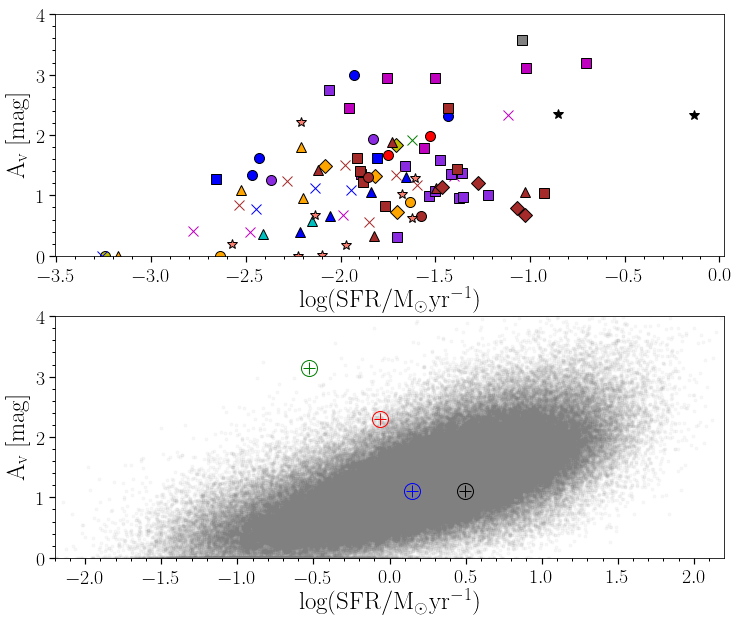}
    \caption{Relation between A$_v$ and SFR for: upper panel) the 91 SQ HII regions. All the points in the figure have the same colours and markers as Fig.~\ref{fig:regions}; and lower panel) a sample of 209276 SDSS star-forming galaxies corrected for aperture effects found in \cite{2017A&A...599A..71D}. Black, blue, red, and green $\oplus$ represent the position of the total SFR and the mean A$_v$ for all the SQ HII regions, NI, SQA, and SQB respectively. Grey points show the values for the sample of 209276 SDSS star-forming galaxies.}
    \label{fig:AvSFR}
\end{figure}

Figure~\ref{fig:AvSFR} shows the relation between A$_v$ and SFR for the sample of 91 SQ HII regions (upper panel) and for the sample of 209276 Sloan Digital Sky Survey (SDSS) star-forming galaxies corrected for aperture effects\footnote{The total SFR for the SDSS star-forming galaxies was derived using the empirical aperture corrections based on the Calar Alto Legacy Integral Field Area (CALIFA) survey defined by \cite{2013A&A...553L...7I,2016ApJ...826...71I}.} from \cite{2017A&A...599A..71D}. We show that, in both SQ HII regions and SDSS star-forming galaxies, the extinction is correlated with the SFR in such a way that the higher the SFR, the higher the extinction (with their dispersion), also spanning the same range of A$_v$ values. Therefore, the most massive galaxies and regions present more extinction. For the 91 SQ HII regions, we parameterised the correlation between SFR and A$_{V}$ by using Pearson's coefficient. We have found that there is a moderate positive correlation (0.5). It can be expected that the SQ HII regions are shifted in this diagram since they have lower SFR values. In the SDSS star-forming galaxies, each extended galaxy is considered as a single point in the A$_v$ versus SFR diagram. We note that when considering normal galaxies (i.e. not reduced to a single point) the gas is concentrated in the central zones whilst in SQ the gas is not located in the central areas of the galaxies. In our case, when we considered the total SFR and the A$_v$ median for all SQ and NI, both are located in the same part as the footprint of SDSS star-forming galaxies. On the one hand, we can appreciate that both SQA and SQB are outliers in this diagram, since they have more A$_v$ than galaxies with the same SFR. Besides, the SQ HII regions are displaced elsewhere since there is an interaction and, thus, their position depends on whether the HI gas is moving or not. The regions with more extinction and SFR are outside the galaxies because the interactions have dispersed the gas to the peripheral zones. Henceforth, the extinction in the SQ HII regions does not follow the general trend observed in spiral galaxies. Generally, the inner-outer diminishing extinction pattern that occurs in spiral galaxies disappears giving rise to a trend dominated by the successive interactions. All the above is in accordance with the interpretation of \cite{2001A&A...377..812V} concerning the distribution of HI in compact groups. SQ is part of Phase 3a according to the evolutionary sequence from the HI distribution of \cite{2001A&A...377..812V}. They say that in this phase the HI has been stripped out entirely, or almost, from the disc of the galaxies that belong to the compact groups, suggesting that the HI could be found in the tails produced by the interactions that the galaxies suffered.

\begin{figure*}[h!]
    \centering
    \includegraphics[width=0.49\textwidth]{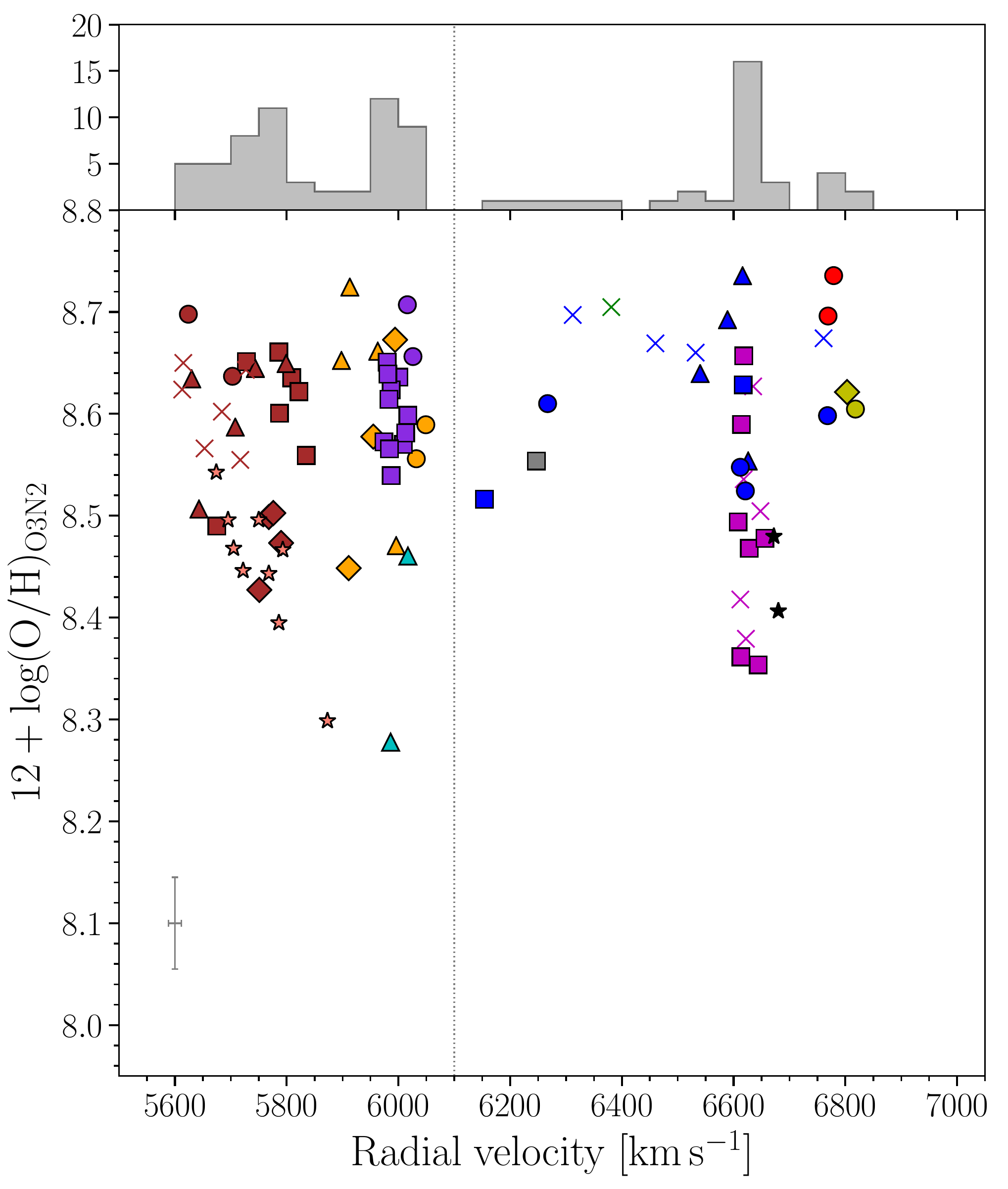}
    \includegraphics[width=0.49\textwidth]{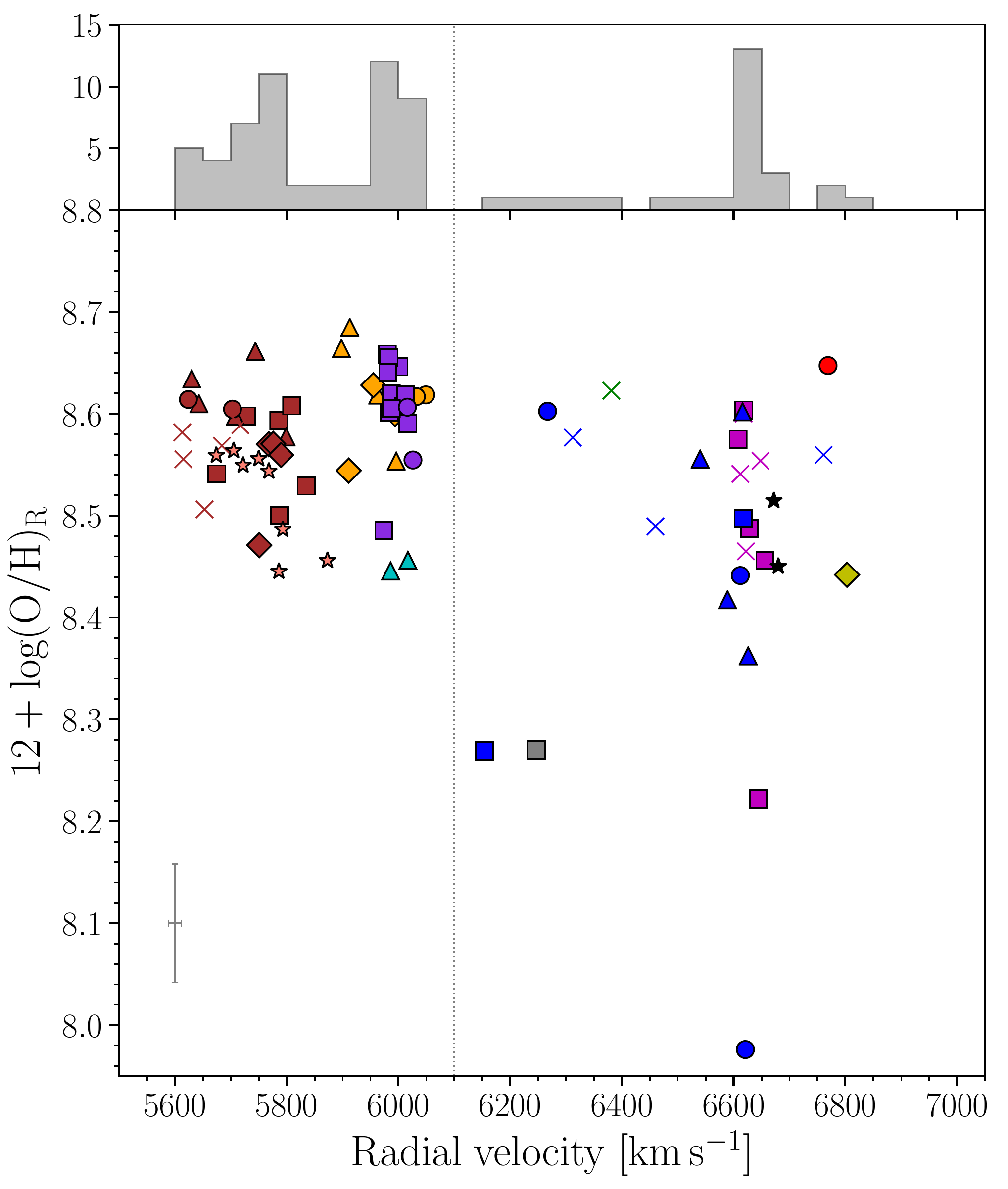}\\
    \includegraphics[width=0.49\textwidth]{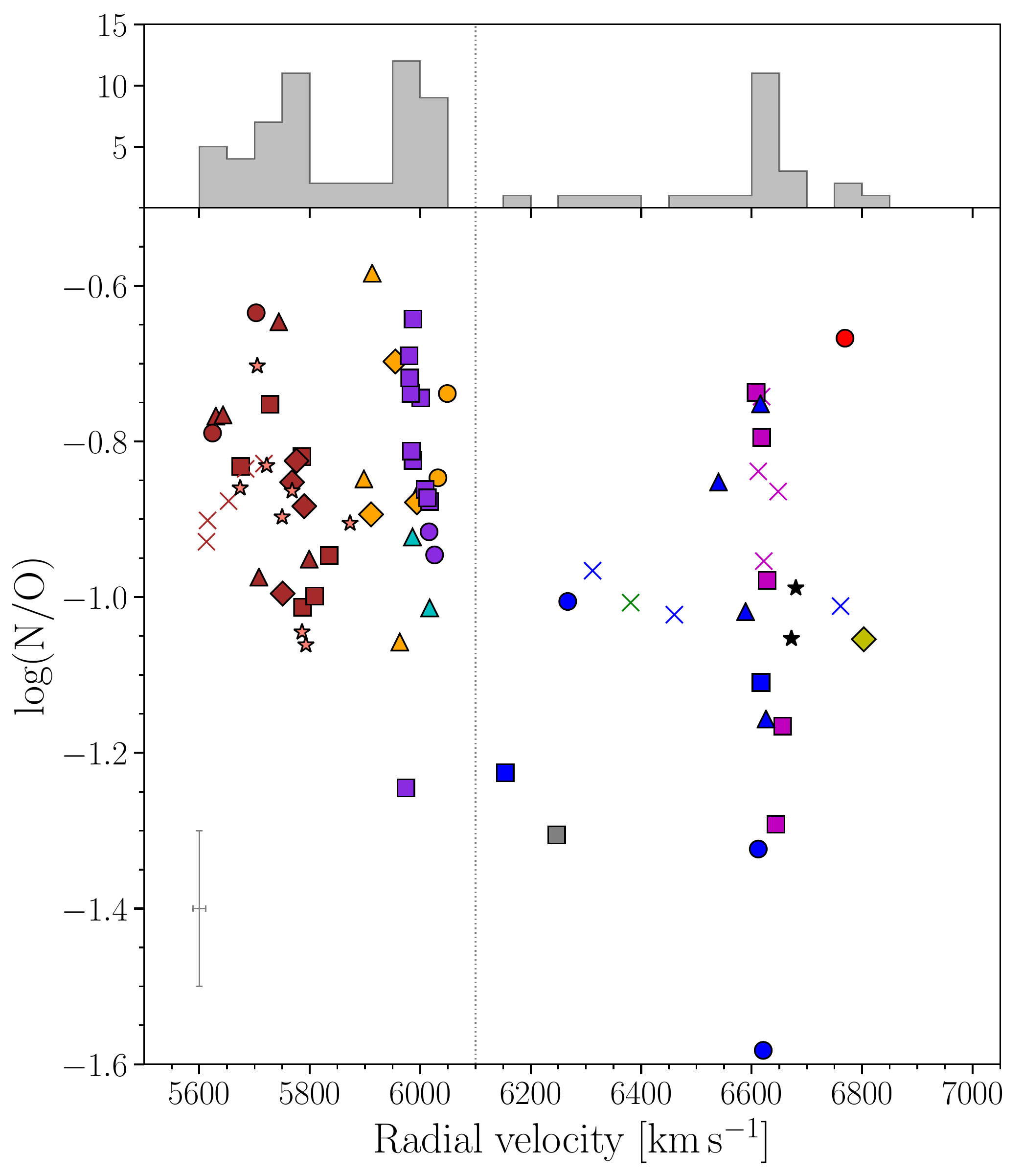}
    \includegraphics[width=0.49\textwidth]{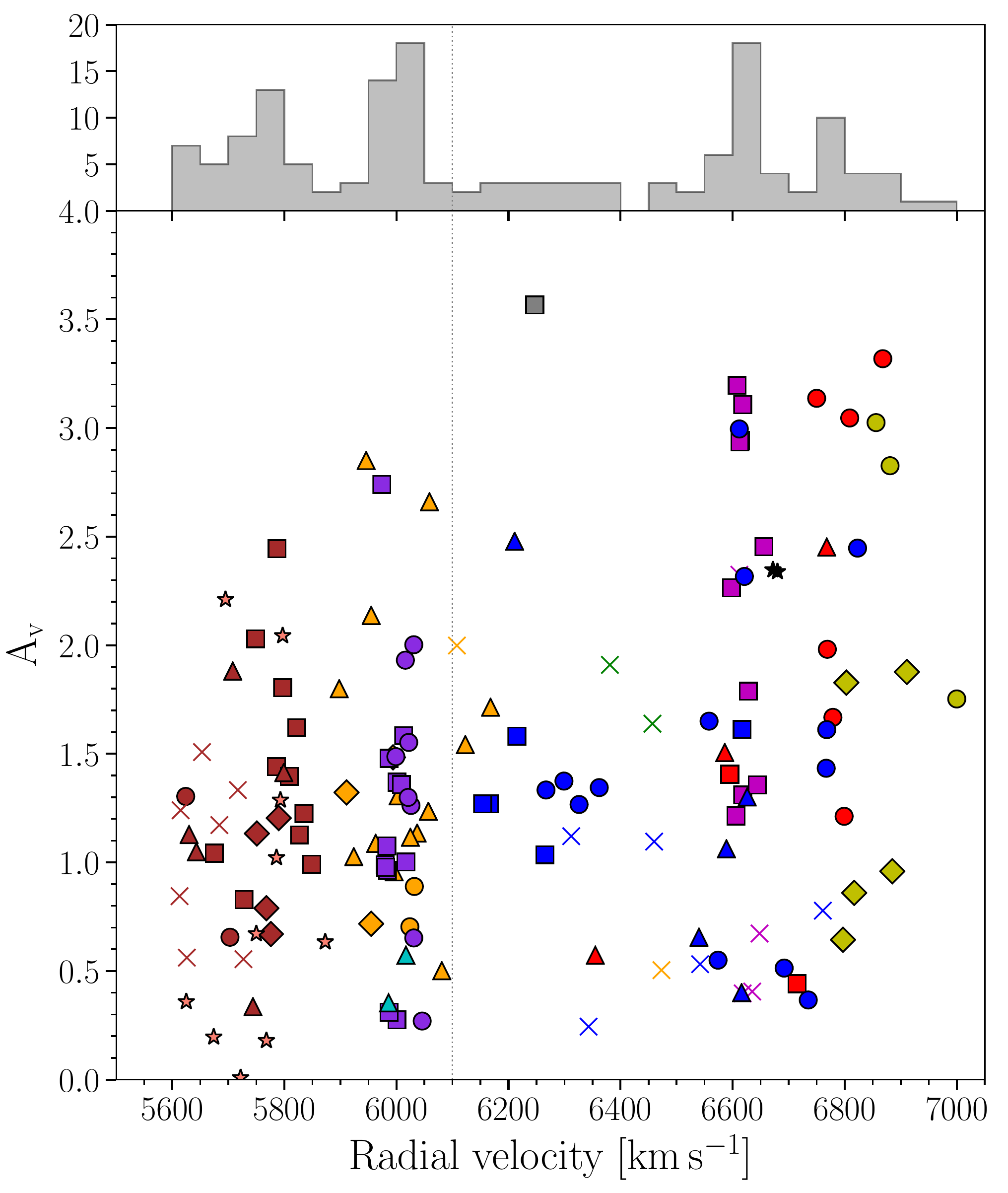}
    \caption{i) Upper left panel: Relation between 12+log(O/H)$_{O3N2}$ and radial velocity for the HII regions; ii) Upper right panel: Relation between 12+log(O/H)$_{R}$ and radial velocity for the HII regions; iii) Lower left panel: Relation between log(N/O) and radial velocity for the HII regions; iv) Lower right panel: Relation between A$_v$ and radial velocity for the H$\alpha$ emission regions. All the points in the figures have the same colours and markers as Fig.~\ref{fig:regions}. The vertical black dashed line corresponds to the value at radial velocity = 6100 $km\, s^{-1}$. Above each figure its distribution of the radial velocity for the SQ HII regions is represented. The lower left crosses indicate the typical error of the two parameters.}
    \label{fig:OHNO_velo}
\end{figure*}

The analysis of the chemical abundances of oxygen and N/O for our sample of SQ HII regions shows that the range of oxygen abundance and N/O are between solar and a fourth of solar approximately. The lowest values are found in region 59b (12+log(O/H)$\simeq$8 and log(N/O)$\simeq$-1.6). Figure~\ref{fig:OHNO_velo} shows the relation between metallicity (upper panels) or N/O (left lower panel) and the radial velocity for the sample of HII SQ regions. Also, in the right lower panel, we show the relation between A$_v$ and the radial velocity for all the SQ emission regions. The presence of an inner-outer radial metallicity gradient along the tail is clearly visible for NIs and it will be commented on below. The YTT presents a median oxygen abundance of $\sim$8.55 that agrees with \cite{2004ApJ...605L..17M}. In the case of the regions in NSQA, a hint for the existence of a radial variation of the oxygen abundance and N/O ratio has been seen, though the complexity of this system prevents us from associating this variation with a spiral arm. 

The SQ HII regions are mostly metal-rich (12+log(O/H)$\gtrsim$8.5). All eight metal-poor regions found here are located in the HV sub-sample and none in the LV sub-sample. In the case of the N/O, we found ten regions with log(N/O) lower than half solar value. Ninety percent of the regions with N/O $\leq\,{{1}\over{2}}$(N/O)$_\odot$ are located in the HV sub-sample. All the regions in the LV sub-sample (except one) have log(N/O) values higher than half solar. This might indicate that at least two chemically different gas components cohabit in SQ, one metal-rich in the LV sub-sample and in most regions of YTT, and one metal-poor in LSSR and in a few YTT regions.

The oxygen abundance values obtained from the empirical calibrations for the SQ HII regions were compared with the predictions of theoretical models from \textsc{HIIChemistry} \citep{2014MNRAS.441.2663P}. Consistency has been found between the theoretical and empirical oxygen abundances within the errors. The mean of the O/H differences between O3N2 and \textsc{HIIChemistry} abundance derivations is found to be -0.05 dex, with a 1$\sigma$ rms of 0.05 dex. A similar comparison carried out between R and \textsc{HIIChemistry} O/H derivations gives a mean value of the difference of -0.07 dex, with a 1$\sigma$ rms of 0.11 dex. As for the N/O ratio, the distribution of the differences between N/O from \cite{2016MNRAS.457.3678P} and from \textsc{HIIChemistry} gives a mean of -0.19 dex, with a 1$\sigma$ rms of 0.10 dex.

\begin{figure*}[h!]
    \centering
    \includegraphics[width=.9\textwidth]{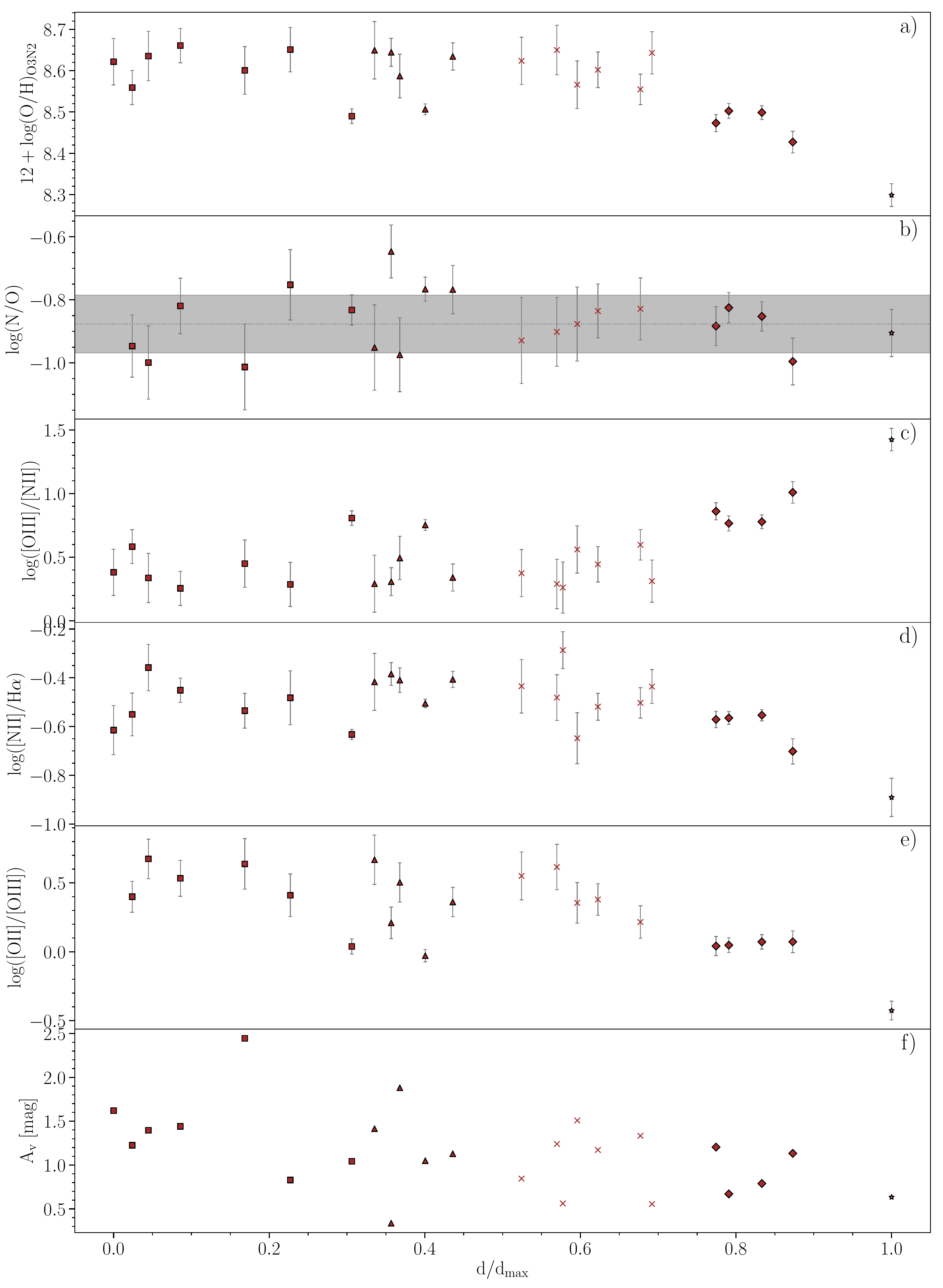}
    \caption{From top to bottom: Radial gradients for 12+log(O/H), log(N/O), log([\ion{O}{iii}]/[\ion{N}{ii}]), log([\ion{N}{ii}]/H$\alpha$), log([\ion{O}{ii}]/[\ion{O}{iii}]), and A$_v$ along the NI tails.}
    \label{fig:NI_gradOH}
\end{figure*}

\begin{figure}[h!]
    \centering
    \includegraphics[width=\columnwidth]{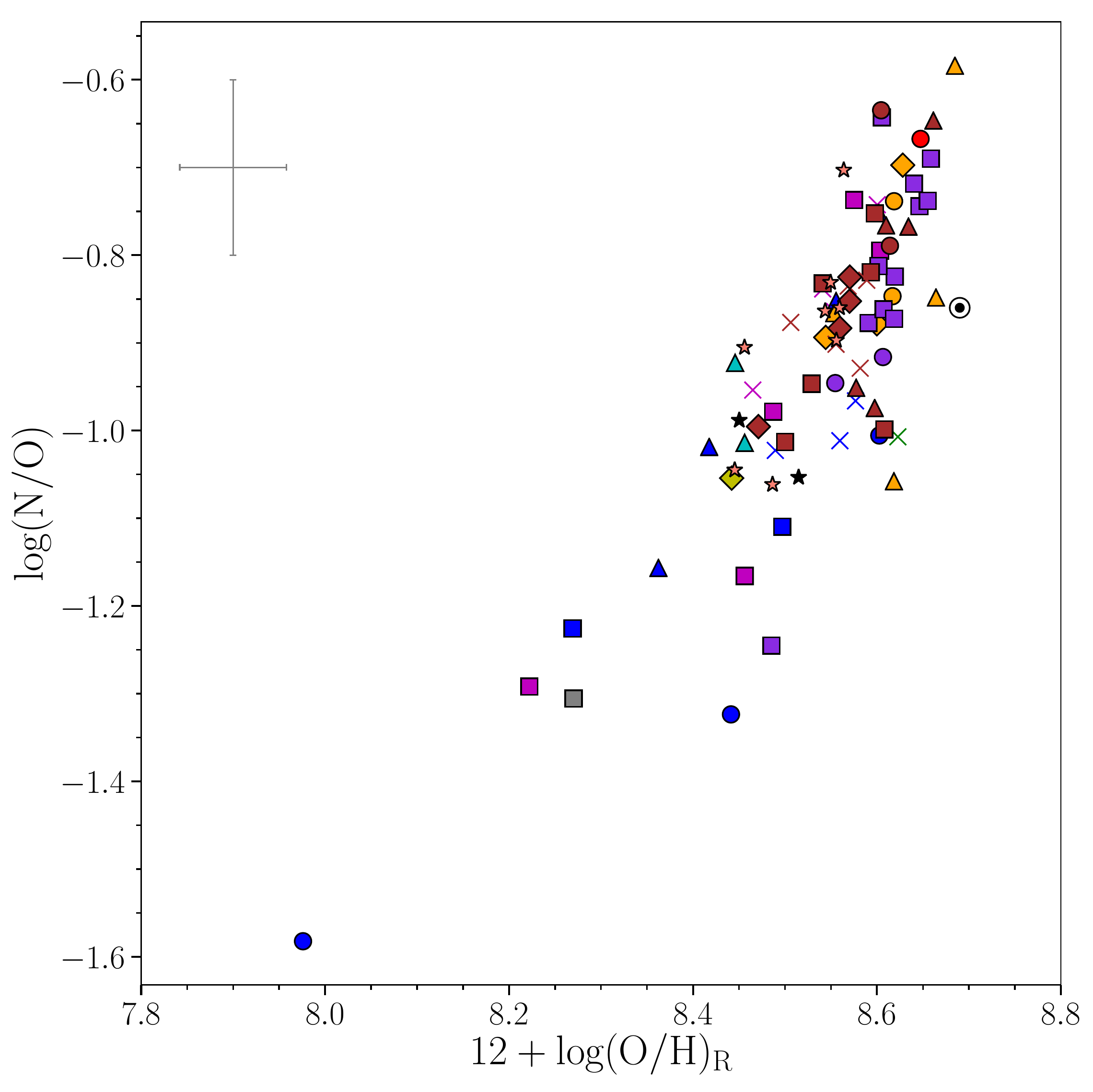}
    \caption{Relation between log(N/O) and 12+log(O/H)$_{R}$. All the points in the figures have the same colours and markers as Fig.~\ref{fig:regions}. The upper left cross indicates the typical error of both parameters. The marker $\odot$ shows the position of the solar value in this diagram.}
    \label{fig:NOOH}
\end{figure}

In this work, we studied the radial gradients for several properties of the spiral arm associated with the NIs (i.e. NI1, NI2, NI4, and NI5 strands) and SDR. Taking into account that the distance of SQ is 88.6 Mpc, we defined the distance d as the projected distance from one region to the next, and d$_{max}$ as the projected distance between the first region (the northernmost region of NI1, \#46) to the last region (\#172). As we can see in Fig.~\ref{fig:NI_gradOH}, the oxygen abundance in NI1, NI2, and NI4 appears to be constant. This is compatible with the results found in a spiral arm at a constant distance from the nucleus. Then, the metallicity value decreases as we approach the NI5 and SDR regions. We found a similar result when we studied log([\ion{N}{ii}]$\lambda$6584/H$\alpha$). We did not find a N/O gradient, because the values are almost constant (solar value). Thus, the metal-rich material should come from the inner part of the galaxy. In the case of the ratio log([\ion{O}{iii}]/H$\beta$), we found lower values in NI1 than in SDR. We obtained an A$_v$ gradient with higher values in NI1 than in SDR. All the properties studied in Fig.~\ref{fig:NI_gradOH} are consistent with the presence of an inner-outer variation along the NI tails in the line of the results obtained by \cite{2012A&A...539A.127I} and \cite{2014MNRAS.442..495R}. Finally, the relation between log(N/O) and 12+log(O/H) for the SQ HII region is presented in Fig.~\ref{fig:NOOH}. A secondary production of nitrogen can be seen in the behaviour of the relation between log(N/O) and 12+log(O/H) for the HII regions belonging to the arms. In Fig.~\ref{fig:NOOH} we can see that almost ten HII regions present metallicity similar to that of the Large Magellanic Cloud (12+log(O/H)$\lesssim$8.45), and log(N/O) values typical of regions belonging to the outer part of galactic discs (log(N/O)$\lesssim$-1). Further deep and higher spectral observations will help to understand the nature of the different chemical, kinematical, and structural components unveiled in this work.

\clearpage

\clearpage

\begin{acknowledgements}
We thank the anonymous referee for very constructive comments and suggestions that have helped us to improve this manuscript. Based on observations obtained with SITELLE, a joint project of Universit\'e Laval, ABB, Universit\'e de Montr\'eal, and the Canada-France-Hawaii Telescope (CFHT), which is operated by the National Research Council of Canada, the Institut National des Sciences de l'Univers of the Centre National de la Recherche Scientifique of France, and the University of Hawaii. The authors wish to recognise and acknowledge the very significant cultural role that the summit of Mauna Kea has always had within the indigenous Hawaiian community. We are most grateful to have the opportunity to conduct observations from this mountain. SDP, JVM, JIP, CK, EPM, and AAP acknowledge financial support from the Spanish Ministerio de Econom\'ia y Competitividad under grants AYA2013-47742-C4-1-P and AYA2016-79724-C4-4-P, from Junta de Andaluc\'ia Excellence Project PEX2011-FQM-7058, and also acknowledge support from the State Agency for Research of the Spanish MCIU through the `Center of Excellence Severo Ochoa' award for the Instituto de Astrof\'isica de Andaluc\'ia (SEV-2017-0709). LD is grateful to the Natural Sciences and Engineering Research Council of Canada, the Fonds de Recherche du Qu\'ebec, and the Canada Foundation for Innovation for funding. \\

This research made use of Python ({\tt \href{http://www.python.org}{http://www.python.org}}) and IPython \citep{PER-GRA:2007}; APLpy \citep{2012ascl.soft08017R}; Numpy \citep{2011arXiv1102.1523V}; Pandas \citep{mckinneyprocscipy2010}; of Matplotlib \citep{Hunter:2007}, a suite of open-source Python modules that provides a framework for creating scientific plots. This research made use of Astropy, a community-developed core Python package for Astronomy \citep{2013A&A...558A..33A}. The Astropy web site is {\tt \href{http://www.astropy.org/}{http://www.astropy.org}}. This research made use of astrodendro, a Python package to compute dendrograms of astronomical data ({\tt \href{http://www.dendrograms.org/}{http://www.dendrograms.org/}})\\
\end{acknowledgements}

\bibliography{SQ_publication.bib}

\begin{thebibliography}{60}
\expandafter\ifx\csname natexlab\endcsname\relax\def\natexlab#1{#1}\fi

\bibitem[{{Allen} {et~al.}(2008){Allen}, {Groves}, {Dopita}, {Sutherland}, \&
  {Kewley}}]{2008ApJS..178...20A}
{Allen}, M.~G., {Groves}, B.~A., {Dopita}, M.~A., {Sutherland}, R.~S., \&
  {Kewley}, L.~J. 2008, \apjs, 178, 20

\bibitem[{{Allen} \& {Hartsuiker}(1972)}]{1972Natur.239..324A}
{Allen}, R.~J. \& {Hartsuiker}, J.~W. 1972, \nat, 239, 324

\bibitem[{{Alloin} {et~al.}(1979){Alloin}, {Collin-Souffrin}, {Joly}, \&
  {Vigroux}}]{1979A&A....78..200A}
{Alloin}, D., {Collin-Souffrin}, S., {Joly}, M., \& {Vigroux}, L. 1979, \aap,
  78, 200

\bibitem[{{Asplund} {et~al.}(2009){Asplund}, {Grevesse}, {Sauval}, \&
  {Scott}}]{2009ARA&A..47..481A}
{Asplund}, M., {Grevesse}, N., {Sauval}, A.~J., \& {Scott}, P. 2009, \araa, 47,
  481

\bibitem[{{Astropy Collaboration} {et~al.}(2013){Astropy Collaboration},
  {Robitaille}, {Tollerud}, {Greenfield}, {Droettboom}, {Bray}, {Aldcroft},
  {Davis}, {Ginsburg}, {Price-Whelan}, {Kerzendorf}, {Conley}, {Crighton},
  {Barbary}, {Muna}, {Ferguson}, {Grollier}, {Parikh}, {Nair}, {Unther},
  {Deil}, {Woillez}, {Conseil}, {Kramer}, {Turner}, {Singer}, {Fox}, {Weaver},
  {Zabalza}, {Edwards}, {Azalee Bostroem}, {Burke}, {Casey}, {Crawford},
  {Dencheva}, {Ely}, {Jenness}, {Labrie}, {Lim}, {Pierfederici}, {Pontzen},
  {Ptak}, {Refsdal}, {Servillat}, \& {Streicher}}]{2013A&A...558A..33A}
{Astropy Collaboration}, {Robitaille}, T.~P., {Tollerud}, E.~J., {et~al.} 2013,
  \aap, 558, A33

\bibitem[{{Baldwin} {et~al.}(1981){Baldwin}, {Phillips}, \&
  {Terlevich}}]{1981PASP...93....5B}
{Baldwin}, J.~A., {Phillips}, M.~M., \& {Terlevich}, R. 1981, \pasp, 93, 5

\bibitem[{{Berg} {et~al.}(2012){Berg}, {Skillman}, {Marble}, {van Zee},
  {Engelbracht}, {Lee}, {Kennicutt}, {Calzetti}, {Dale}, \&
  {Johnson}}]{2012ApJ...754...98B}
{Berg}, D.~A., {Skillman}, E.~D., {Marble}, A.~R., {et~al.} 2012, \apj, 754, 98

\bibitem[{{Cardelli} {et~al.}(1989){Cardelli}, {Clayton}, \&
  {Mathis}}]{1989ApJ...345..245C}
{Cardelli}, J.~A., {Clayton}, G.~C., \& {Mathis}, J.~S. 1989, \apj, 345, 245

\bibitem[{{Croxall} {et~al.}(2009){Croxall}, {van Zee}, {Lee}, {Skillman},
  {Lee}, {C{\^o}t{\'e}}, {Kennicutt}, \& {Miller}}]{2009ApJ...705..723C}
{Croxall}, K.~V., {van Zee}, L., {Lee}, H., {et~al.} 2009, \apj, 705, 723

\bibitem[{{de Mello} {et~al.}(2012){de Mello}, {Urrutia-Viscarra}, {Mendes de
  Oliveira}, {Torres-Flores}, {Carrasco}, \& {Cypriano}}]{2012MNRAS.426.2441D}
{de Mello}, D.~F., {Urrutia-Viscarra}, F., {Mendes de Oliveira}, C., {et~al.}
  2012, \mnras, 426, 2441

\bibitem[{{D{\'\i}az} {et~al.}(2000){D{\'\i}az}, {Castellanos}, {Terlevich}, \&
  {Luisa Garc{\'\i}a-Vargas}}]{2000MNRAS.318..462D}
{D{\'\i}az}, A.~I., {Castellanos}, M., {Terlevich}, E., \& {Luisa
  Garc{\'\i}a-Vargas}, M. 2000, \mnras, 318, 462

\bibitem[{{Drissen} {et~al.}(2019){Drissen}, {Martin}, {Rousseau-Nepton},
  {Robert}, {Martin}, {Baril}, {Prunet}, {Joncas}, {Thibault}, {Brousseau},
  {Mandar}, {Grandmont}, {Yee}, \& {Simard}}]{2019MNRAS.485.3930D}
{Drissen}, L., {Martin}, T., {Rousseau-Nepton}, L., {et~al.} 2019, \mnras, 485,
  3930

\bibitem[{{Duarte Puertas} {et~al.}(2019){Duarte Puertas},
  {Iglesias-P{\'a}ramo}, {Vilchez}, {Drissen}, {Kehrig}, \&
  {Martin}}]{2019A&A...629A.102D}
{Duarte Puertas}, S., {Iglesias-P{\'a}ramo}, J., {Vilchez}, J.~M., {et~al.}
  2019, \aap, 629, A102

\bibitem[{{Duarte Puertas} {et~al.}(2017){Duarte Puertas}, {Vilchez},
  {Iglesias-P{\'a}ramo}, {Kehrig}, {P{\'e}rez-Montero}, \&
  {Rosales-Ortega}}]{2017A&A...599A..71D}
{Duarte Puertas}, S., {Vilchez}, J.~M., {Iglesias-P{\'a}ramo}, J., {et~al.}
  2017, \aap, 599, A71

\bibitem[{{Duc} {et~al.}(2018){Duc}, {Cuillandre}, \&
  {Renaud}}]{2018MNRAS.475L..40D}
{Duc}, P.-A., {Cuillandre}, J.-C., \& {Renaud}, F. 2018, \mnras, 475, L40

\bibitem[{{Duc} {et~al.}(2014){Duc}, {Paudel}, {McDermid}, {Cuillandre},
  {Serra}, {Bournaud}, {Cappellari}, \& {Emsellem}}]{2014MNRAS.440.1458D}
{Duc}, P.-A., {Paudel}, S., {McDermid}, R.~M., {et~al.} 2014, \mnras, 440, 1458

\bibitem[{{Fedotov} {et~al.}(2015){Fedotov}, {Gallagher}, {Durrell}, {Bastian},
  {Konstantopoulos}, {Charlton}, {Johnson}, \& {Chandar}}]{2015MNRAS.449.2937F}
{Fedotov}, K., {Gallagher}, S.~C., {Durrell}, P.~R., {et~al.} 2015, \mnras,
  449, 2937

\bibitem[{{Fedotov} {et~al.}(2011){Fedotov}, {Gallagher}, {Konstantopoulos},
  {Chandar}, {Bastian}, {Charlton}, {Whitmore}, \&
  {Trancho}}]{2011AJ....142...42F}
{Fedotov}, K., {Gallagher}, S.~C., {Konstantopoulos}, I.~S., {et~al.} 2011,
  \aj, 142, 42

\bibitem[{{Grandmont} {et~al.}(2012){Grandmont}, {Drissen}, {Mandar},
  {Thibault}, \& {Baril}}]{2012SPIE.8446E..0UG}
{Grandmont}, F., {Drissen}, L., {Mandar}, J., {Thibault}, S., \& {Baril}, M.
  2012, in Society of Photo-Optical Instrumentation Engineers (SPIE) Conference
  Series, Vol. 8446, \procspie, 84460U

\bibitem[{{Hickson}(1982)}]{1982ApJ...255..382H}
{Hickson}, P. 1982, \apj, 255, 382

\bibitem[{Hunter(2007)}]{Hunter:2007}
Hunter, J.~D. 2007, Computing In Science \& Engineering, 9, 90

\bibitem[{{Iglesias-P{\'a}ramo} {et~al.}(2012){Iglesias-P{\'a}ramo},
  {L{\'o}pez-Mart{\'{\i}}n}, {V{\'{\i}}lchez}, {Petropoulou}, \&
  {Sulentic}}]{2012A&A...539A.127I}
{Iglesias-P{\'a}ramo}, J., {L{\'o}pez-Mart{\'{\i}}n}, L., {V{\'{\i}}lchez},
  J.~M., {Petropoulou}, V., \& {Sulentic}, J.~W. 2012, \aap, 539, A127

\bibitem[{{Iglesias-P{\'a}ramo} {et~al.}(2013){Iglesias-P{\'a}ramo},
  {V{\'{\i}}lchez}, {Galbany}, {S{\'a}nchez}, {Rosales-Ortega}, {Mast},
  {Garc{\'{\i}}a-Benito}, {Husemann}, {Aguerri}, {Alves}, {Bekerait{\'e}},
  {Bland-Hawthorn}, {Catal{\'a}n-Torrecilla}, {de Amorim}, {de
  Lorenzo-C{\'a}ceres}, {Ellis}, {Falc{\'o}n-Barroso}, {Flores}, {Florido},
  {Gallazzi}, {Gomes}, {Gonz{\'a}lez Delgado}, {Haines},
  {Hern{\'a}ndez-Fern{\'a}ndez}, {Kehrig}, {L{\'o}pez-S{\'a}nchez},
  {Lyubenova}, {Marino}, {Moll{\'a}}, {Monreal-Ibero}, {Mour{\~a}o},
  {Papaderos}, {Rodrigues}, {S{\'a}nchez-Bl{\'a}zquez}, {Spekkens},
  {Stanishev}, {van de Ven}, {Walcher}, {Wisotzki}, {Zibetti}, \&
  {Ziegler}}]{2013A&A...553L...7I}
{Iglesias-P{\'a}ramo}, J., {V{\'{\i}}lchez}, J.~M., {Galbany}, L., {et~al.}
  2013, \aap, 553, L7

\bibitem[{{Iglesias-P{\'a}ramo} {et~al.}(2016){Iglesias-P{\'a}ramo},
  {V{\'{\i}}lchez}, {Rosales-Ortega}, {S{\'a}nchez}, {Duarte Puertas},
  {Petropoulou}, {Gil de Paz}, {Galbany}, {Moll{\'a}},
  {Catal{\'a}n-Torrecilla}, {Castillo Morales}, {Mast}, {Husemann},
  {Garc{\'{\i}}a-Benito}, {Mendoza}, {Kehrig}, {P{\'e}rez-Montero},
  {Papaderos}, {Gomes}, {Walcher}, {Gonz{\'a}lez Delgado}, {Marino},
  {L{\'o}pez-S{\'a}nchez}, {Ziegler}, {Flores}, \&
  {Alves}}]{2016ApJ...826...71I}
{Iglesias-P{\'a}ramo}, J., {V{\'{\i}}lchez}, J.~M., {Rosales-Ortega}, F.~F.,
  {et~al.} 2016, \apj, 826, 71

\bibitem[{{Kauffmann} {et~al.}(2003){Kauffmann}, {Heckman}, {Tremonti},
  {Brinchmann}, {Charlot}, {White}, {Ridgway}, {Brinkmann}, {Fukugita}, {Hall},
  {Ivezi{\'c}}, {Richards}, \& {Schneider}}]{2003MNRAS.346.1055K}
{Kauffmann}, G., {Heckman}, T.~M., {Tremonti}, C., {et~al.} 2003, \mnras, 346,
  1055

\bibitem[{{Kennicutt}(1998)}]{1998ARA&A..36..189K}
{Kennicutt}, Jr., R.~C. 1998, \araa, 36, 189

\bibitem[{{Kewley} {et~al.}(2001){Kewley}, {Dopita}, {Sutherland}, {Heisler},
  \& {Trevena}}]{2001ApJ...556..121K}
{Kewley}, L.~J., {Dopita}, M.~A., {Sutherland}, R.~S., {Heisler}, C.~A., \&
  {Trevena}, J. 2001, \apj, 556, 121

\bibitem[{{Konstantopoulos} {et~al.}(2014){Konstantopoulos}, {Appleton},
  {Guillard}, {Trancho}, {Cluver}, {Bastian}, {Charlton}, {Fedotov},
  {Gallagher}, {Smith}, \& {Struck}}]{2014ApJ...784....1K}
{Konstantopoulos}, I.~S., {Appleton}, P.~N., {Guillard}, P., {et~al.} 2014,
  \apj, 784, 1

\bibitem[{{Lee-Waddell} {et~al.}(2018){Lee-Waddell}, {Madrid}, {Spekkens},
  {Donzelli}, {Koribalski}, {Serra}, \& {Cannon}}]{2018MNRAS.480.2719L}
{Lee-Waddell}, K., {Madrid}, J.~P., {Spekkens}, K., {et~al.} 2018, \mnras, 480,
  2719

\bibitem[{{Lelli} {et~al.}(2015){Lelli}, {Duc}, {Brinks}, {Bournaud},
  {McGaugh}, {Lisenfeld}, {Weilbacher}, {Boquien}, {Revaz}, {Braine},
  {Koribalski}, \& {Belles}}]{2015A&A...584A.113L}
{Lelli}, F., {Duc}, P.-A., {Brinks}, E., {et~al.} 2015, \aap, 584, A113

\bibitem[{{Lisenfeld} {et~al.}(2002){Lisenfeld}, {Braine}, {Duc}, {Leon},
  {Charmandaris}, \& {Brinks}}]{2002A&A...394..823L}
{Lisenfeld}, U., {Braine}, J., {Duc}, P.-A., {et~al.} 2002, \aap, 394, 823

\bibitem[{{Martin} {et~al.}(2015){Martin}, {Drissen}, \&
  {Joncas}}]{2015ASPC..495..327M}
{Martin}, T., {Drissen}, L., \& {Joncas}, G. 2015, in Astronomical Society of
  the Pacific Conference Series, Vol. 495, Astronomical Data Analysis Software
  an Systems XXIV (ADASS XXIV), ed. A.~R. {Taylor} \& E.~{Rosolowsky}, 327

\bibitem[{McKinney(2010)}]{mckinneyprocscipy2010}
McKinney, W. 2010, in Proceedings of the 9th Python in Science Conference, ed.
  S.~van~der Walt \& J.~Millman, 51 -- 56

\bibitem[{{Mendes de Oliveira} {et~al.}(2004){Mendes de Oliveira}, {Cypriano},
  {Sodr{\'e}}, \& {Balkowski}}]{2004ApJ...605L..17M}
{Mendes de Oliveira}, C., {Cypriano}, E.~S., {Sodr{\'e}}, Jr., L., \&
  {Balkowski}, C. 2004, \apjl, 605, L17

\bibitem[{{Moles} {et~al.}(1997){Moles}, {Sulentic}, \&
  {M{\'a}rquez}}]{1997ApJ...485L..69M}
{Moles}, M., {Sulentic}, J.~W., \& {M{\'a}rquez}, I. 1997, \apjl, 485, L69

\bibitem[{{Mould} {et~al.}(2000){Mould}, {Huchra}, {Freedman}, {Kennicutt},
  {Ferrarese}, {Ford}, {Gibson}, {Graham}, {Hughes}, {Illingworth}, {Kelson},
  {Macri}, {Madore}, {Sakai}, {Sebo}, {Silbermann}, \&
  {Stetson}}]{2000ApJ...529..786M}
{Mould}, J.~R., {Huchra}, J.~P., {Freedman}, W.~L., {et~al.} 2000, \apj, 529,
  786

\bibitem[{{O'Donnell}(1994)}]{1994ApJ...422..158O}
{O'Donnell}, J.~E. 1994, \apj, 422, 158

\bibitem[{{Ohyama} {et~al.}(1998){Ohyama}, {Nishiura}, {Murayama}, \&
  {Taniguchi}}]{1998ApJ...492L..25O}
{Ohyama}, Y., {Nishiura}, S., {Murayama}, T., \& {Taniguchi}, Y. 1998, \apjl,
  492, L25

\bibitem[{{Osterbrock}(1989)}]{1989agna.book.....O}
{Osterbrock}, D.~E. 1989, {Astrophysics of gaseous nebulae and active galactic
  nuclei}

\bibitem[{{Pagel} {et~al.}(1979){Pagel}, {Edmunds}, {Blackwell}, {Chun}, \&
  {Smith}}]{1979MNRAS.189...95P}
{Pagel}, B.~E.~J., {Edmunds}, M.~G., {Blackwell}, D.~E., {Chun}, M.~S., \&
  {Smith}, G. 1979, \mnras, 189, 95

\bibitem[{P\'erez \& Granger(2007)}]{PER-GRA:2007}
P\'erez, F. \& Granger, B.~E. 2007, Computing in Science and Engineering, 9, 21

\bibitem[{{P{\'e}rez-Montero}(2014)}]{2014MNRAS.441.2663P}
{P{\'e}rez-Montero}, E. 2014, \mnras, 441, 2663

\bibitem[{{P{\'e}rez-Montero} \& {Contini}(2009)}]{2009MNRAS.398..949P}
{P{\'e}rez-Montero}, E. \& {Contini}, T. 2009, \mnras, 398, 949

\bibitem[{{Pilyugin} \& {Grebel}(2016)}]{2016MNRAS.457.3678P}
{Pilyugin}, L.~S. \& {Grebel}, E.~K. 2016, \mnras, 457, 3678

\bibitem[{{Renaud} {et~al.}(2010){Renaud}, {Appleton}, \&
  {Xu}}]{2010ApJ...724...80R}
{Renaud}, F., {Appleton}, P.~N., \& {Xu}, C.~K. 2010, \apj, 724, 80

\bibitem[{{Robitaille} \& {Bressert}(2012)}]{2012ascl.soft08017R}
{Robitaille}, T. \& {Bressert}, E. 2012, {APLpy: Astronomical Plotting Library
  in Python}, Astrophysics Source Code Library

\bibitem[{{Rodr{\'{\i}}guez-Baras} {et~al.}(2018){Rodr{\'{\i}}guez-Baras},
  {D{\'{\i}}az}, {Rosales-Ortega}, \& {S{\'a}nchez}}]{2018A&A...609A.102R}
{Rodr{\'{\i}}guez-Baras}, M., {D{\'{\i}}az}, A.~I., {Rosales-Ortega}, F.~F., \&
  {S{\'a}nchez}, S.~F. 2018, \aap, 609, A102

\bibitem[{{Rodr{\'{\i}}guez-Baras} {et~al.}(2014){Rodr{\'{\i}}guez-Baras},
  {Rosales-Ortega}, {D{\'{\i}}az}, {S{\'a}nchez}, \&
  {Pasquali}}]{2014MNRAS.442..495R}
{Rodr{\'{\i}}guez-Baras}, M., {Rosales-Ortega}, F.~F., {D{\'{\i}}az}, A.~I.,
  {S{\'a}nchez}, S.~F., \& {Pasquali}, A. 2014, \mnras, 442, 495

\bibitem[{{S{\'a}nchez} {et~al.}(2014){S{\'a}nchez}, {Rosales-Ortega},
  {Iglesias-P{\'a}ramo}, {Moll{\'a}}, {Barrera-Ballesteros}, {Marino},
  {P{\'e}rez}, {S{\'a}nchez-Blazquez}, {Gonz{\'a}lez Delgado}, {Cid Fernandes},
  {de Lorenzo-C{\'a}ceres}, {Mendez-Abreu}, {Galbany}, {Falcon-Barroso},
  {Miralles-Caballero}, {Husemann}, {Garc{\'{\i}}a-Benito}, {Mast}, {Walcher},
  {Gil de Paz}, {Garc{\'{\i}}a-Lorenzo}, {Jungwiert}, {V{\'{\i}}lchez},
  {J{\'{\i}}lkov{\'a}}, {Lyubenova}, {Cortijo-Ferrero}, {D{\'{\i}}az},
  {Wisotzki}, {M{\'a}rquez}, {Bland-Hawthorn}, {Ellis}, {van de Ven}, {Jahnke},
  {Papaderos}, {Gomes}, {Mendoza}, \&
  {L{\'o}pez-S{\'a}nchez}}]{2014A&A...563A..49S}
{S{\'a}nchez}, S.~F., {Rosales-Ortega}, F.~F., {Iglesias-P{\'a}ramo}, J.,
  {et~al.} 2014, \aap, 563, A49

\bibitem[{{Schlegel} {et~al.}(1998){Schlegel}, {Finkbeiner}, \&
  {Davis}}]{1998ApJ...500..525S}
{Schlegel}, D.~J., {Finkbeiner}, D.~P., \& {Davis}, M. 1998, \apj, 500, 525

\bibitem[{{Stephan}(1877)}]{1877MNRAS..37..334S}
{Stephan}, M. 1877, \mnras, 37, 334

\bibitem[{{Storey} \& {Hummer}(1995)}]{1995MNRAS.272...41S}
{Storey}, P.~J. \& {Hummer}, D.~G. 1995, \mnras, 272, 41

\bibitem[{{Sulentic} {et~al.}(2001){Sulentic}, {Rosado}, {Dultzin-Hacyan},
  {Verdes-Montenegro}, {Trinchieri}, {Xu}, \& {Pietsch}}]{2001AJ....122.2993S}
{Sulentic}, J.~W., {Rosado}, M., {Dultzin-Hacyan}, D., {et~al.} 2001, \aj, 122,
  2993

\bibitem[{{Tsamis} {et~al.}(2011){Tsamis}, {Walsh}, {V{\'\i}lchez}, \&
  {P{\'e}quignot}}]{2011MNRAS.412.1367T}
{Tsamis}, Y.~G., {Walsh}, J.~R., {V{\'\i}lchez}, J.~M., \& {P{\'e}quignot}, D.
  2011, \mnras, 412, 1367

\bibitem[{{Van Der Walt} {et~al.}(2011){Van Der Walt}, {Colbert}, \&
  {Varoquaux}}]{2011arXiv1102.1523V}
{Van Der Walt}, S., {Colbert}, S.~C., \& {Varoquaux}, G. 2011, ArXiv e-prints

\bibitem[{{van Zee} \& {Haynes}(2006)}]{2006ApJ...636..214V}
{van Zee}, L. \& {Haynes}, M.~P. 2006, \apj, 636, 214

\bibitem[{{Verdes-Montenegro} {et~al.}(2001){Verdes-Montenegro}, {Yun},
  {Williams}, {Huchtmeier}, {Del Olmo}, \& {Perea}}]{2001A&A...377..812V}
{Verdes-Montenegro}, L., {Yun}, M.~S., {Williams}, B.~A., {et~al.} 2001, \aap,
  377, 812

\bibitem[{{Williams} {et~al.}(2002){Williams}, {Yun}, \&
  {Verdes-Montenegro}}]{2002AJ....123.2417W}
{Williams}, B.~A., {Yun}, M.~S., \& {Verdes-Montenegro}, L. 2002, \aj, 123,
  2417

\bibitem[{{Xu} {et~al.}(1999){Xu}, {Sulentic}, \&
  {Tuffs}}]{1999ApJ...512..178X}
{Xu}, C., {Sulentic}, J.~W., \& {Tuffs}, R. 1999, \apj, 512, 178

\bibitem[{{Xu} {et~al.}(2005){Xu}, {Iglesias-P{\'a}ramo}, {Burgarella}, {Rich},
  {Neff}, {Lauger}, {Barlow}, {Bianchi}, {Byun}, {Forster}, {Friedman},
  {Heckman}, {Jelinsky}, {Lee}, {Madore}, {Malina}, {Martin}, {Milliard},
  {Morrissey}, {Schiminovich}, {Siegmund}, {Small}, {Szalay}, {Welsh}, \&
  {Wyder}}]{2005ApJ...619L..95X}
{Xu}, C.~K., {Iglesias-P{\'a}ramo}, J., {Burgarella}, D., {et~al.} 2005, \apjl,
  619, L95

\end{thebibliography}

\begin{appendix}
\section{The Stephan's Quintet, zone by zone}
\label{append:app1}

\begin{itemize}
\item YTT:
The young tidal tail (YTT) is located between the AGN galaxy NGC 7319 and the older intruder, NGC7320c. YTT is composed of 17 H$\alpha$ emission regions. We divided the area into two sub-zones: north (YTTN) and south (YTTS). YTTN corresponds to the trace of a possible filament that points towards the rest of YTT, as can be seen in the upper left panel of Fig. 1 in \cite{2018MNRAS.475L..40D}. 

YTTN consists of 6 H$\alpha$ emission regions and YTTS of 11. The average radial velocity in YTT is $\sim$6620 $km\, s^{-1}$. On average, the A$_v$ extinction in YTTS is higher than in YTTN (2.3 and 0.7, respectively).

Of the 17 initial regions, 12 are HII regions, 1 is composite, and the rest are not found in the BPT diagram.\footnote{Only the H$\alpha$ emission regions' spectra showing the four BPT emission line (i.e. [\ion{O}{iii}]$\lambda$5007, H$\beta$, [\ion{N}{ii}]$\lambda$6584, and H$\alpha$) fluxes with S/N $\geq$ 3 have been spectroscopically classified. This applies to all the other zones describe in Appendix~\ref{append:app1}.} Except for region 15 (SQB) the remaining regions have a relatively low log(SFR/M$_\odot$ yr$^{-1}$) (< -1.2) reaching values up to $\sim$-2.8 (reg 12). In SQB, the log(SFR/M$_\odot$ yr$^{-1}$) is $\sim$-0.7. The average metallicity in YTT is $\sim$8.52 ($\sim$70\% solar value). According to \cite{2004ApJ...605L..17M}, the average metallicity using the N2 calibrator is $\sim$8.58 (we derive $\sim$8.6 when using this calibrator). Also, we found a solar value for N/O in YTT, on average.

\item NGC 7319:\\
  - North lobe:
Zone located at the north of the active galactic nuclei NGC 7319 and composed of seven H$\alpha$ emission regions, of which only two are HII regions. The average radial velocity is $\sim$6780 $km\, s^{-1}$. Although there is an absence of molecular gas, the average A$_v$ extinction is high ($\sim$3 mag), where the minimum is $\sim$1.2 and the maximum is 4.2. The average log(SFR/M$_\odot$ yr$^{-1}$) in both HII regions is -1.64 and the metallicity is solar. N/O is higher than solar value ($\sim$-0.65).\\
  - NGC 7319 arm:
We detected three H$\alpha$ emission regions in this zone. The radial velocity covers a range from 6600 to 6715 $km\, s^{-1}$. The average A$_v$ extinction is 0.9 mag. We did not find any region in the BPT diagram.\\
  - NGC 7319 N:
In NGC 7319 nucleii, we found three H$\alpha$ emission regions with radial velocities between 6690 and 6355 $km\, s^{-1}$. As expected, all regions are AGN-like in the BPT diagram. The average A$_v$ extinction is $\sim$1.5 mag, showing a double core (1.5 and 2.5 mag) that would imply we detected an outflow. 

\item NSQA:
North of SQA, consisting of 13 H$\alpha$ emission regions, 11 of which are HII regions. The average radial velocity is $\sim$5970 $km\, s^{-1}$. The average A$_v$ extinction is $\sim$1 mag. The mean log(SFR/M$_\odot$ yr$^{-1}$) is -1.5, where the maximum is placed in the region 129 (-1.22) and the minimum in the region 147 (-2.1). On average, the regions show a metallicity value of $\sim$8.6 and a N/O of $\sim$-0.8.

\item SDR:
SDR is located at the south of NI and is composed of 13 H$\alpha$ regions, of which 9 appear in the BPT diagram, being all HII regions. The average radial velocity is $\sim$5720 $km\, s^{-1}$, with a range of radial velocities from $\sim$5620 to $\sim$5870 $km\, s^{-1}$. We found an average A$_v$ value lower than $\sim$0.65 mag. The mean log(SFR/M$_\odot$ yr$^{-1}$) is $\sim$-2, the metallicity is $\sim$65\% of the solar value ($\sim$8.5) and the N/O is -0.9.

\item SQ-A:
Divided into two zones with a radial velocity of $\sim$6670 $km\, s^{-1}$. Both are HII regions. The A$_v$ extinction in both regions is $\sim$2.3 mag. In paper I, Fig. 6 (upper panel) we can see two well-differentiated zones. Region 122 (north) has a higher log(SFR/M$_\odot$ yr$^{-1}$) than region 121 (south). Region 122 has a log(SFR/M$_\odot$ yr$^{-1}$) = -0.13, while region 121 has -0.85. The sum of both log(SFR/M$_\odot$ yr$^{-1}$) is -0.06. Region 122 is traditionally associated with starburst A (e.g. the placement of the spectroscopic slit of SQA by \citealt{2014ApJ...784....1K}). O/H and N/O are similar for both regions, $\sim$8.45 y $\sim$-1, respectively.

\item Hs:
The high velocity strands (Hs) encompass sub-zones H1 and H2. The H1 strand is composed of four H$\alpha$ regions and H2 has five. Paper I shows that the H2 strand connects SQA with the high radial velocity region detected in paper I for SQ (reg 111. v$\sim$7000 $km\, s^{-1}$). The H1 strand links region 111 with the shocked strands (Shs). Two of the H$\alpha$ emission regions are HII regions (one per filament) while the other one is composite. Both log(SFR/M$_\odot$ yr$^{-1}$) are relatively low ($\sim$-1.7 and $\sim$-3.2). The metallicity is $\sim$8.6 and the N/O is lower than the solar value.

\item Ls:
The Ls strands (low velocity strands) encompass sub-zones L1, L2, L3, and L4. The L1 strand is composed of 23 H$\alpha$ regions, L2 of 4, L3 of 3, and the L4 strand of 3. The average radial velocity in this zone is $\sim$6000 $km\, s^{-1}$ and its average A$_v$ extinction is $\sim$0.9 mag. We did not find any region from the L3 strand in the BPT diagram. Conversely, strands L1, L2, and L4 have 16, 2, and 3 regions in the BPT diagram. All regions detected in L2 and L4 are HII regions, while in the L1 strand only four are HII region, and the rest are consistent with fast shock ionisation without a precursor for solar metallicity and low density with velocities between 175 and 300 km s$^{-1}$ using the models from \cite{2008ApJS..178...20A}. The average log(SFR/M$_\odot$ yr$^{-1}$) for Ls is $\sim$-2.15 and the metallicity is $\sim$80\% of the solar value ($\sim$8.6), while the N/O is solar.

\item SSQA:
SSQA is composed of nine H$\alpha$ emission regions with an average radial velocity of $\sim$6020 $km\, s^{-1}$. The average A$_v$ extinction is $\sim$1.3 mag. Two regions are SF and five are composite. Both regions present a metallicity value of $\sim$8.6 and a N/O of $\sim$-0.9.

\item NW:
NW is a tidal tail located at the north of NSQA that seems to link with NSQA \citep{2010ApJ...724...80R,2018MNRAS.475L..40D}. NW is composed of five H$\alpha$ emission regions with an average radial velocity of 6030 $km\, s^{-1}$. The A$_v$ extinction is low ($\sim$0.4 mag). Two H$\alpha$ emission regions are HII regions. The average log(SFR/M$_\odot$ yr$^{-1}$) is low (-2.3). Additionally, the metallicity in this region is half-solar (8.45) and its N/O is lower than the solar value ($\sim$-1).

\item Shs:
The shock strands (Shs) encompass strands Sh1, Sh2, Sh3, and Sh4. Shs connect the Hs and Ls zones. The Sh1 and Sh4 strands have 9 regions each, while the Sh2 and Sh3 strands have 10 and 22 regions, respectively. The radial velocities are between 6000 $km\, s^{-1}$ and 6850 $km\, s^{-1}$. The A$_v$ extinction mean values in Sh1, Sh2, Sh3, and Sh4 are 1.2, 0.8, 1.2, and 0.8 mag, respectively. Fourteen regions in Shs are found in the BPT diagram, being all HII regions. The mean log(SFR/M$_\odot$ yr$^{-1}$) is lower than $\sim$-1.9. The metallicity in the different strands varies slightly. Strands Sh1 and Sh4 have a mean metallicity of $\sim$8.65, even though region 104 has a half-solar metallicity value. The average metallicity in strands Sh2 and Sh3 is 8.45, where the minimum metallicity value derived for strand Sh2 has a half-solar value ($\sim$8.4) while the minimum in strand Sh3 is almost a quarter of the solar value ($\sim$8, region 59). The mean N/O is lower than the solar value. In strands Sh1, Sh2, and Sh4 the average log(N/O) is $\sim$-1, while in strand Sh3 it is $\sim$-1.3. In Sh3, the lowest N/O derived is $\sim$-1.6 (region 59).

\item NG:
NG is a new SQ M82-like dwarf galaxy. NG has an average radial velocity of $\sim$6250 $km\, s^{-1}$ and presents a radial velocity gradient of +-60 $km\, s^{-1}$. The A$_v$ extinction is 3.5 mag. NG is a star-forming galaxy according to the BPT diagram. Its log(SFR/M$_\odot$ yr$^{-1}$) is -1.05. Both metallicity and N/O in this galaxy are slightly lower than the half-solar value (8.27 and -1.31, respectively).

\item Bridge:
As discussed in paper I, the emission detected in the bridge presents a low H$\alpha$ surface emission. Two H$\alpha$ emission regions were detected in this zone with a mean radial velocity of $\sim$6420 $km\, s^{-1}$ and a mean A$_v$ extinction of $\sim$1.8 mag. One region is composite and the other one is an HII region. The log(SFR/M$_\odot$ yr$^{-1}$) in the detected HII region (region 35) is -1.6, with a metallicity that is almost solar ($\sim$8.7) and N/O -1.

\item NI:
In order to study each zone in more detail, we divided the new intruder (NI) into five sub-zones. We found 29 H$\alpha$ emission regions in NI. The radial velocity covers the range between $\sim$5600 and $\sim$5850 $km\, s^{-1}$. The average A$_v$ extinction is lower than 1.2 mag. The regions contaminated by the LSSR in NI (regions from NI1) are the regions with the highest A$_v$ values. Twenty-five regions are found in the BPT diagram. Except for one region in zone NI4, the other are HII regions. The log(SFR/M$_\odot$ yr$^{-1}$) covers a range from $\sim$-0.9 (region 117) to $\sim$-2.5 (region 149), the average value being $\sim$-1.7. The mean metallicity is $\sim$8.6 and the average N/O has a solar value.
\end{itemize}

\section{Oxygen abundance using the R calibrator}
\label{append:app3}

\begin{figure*}
    \centering
    \includegraphics[width=.9\textwidth]{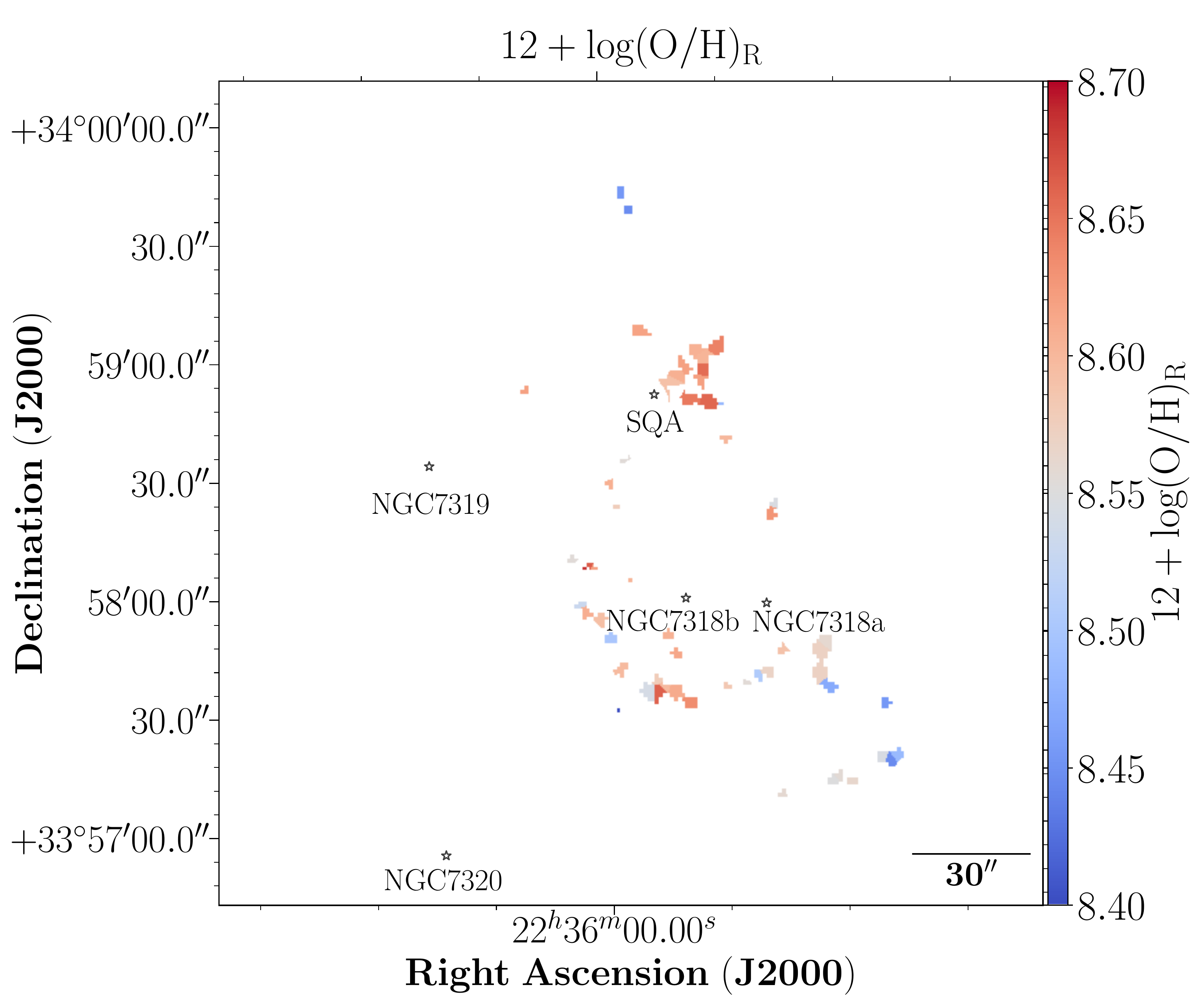}
    \includegraphics[width=.9\textwidth]{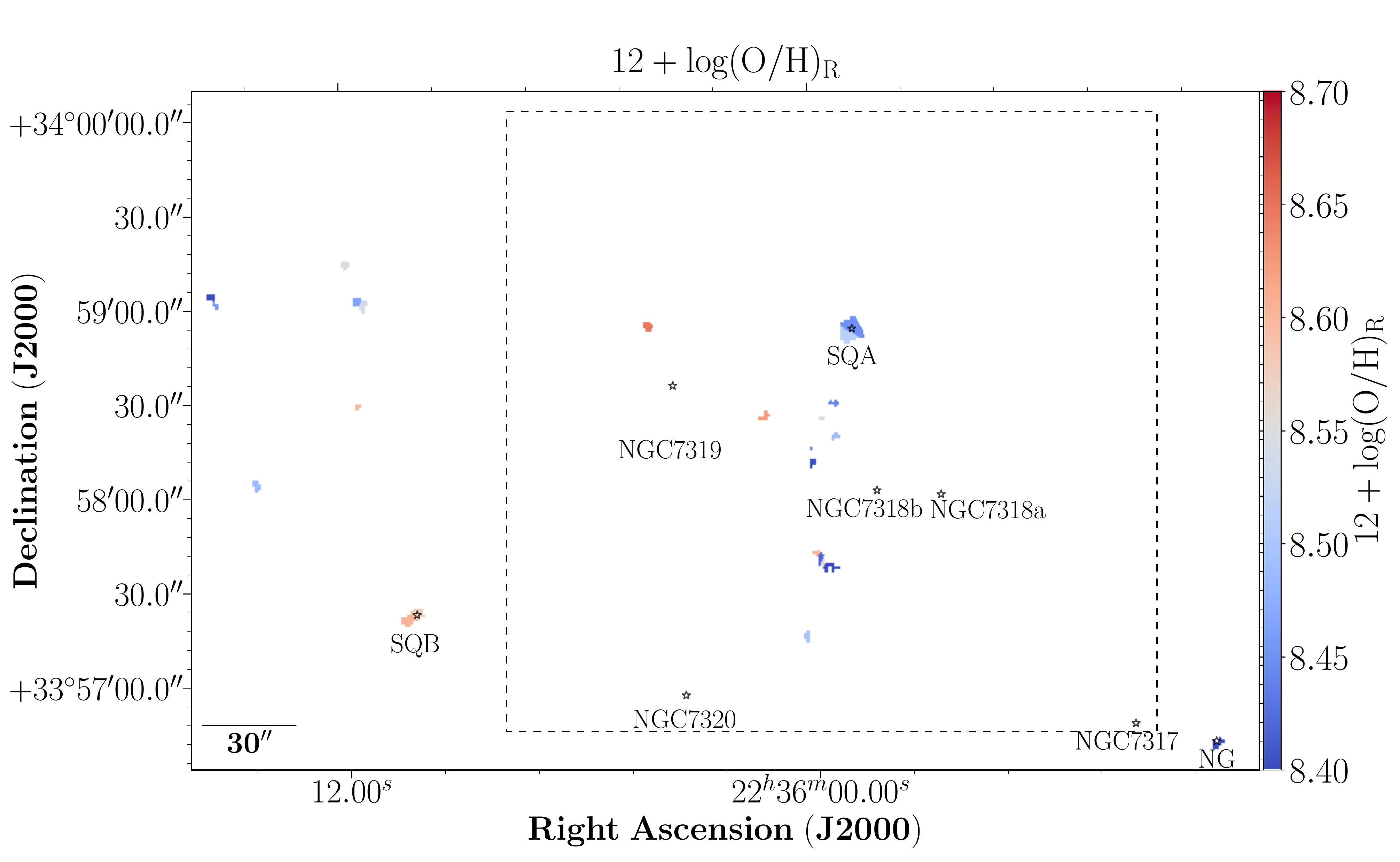}
    \caption{Stephan's Quintet spatial map colour coded according to their 12+log(O/H) derived using the R calibrator from \cite{2016MNRAS.457.3678P} calibrator for LV sub-sample (upper left panel) and HV sub-sample (upper right panel).}
    \label{fig:metall_pi}
\end{figure*}

In Fig.~\ref{fig:metall_pi} we show the oxygen abundance maps for the LV (upper panel) and HV (lower panel) sub-samples using the R calibrator from \cite{2016MNRAS.457.3678P}. Section~\ref{subsec:SFR} provides more detail.

\section{Additional table}
\label{append:app4}
\begin{table*}
\setlength{\tabcolsep}{5pt}
\tiny
\caption{Reddening corrected line fluxes relative to H$\beta$ = 100.}
\label{table:table3}
\begin{tabular}{c | c c c c c c c c c c }
\hline\hline  \\[-2ex]
(1) & (2) & (3) & (4) & (5) & (6) & (7) & (8) & (9) & (10) & (11) \\[0.5ex] 
Region & [\ion{O}{ii}]$\lambda$3727 & [\ion{O}{iii}]$\lambda$5007 & H$\alpha$ & [\ion{N}{ii}]$\lambda$6584 & [\ion{S}{ii}]$\lambda$6716 & [\ion{S}{ii}]$\lambda$6731 & A$_v$ & Vel & Zone & Subzone\\[0.5ex]
ID &  &  &  &  &  &  &  & (mag) &  &  \\[0.5ex]
\hline\\[-2ex]
1* & -- & -- & -- & -- & -- & -- & -- & -- & NGC7320c & NGC7320c\\[0.5ex]
2 & 341.8$\pm$87.3 & 188.0$\pm$31.8 & 286.0$\pm$43.3 & 30.5$\pm$10.1 & -- & -- & 1.4 & 1 & YTT & YTT\\[0.5ex]
3 & 486.2$\pm$186.5 & 184.6$\pm$65.7 & 286.0$\pm$91.4 & 75.3$\pm$19.7 & -- & -- & 2.5 & 1 & YTT & YTT\\[0.5ex]
4 & 236.0$\pm$66.5 & 165.4$\pm$22.8 & 286.0$\pm$33.0 & 62.7$\pm$6.1 & -- & -- & 1.8 & 1 & YTT & YTT\\[0.5ex]
5 & -- & 80.9$\pm$35.6 & 249.2$\pm$79.5 & 182.6$\pm$40.1 & -- & -- & 0.0 & 1 & YTT & YTT\\[0.5ex]
6 & -- & 298.9$\pm$107.8 & 286.0$\pm$99.4 & 51.4$\pm$14.8 & -- & -- & 2.9 & 1 & YTT & YTT\\[0.5ex]
7 & -- & -- & 276.0$\pm$88.2 & -- & -- & -- & 0.0 & 1 & YTT & YTT\\[0.5ex]
8 & -- & 75.7$\pm$26.5 & 286.0$\pm$63.6 & 70.9$\pm$13.8 & -- & -- & 2.9 & 1 & YTT & YTT\\[0.5ex]
9 & 191.4$\pm$53.4 & 152.0$\pm$24.7 & 286.0$\pm$39.2 & 75.6$\pm$9.2 & -- & -- & 0.7 & 1 & YTT & YTT\\[0.5ex]
10 & 177.4$\pm$30.0 & 244.2$\pm$19.8 & 286.0$\pm$21.3 & 47.9$\pm$4.2 & -- & -- & 1.3 & 1 & YTT & YTT\\[0.5ex]
11 & 143.7$\pm$44.1 & 136.3$\pm$29.2 & 286.0$\pm$51.7 & 85.5$\pm$15.1 & -- & -- & 0.4 & 1 & YTT & YTT\\[0.5ex]
12 & -- & 62.2$\pm$24.4 & 286.0$\pm$67.4 & 76.8$\pm$20.2 & -- & -- & 0.4 & 1 & YTT & YTT\\[0.5ex]
13 & 153.6$\pm$42.0 & 240.5$\pm$24.7 & 286.0$\pm$26.4 & 62.8$\pm$4.2 & -- & -- & 2.3 & 1 & YTT & YTT\\[0.5ex]
14 & 193.8$\pm$60.2 & 66.6$\pm$13.9 & 286.0$\pm$35.6 & 102.7$\pm$9.6 & -- & -- & 3.1 & 1 & YTT & YTT\\[0.5ex]
15 & 123.7$\pm$26.2 & 155.4$\pm$18.8 & 286.0$\pm$28.7 & 71.5$\pm$5.5 & -- & -- & 3.2 & 1 & YTT & YTT\\[0.5ex]
16 & -- & -- & 286.0$\pm$97.0 & 114.4$\pm$19.9 & -- & -- & 1.3 & 1 & YTT & YTT\\[0.5ex]
17 & -- & -- & 286.0$\pm$61.5 & 128.7$\pm$16.8 & -- & -- & 2.3 & 1 & YTT & YTT\\[0.5ex]
18 & -- & -- & 286.0$\pm$77.3 & 117.9$\pm$26.3 & -- & -- & 1.4 & 1 & NGC7319 & NGC7319 Arm\\[0.5ex]
19 & 167.0$\pm$69.6 & -- & 286.0$\pm$83.8 & 111.4$\pm$15.8 & -- & -- & 1.2 & 1 & YTT & YTT\\[0.5ex]
20 & -- & -- & 286.0$\pm$61.9 & 106.8$\pm$13.6 & -- & -- & 4.2 & 1 & NGC7319 & North lobe\\[0.5ex]
21 & -- & -- & 286.0$\pm$71.2 & 115.3$\pm$13.2 & -- & -- & 3.0 & 1 & NGC7319 & North lobe\\[0.5ex]
22$^\dagger$ & -- & -- & 286.0$\pm$117.6 & 163.7$\pm$66.4 & -- & -- & 1.9 & 1 & NGC7319 & NGC7319 Arm\\[0.5ex]
23 & -- & -- & 286.0$\pm$78.6 & 105.8$\pm$22.0 & -- & -- & 3.3 & 1 & NGC7319 & North lobe\\[0.5ex]
24 & 141.6$\pm$39.5 & 56.1$\pm$19.7 & 286.0$\pm$62.0 & 115.8$\pm$18.4 & -- & -- & 2.0 & 1 & NGC7319 & North lobe\\[0.5ex]
25 & -- & 48.9$\pm$17.4 & 286.0$\pm$58.2 & 135.3$\pm$21.1 & -- & -- & 1.7 & 1 & NGC7319 & North lobe\\[0.5ex]
26 & 507.4$\pm$151.0 & 951.4$\pm$280.0 & 286.0$\pm$85.6 & 385.0$\pm$36.0 & -- & -- & 2.5 & 1 & NGC7319 & NGC7319 Nucl.\\[0.5ex]
27 & 278.0$\pm$95.3 & 793.7$\pm$268.3 & 286.0$\pm$99.5 & 289.1$\pm$40.7 & -- & -- & 1.5 & 1 & NGC7319 & NGC7319 Nucl.\\[0.5ex]
28 & -- & -- & 286.0$\pm$85.9 & 100.4$\pm$25.3 & -- & -- & 0.4 & 1 & NGC7319 & NGC7319 Arm\\[0.5ex]
29 & 259.4$\pm$110.7 & 756.2$\pm$257.4 & 286.0$\pm$103.7 & 252.5$\pm$55.6 & -- & -- & 0.6 & 1 & NGC7319 & NGC7319 Nucl.\\[0.5ex]
30 & -- & 119.5$\pm$53.0 & 286.0$\pm$102.5 & 170.5$\pm$39.8 & -- & -- & 3.1 & 1 & NGC7319 & North lobe\\[0.5ex]
31$^\dagger$ & 172.9$\pm$86.6 & 298.2$\pm$116.1 & 286.0$\pm$111.8 & -- & -- & -- &0.3&0& Ls & L1\\[0.5ex]
32 & -- & -- & 286.0$\pm$90.3 & 133.2$\pm$40.5 & -- & -- & 1.2 & 1 & NGC7319 & North lobe\\[0.5ex]
33 & 156.5$\pm$46.4 & 104.6$\pm$23.5 & 256.4$\pm$51.1 & 97.7$\pm$25.4 & -- & -- & 0.0 & 0 & Ls & L2\\[0.5ex]
34 & -- & -- & 283.1$\pm$70.2 & -- & -- & -- & 0.0 & 1 & NW & NW\\[0.5ex]
35 & 491.4$\pm$129.1 & 65.7$\pm$20.3 & 286.0$\pm$62.1 & 144.7$\pm$31.9 & -- & -- & 1.9 & 1 & Bridge & Bridge\\[0.5ex]
36 & -- & 165.4$\pm$57.4 & 265.2$\pm$94.6 & -- & -- & -- & 0.0 & 0 & Ls & L2\\[0.5ex]
37 & 192.5$\pm$47.4 & 154.4$\pm$30.8 & 274.5$\pm$52.2 & -- & -- & -- & 0.0 & 0 & NW & NW\\[0.5ex]
38 & -- & 168.5$\pm$56.4 & 286.0$\pm$93.7 & -- & -- & -- & 1.0 & 0 & NIs & NI1\\[0.5ex]
39 & 187.5$\pm$53.7 & 188.8$\pm$40.7 & 286.0$\pm$57.8 & 73.0$\pm$18.5 & -- & -- & 1.0 & 0 & Ls & L1\\[0.5ex]
40 & 541.6$\pm$149.6 & 141.0$\pm$29.5 & 286.0$\pm$56.0 & 183.3$\pm$31.0 & -- & -- & 1.6 & 1 & Bridge & Bridge\\[0.5ex]
41 & -- & 197.1$\pm$75.1 & 286.0$\pm$98.9 & -- & -- & -- & 0.7 & 0 & Ls & L2\\[0.5ex]
42 & 259.4$\pm$68.6 & 107.9$\pm$24.8 & 286.0$\pm$51.2 & 80.5$\pm$16.2 & 40.8$\pm$13.0 & -- & 1.2 & 0 & NIs & NI1\\[0.5ex]
43 & -- & 139.6$\pm$56.8 & 286.0$\pm$98.4 & 129.8$\pm$32.4 & -- & -- & 1.3 & 0 & Ls & L1\\[0.5ex]
44 & -- & -- & 286.0$\pm$89.5 & 115.8$\pm$31.2 & -- & -- & 1.2 & 0 & Ls & L1\\[0.5ex]
45 & 428.9$\pm$140.6 & 95.3$\pm$36.8 & 286.0$\pm$83.3 & 125.4$\pm$27.5 & 104.1$\pm$25.7 & -- & 1.4 & 0 & NIs & NI1\\[0.5ex]
46 & -- & 58.4$\pm$20.3 & 286.0$\pm$55.4 & 69.4$\pm$16.1 & 61.1$\pm$17.1 & -- & 1.6 & 0 & NIs & NI1\\[0.5ex]
47 & 121.7$\pm$42.9 & 53.3$\pm$19.7 & 249.2$\pm$61.0 & 135.9$\pm$37.1 & -- & -- & 0.0 & 0 & Ls & L1\\[0.5ex]
48 & 282.9$\pm$115.8 & 91.3$\pm$34.1 & 286.0$\pm$85.0 & 136.2$\pm$42.3 & -- & -- & 1.8 & 0 & Ls & L1\\[0.5ex]
49 & 361.7$\pm$101.6 & -- & 286.0$\pm$71.9 & -- & 121.3$\pm$33.8 & -- & 1.8 & 0 & NIs & NI1\\[0.5ex]
50 & 567.7$\pm$193.0 & 88.6$\pm$33.2 & 286.0$\pm$81.4 & 141.4$\pm$36.0 & -- & -- & 1.1 & 0 & Ls & L1\\[0.5ex]
51$^{a}$ & 323.2$\pm$126.3 & -- & 286.0$\pm$85.0 & -- & -- & -- & 1.6 & 0 & Shs & Sh1\\[0.5ex]
51$^{b}$ & 507.3$\pm$159.4 & 79.3$\pm$27.4 & 286.0$\pm$80.6 & 99.1$\pm$33.5 & -- & -- & 1.6 & 1 & Shs & Sh1\\[0.5ex]
52 & 207.2$\pm$54.6 & 63.8$\pm$18.3 & 286.0$\pm$51.6 & 101.3$\pm$11.6 & 69.9$\pm$10.1 & -- & 1.4 & 0 & NIs & NI1\\[0.5ex]
53 & 441.0$\pm$180.1 & 112.0$\pm$45.1 & 220.0$\pm$77.2 & 221.0$\pm$63.0 & 153.3$\pm$54.2 & -- & 0.0 & 0 & Ls & L1\\[0.5ex]
54 & 327.7$\pm$113.8 & 98.9$\pm$34.3 & 286.0$\pm$80.8 & 192.1$\pm$43.9 & -- & -- & 1.0 & 0 & Ls & L1\\[0.5ex]
55$^\dagger$ & 275.4$\pm$129.0 & -- & 286.0$\pm$102.8 & 199.4$\pm$76.9 & -- & -- &0.6&0& Ls & L1\\[0.5ex]
56$^{\triangledown}$ & -- & -- & 218.9$\pm$73.6 & -- & -- & -- & 0.1 & 1 & Shs & Sh3\\[0.5ex]
57 & -- & -- & 281.5$\pm$107.0 & 516.7$\pm$170.7 & -- & -- & 0.0 & 0 & Ls & L1\\[0.5ex]
58$^{a}$ & 488.9$\pm$144.3 & 146.5$\pm$38.4 & 286.0$\pm$78.6 & 337.9$\pm$75.0 & 181.1$\pm$44.7 & -- & 1.1 & 0 & Ls & L1\\[0.5ex]
58$^{b}$ & -- & 133.2$\pm$75.3 & 286.0$\pm$152.5 & 88.2$\pm$37.1 & -- & -- & 1.9 & 1 & Shs & Sh3\\[0.5ex]
\hline\\[-2ex]
\end{tabular}
\tablefoot{Continued.}
\end{table*}

\newpage
\clearpage

\begin{table*}
\setlength{\tabcolsep}{5pt}
\tiny
\begin{tabular}{c | c c c c c c c c c c }
\hline\hline  \\[-2ex]
(1) & (2) & (3) & (4) & (5) & (6) & (7) & (8) & (9) & (10) & (11) \\[0.5ex] 
Region & [\ion{O}{ii}]$\lambda$3727 & [\ion{O}{iii}]$\lambda$5007 & H$\alpha$ & [\ion{N}{ii}]$\lambda$6584 & [\ion{S}{ii}]$\lambda$6716 & [\ion{S}{ii}]$\lambda$6731 & A$_v$ & Vel & Zone & Subzone\\[0.5ex]
ID &  &  &  &  &  &  &  & (mag) &  &  \\[0.5ex]
\hline\\[-2ex]
59$^{a}$ & 155.5$\pm$34.4 & 69.6$\pm$15.4 & 253.1$\pm$48.9 & 208.1$\pm$48.9 & 133.3$\pm$30.7 & -- & 0.0 & 0 & Ls & L1\\[0.5ex]
59$^{b}$ & 1009.2$\pm$250.4 & 67.9$\pm$26.3 & 286.0$\pm$86.9 & 39.1$\pm$14.3 & -- & -- & 2.3 & 1 & Shs & Sh3\\[0.5ex]
60$^{a}$ & 211.9$\pm$77.3 & 49.5$\pm$17.1 & 270.8$\pm$57.5 & 292.6$\pm$78.6 & -- & -- & 0.0 & 0 & Ls & L1\\[0.5ex]
60$^{b}$ & 997.2$\pm$355.3 & 145.4$\pm$46.8 & 286.0$\pm$91.0 & 99.6$\pm$26.9 & -- & -- & 3.0 & 1 & Shs & Sh3\\[0.5ex]
61 & 183.9$\pm$67.6 & -- & 286.0$\pm$91.5 & 234.0$\pm$58.7 & 129.6$\pm$45.4 & -- & 0.3 & 0 & SSQA & SSQA\\[0.5ex]
62$^{\triangledown}$ & 194.9$\pm$51.6 & -- & 265.3$\pm$48.1 & 74.1$\pm$21.1 & -- & -- & 0.0 & 1 & Shs & Sh1\\[0.5ex]
63 & 256.8$\pm$98.8 & -- & 286.0$\pm$83.1 & 130.6$\pm$48.4 & -- & -- & 1.0 & 1 & Shs & Sh1\\[0.5ex]
64 & 171.3$\pm$41.6 & 37.7$\pm$13.6 & 286.0$\pm$50.3 & 104.4$\pm$35.4 & -- & -- & 0.4 & 1 & Shs & Sh2\\[0.5ex]
65 & 327.9$\pm$132.4 & 82.2$\pm$32.4 & 286.0$\pm$77.9 & 83.4$\pm$13.7 & 47.7$\pm$12.8 & -- & 2.4 & 0 & NIs & NI1\\[0.5ex]
66$^{a}$ & 728.8$\pm$301.5 & 135.1$\pm$41.8 & 286.0$\pm$80.2 & 260.7$\pm$55.8 & 85.1$\pm$22.0 & -- & 2.9 & 0 & Ls & L1\\[0.5ex]
66$^{b}$ & 675.8$\pm$312.9 & -- & 286.0$\pm$104.6 & -- & -- & -- & 1.3 & 1 & Shs & Sh3\\[0.5ex]
67 & -- & 70.8$\pm$26.2 & 286.0$\pm$75.6 & 209.0$\pm$44.7 & 122.2$\pm$35.2 & -- & 2.1 & 0 & Ls & L1\\[0.5ex]
68$^{a}$ & 305.2$\pm$82.8 & 66.8$\pm$22.2 & 286.0$\pm$60.8 & 246.8$\pm$48.2 & 85.6$\pm$20.2 & -- & 1.1 & 0 & Ls & L1\\[0.5ex]
68$^{b}$ & 416.7$\pm$138.6 & -- & 286.0$\pm$106.9 & -- & -- & -- & 0.6 & 1 & Shs & Sh3\\[0.5ex]
69$^{a}$ & -- & 93.5$\pm$37.5 & 286.0$\pm$97.8 & 168.9$\pm$54.3 & 152.6$\pm$49.5 & -- & 0.5 & 0 & Ls & L1\\[0.5ex]
69$^{b}$$^\dagger$ & 318.9$\pm$147.6 & -- & 273.6$\pm$107.6 & 126.9$\pm$58.1 & -- & -- &0.0&1& Shs & Sh4\\[0.5ex]
70 & 330.5$\pm$102.1 & 56.1$\pm$20.8 & 286.0$\pm$56.0 & 125.6$\pm$16.6 & 82.1$\pm$15.7 & -- & 1.9 & 0 & SSQA & SSQA\\[0.5ex]
71 & 145.2$\pm$50.9 & 45.2$\pm$15.3 & 264.0$\pm$48.9 & 209.7$\pm$36.4 & 90.4$\pm$26.9 & -- & 0.0 & 0 & SSQA & SSQA\\[0.5ex]
72 & 166.5$\pm$64.0 & 78.4$\pm$33.7 & 259.0$\pm$82.2 & -- & -- & -- & 0.0 & 1 & Shs & Sh1\\[0.5ex]
73 & -- & 44.5$\pm$15.7 & 280.0$\pm$53.5 & -- & -- & -- & 0.0 & 1 & Shs & Sh1\\[0.5ex]
74 & 280.5$\pm$67.1 & 33.3$\pm$12.3 & 286.0$\pm$51.0 & 66.8$\pm$20.5 & -- & -- & 1.1 & 1 & Shs & Sh2\\[0.5ex]
75 & 159.2$\pm$47.1 & 63.8$\pm$19.7 & 286.0$\pm$50.9 & 94.3$\pm$24.0 & 93.3$\pm$24.6 & -- & 0.8 & 0 & NIs & NI1\\[0.5ex]
76$^{a}$ & 369.7$\pm$98.8 & 56.8$\pm$15.9 & 286.0$\pm$54.8 & 118.2$\pm$30.2 & 83.0$\pm$22.6 & -- & 1.1 & 0 & Shs & Sh4\\[0.5ex]
76$^{b}$ & 416.1$\pm$138.5 & 65.1$\pm$23.3 & 286.0$\pm$76.4 & 114.3$\pm$38.9 & -- & -- & 0.8 & 1 & Shs & Sh4\\[0.5ex]
77 & 204.6$\pm$62.2 & 64.0$\pm$19.4 & 286.0$\pm$61.2 & 86.8$\pm$29.0 & -- & -- & 0.7 & 1 & Shs & Sh2\\[0.5ex]
78 & 370.1$\pm$121.6 & 78.9$\pm$26.7 & 286.0$\pm$76.8 & 42.8$\pm$14.5 & 82.2$\pm$18.7 & -- & 1.3 & 0 & Shs & Sh1\\[0.5ex]
79$^{a}$ & 496.8$\pm$184.0 & -- & 286.0$\pm$103.0 & 135.5$\pm$49.2 & -- & -- & 1.3 & 0 & Shs & Sh3\\[0.5ex]
79$^{b}$ & 703.9$\pm$230.4 & 112.2$\pm$34.7 & 286.0$\pm$74.0 & -- & -- & -- & 2.4 & 1 & Shs & Sh3\\[0.5ex]
80$^{a}$ & -- & -- & 286.0$\pm$74.6 & 152.3$\pm$47.2 & -- & -- & 0.2 & 0 & Shs & Sh4\\[0.5ex]
80$^{b}$$^{\triangledown}$ & -- & -- & 147.4$\pm$48.2 & -- & -- & -- & 0.0 & 1 & Shs & Sh4\\[0.5ex]
81$^{a}$ & -- & 179.3$\pm$60.2 & 286.0$\pm$95.6 & 123.1$\pm$43.2 & -- & -- & 1.5 & 0 & Ls & L1\\[0.5ex]
81$^{b}$ & -- & 61.5$\pm$24.6 & 249.2$\pm$73.1 & -- & -- & -- & 0.0 & 1 & Shs & Sh4\\[0.5ex]
82 & 244.7$\pm$59.4 & 159.3$\pm$27.0 & 286.0$\pm$42.2 & 57.1$\pm$10.9 & 42.5$\pm$11.4 & -- & 0.6 & 0 & NW & NW\\[0.5ex]
83$^{a}$ & 570.6$\pm$243.3 & -- & 286.0$\pm$110.3 & 98.8$\pm$36.2 & -- & -- & 2.5 & 0 & Shs & Sh2\\[0.5ex]
83$^{b}$ & 385.5$\pm$88.1 & 81.4$\pm$20.1 & 286.0$\pm$62.3 & 58.3$\pm$17.1 & -- & -- & 1.3 & 1 & Shs & Sh2\\[0.5ex]
84 & 452.9$\pm$138.0 & -- & 286.0$\pm$66.6 & -- & -- & -- & 1.4 & 1 & Shs & Sh3\\[0.5ex]
85 & -- & -- & 286.0$\pm$84.7 & 82.7$\pm$29.2 & -- & -- & 0.4 & 1 & Shs & Sh3\\[0.5ex]
86$^{a}$ & 590.7$\pm$271.9 & 133.8$\pm$53.5 & 286.0$\pm$106.7 & 304.7$\pm$74.8 & 110.7$\pm$32.5 & -- & 2.7 & 0 & Ls & L1\\[0.5ex]
86$^{b}$$^\dagger$ & 337.3$\pm$152.9 & -- & 260.2$\pm$117.4 & 157.2$\pm$76.1 & -- & -- &0.0&1& Shs & Sh3\\[0.5ex]
87$^{a}$ & 376.4$\pm$125.2 & -- & 286.0$\pm$90.3 & -- & -- & -- & 1.4 & 0 & Shs & Sh3\\[0.5ex]
87$^{b}$$^\dagger$ & -- & 67.7$\pm$41.0 & 286.0$\pm$118.5 & -- & -- & -- & 3.1 & 1 & Shs & Sh3\\[0.5ex]
88 & -- & -- & 286.0$\pm$82.0 & 192.1$\pm$45.9 & -- & -- & 1.7 & 1 & Ls & L1\\[0.5ex]
89$^{a}$ & 265.2$\pm$105.5 & -- & 286.0$\pm$85.1 & -- & -- & -- & 0.5 & 0 & Shs & Sh4\\[0.5ex]
89$^{b}$$^\dagger$ & -- & -- & 256.9$\pm$101.7 & 128.2$\pm$58.7 & -- & -- &0.0&1& Shs & Sh4\\[0.5ex]
90 & 333.8$\pm$106.6 & 110.4$\pm$29.3 & 286.0$\pm$59.5 & 139.5$\pm$18.5 & 57.8$\pm$15.5 & -- & 1.6 & 0 & SSQA & SSQA\\[0.5ex]
91 & 307.3$\pm$95.5 & 65.7$\pm$24.5 & 286.0$\pm$67.5 & 100.8$\pm$26.0 & -- & -- & 1.3 & 0 & SSQA & SSQA\\[0.5ex]
92 & 838.2$\pm$364.9 & 397.8$\pm$130.9 & 286.0$\pm$93.5 & 93.4$\pm$22.6 & -- & -- & 1.5 & 0 & SSQA & SSQA\\[0.5ex]
93$^{\triangledown}$ & -- & -- & 218.2$\pm$69.5 & -- & 180.5$\pm$54.2 & -- & 0.0 & 0 & Shs & Sh1\\[0.5ex]
94$^{a}$ & -- & -- & 286.0$\pm$101.7 & 113.3$\pm$31.6 & -- & -- & 2.0 & 0 & NIs & NI1\\[0.5ex]
94$^{b}$ & -- & -- & 231.7$\pm$70.7 & 142.8$\pm$50.7 & -- & -- & 0.0 & 1 & Shs & Sh2\\[0.5ex]
95$^{a}$$^\dagger$ & -- & -- & 237.0$\pm$135.6 & 212.2$\pm$126.6 & -- & -- & 0.0 & 0 & Shs & Sh2\\[0.5ex]
95$^{b}$ & 203.1$\pm$69.4 & -- & 271.6$\pm$64.5 & -- & -- & -- & 0.0 & 1 & Shs & Sh2\\[0.5ex]
96 & 335.4$\pm$104.4 & 61.5$\pm$19.5 & 286.0$\pm$58.7 & 72.9$\pm$19.5 & -- & -- & 1.8 & 1 & Hs & H2\\[0.5ex]
97 & 126.2$\pm$32.1 & 377.8$\pm$62.5 & 286.0$\pm$47.9 & 34.9$\pm$11.9 & -- & -- & 0.4 & 0 & NW & NW\\[0.5ex]
98 & 456.7$\pm$142.8 & 78.3$\pm$22.6 & 286.0$\pm$57.1 & 212.3$\pm$28.8 & 86.2$\pm$20.7 & -- & 2.0 & 0 & SSQA & SSQA\\[0.5ex]
99 & 454.8$\pm$181.5 & 99.6$\pm$32.1 & 286.0$\pm$84.4 & -- & -- & -- & 1.3 & 1 & Shs & Sh1\\[0.5ex]
100$^{a}$ & -- & -- & 232.8$\pm$83.6 & 149.3$\pm$59.3 & -- & -- & 0.0 & 0 & Shs & Sh3\\[0.5ex]
100$^{b}$ & -- & -- & 258.0$\pm$92.3 & -- & -- & -- & 0.0 & 1 & Shs & Sh2\\[0.5ex]
101 & 272.7$\pm$91.1 & -- & 286.0$\pm$89.7 & 183.5$\pm$71.8 & -- & -- & 0.5 & 1 & Shs & Sh3\\[0.5ex]
102$^{a}$ & 424.2$\pm$170.3 & 110.2$\pm$37.3 & 286.0$\pm$89.1 & 120.1$\pm$40.1 & 61.3$\pm$23.8 & -- & 1.3 & 0 & Shs & Sh3\\[0.5ex]
102$^{b}$ & -- & -- & 218.3$\pm$97.1 & -- & -- & -- & 0.0 & 1 & Shs & Sh3\\[0.5ex]
\hline\\[-2ex]
\end{tabular}
\tablefoot{Continued.}
\end{table*}

\newpage
\clearpage

\begin{table*}
\setlength{\tabcolsep}{5pt}
\tiny
\begin{tabular}{c | c c c c c c c c c c }
\hline\hline  \\[-2ex]
(1) & (2) & (3) & (4) & (5) & (6) & (7) & (8) & (9) & (10) & (11) \\[0.5ex] 
Region & [\ion{O}{ii}]$\lambda$3727 & [\ion{O}{iii}]$\lambda$5007 & H$\alpha$ & [\ion{N}{ii}]$\lambda$6584 & [\ion{S}{ii}]$\lambda$6716 & [\ion{S}{ii}]$\lambda$6731 & A$_v$ & Vel & Zone & Subzone\\[0.5ex]
ID &  &  &  &  &  &  &  & (mag) &  &  \\[0.5ex]
\hline\\[-2ex]
103$^{a}$$^\dagger$ & -- & 112.7$\pm$46.0 & 286.0$\pm$103.3 & 263.3$\pm$97.8 & -- & -- & 1.9 & 0 & Ls & L1\\[0.5ex]
103$^{b}$$^\triangledown$ & 253.1$\pm$102.7 & -- & 171.7$\pm$58.9 & -- & -- & -- & 0.0 & 1 & Shs & Sh3\\[0.5ex]
104$^{a}$$^\dagger$ & -- & -- & 228.0$\pm$108.8 & 205.9$\pm$102.7 & -- & -- &0.0&0& Shs & Sh4\\[0.5ex]
104$^{b}$ & 356.4$\pm$101.3 & 52.9$\pm$19.0 & 286.0$\pm$62.0 & 89.4$\pm$23.7 & -- & -- & 1.1 & 1 & Shs & Sh4\\[0.5ex]
105 & 450.5$\pm$174.3 & -- & 286.0$\pm$88.5 & -- & -- & -- & 1.9 & 1 & Hs & H2\\[0.5ex]
106 & 339.7$\pm$126.9 & -- & 286.0$\pm$83.0 & 146.4$\pm$27.1 & -- & -- & 1.3 & 0 & SSQA & SSQA\\[0.5ex]
107 & 311.8$\pm$134.2 & -- & 286.0$\pm$111.6 & 89.5$\pm$43.5 & -- & -- & 1.3 & 1 & Shs & Sh3\\[0.5ex]
108 & 335.3$\pm$113.8 & -- & 286.0$\pm$83.0 & -- & -- & -- & 1.7 & 1 & Shs & Sh3\\[0.5ex]
109$^{\triangledown}$ & -- & 71.4$\pm$25.9 & 221.8$\pm$71.7 & 112.9$\pm$46.7 & -- & -- & 0.0 & 1 & Shs & Sh4\\[0.5ex]
110 & 311.1$\pm$88.3 & -- & 286.0$\pm$58.7 & 89.8$\pm$18.3 & -- & -- & 0.9 & 1 & Hs & H2\\[0.5ex]
111 & 244.1$\pm$93.4 & 270.8$\pm$85.6 & 286.0$\pm$85.0 & -- & -- & -- & 1.8 & 1 & Hs & H1\\[0.5ex]
112 & 228.6$\pm$52.2 & 141.3$\pm$26.4 & 286.0$\pm$44.8 & 102.9$\pm$12.5 & 62.0$\pm$12.0 & -- & 0.9 & 0 & Ls & L2\\[0.5ex]
113 & -- & 103.8$\pm$38.4 & 286.0$\pm$88.4 & 103.5$\pm$33.7 & -- & -- & 1.6 & 1 & Shs & Sh3\\[0.5ex]
114 & 190.1$\pm$60.1 & -- & 258.1$\pm$72.6 & 81.2$\pm$30.0 & -- & -- & 0.0 & 1 & Shs & Sh3\\[0.5ex]
115 & -- & -- & 286.0$\pm$66.6 & 60.6$\pm$20.5 & -- & -- & 0.6 & 1 & Hs & H2\\[0.5ex]
116 & 274.9$\pm$67.3 & -- & 286.0$\pm$54.3 & -- & -- & -- & 1.0 & 1 & Hs & H2\\[0.5ex]
117 & 157.6$\pm$22.5 & 149.5$\pm$18.5 & 286.0$\pm$28.8 & 66.7$\pm$3.2 & 26.9$\pm$2.5 & 27.0$\pm$2.5 & 1.0 & 0 & NIs & NI1\\[0.5ex]
118 & 425.8$\pm$146.5 & -- & 286.0$\pm$68.7 & 40.5$\pm$14.3 & -- & -- & 1.1 & 0 & NIs & NI1\\[0.5ex]
119$^{\triangledown}$ & -- & -- & 248.7$\pm$80.7 & 118.0$\pm$40.3 & -- & -- & 0.0 & 1 & Shs & Sh3\\[0.5ex]
120 & 190.6$\pm$50.0 & 161.6$\pm$29.5 & 286.0$\pm$45.0 & 133.4$\pm$13.5 & 85.2$\pm$12.4 & -- & 0.7 & 0 & SSQA & SSQA\\[0.5ex]
121 & 357.7$\pm$97.4 & 192.0$\pm$37.0 & 286.0$\pm$48.0 & 79.4$\pm$9.5 & -- & -- & 2.3 & 1 & SQA & SQA\\[0.5ex]
122 & 200.7$\pm$33.8 & 204.0$\pm$20.3 & 286.0$\pm$24.7 & 49.0$\pm$3.4 & -- & -- & 2.3 & 1 & SQA & SQA\\[0.5ex]
123 & -- & 121.3$\pm$38.0 & 242.8$\pm$68.2 & 126.9$\pm$42.4 & -- & -- & 0.0 & 1 & Hs & H1\\[0.5ex]
124$^\dagger$ & 305.9$\pm$124.8 & 251.9$\pm$94.3 & 286.0$\pm$104.2 & -- & -- & -- & 1.1 & 0 & NW & NW\\[0.5ex]
125 & 332.9$\pm$124.9 & 75.1$\pm$33.0 & 286.0$\pm$89.8 & 109.5$\pm$29.4 & 85.7$\pm$28.8 & -- & 1.4 & 0 & NIs & NI2\\[0.5ex]
126 & 381.3$\pm$147.9 & 224.7$\pm$57.5 & 286.0$\pm$67.5 & -- & -- & -- & 2.8 & 1 & Hs & H1\\[0.5ex]
127 & 134.3$\pm$34.2 & 83.8$\pm$19.1 & 286.0$\pm$46.7 & 118.0$\pm$12.7 & 92.2$\pm$12.4 & -- & 0.3 & 0 & NIs & NI2\\[0.5ex]
128 & 362.7$\pm$130.0 & 121.4$\pm$45.5 & 286.0$\pm$82.6 & 111.4$\pm$12.9 & 72.6$\pm$12.2 & -- & 1.9 & 0 & NIs & NI2\\[0.5ex]
129 & 246.0$\pm$54.4 & 99.9$\pm$18.0 & 286.0$\pm$38.0 & 99.9$\pm$7.8 & 49.8$\pm$7.3 & -- & 1.0 & 0 & NSQA & NSQA\\[0.5ex]
130 & 100.5$\pm$30.2 & 62.5$\pm$19.9 & 286.0$\pm$56.6 & 83.0$\pm$9.9 & 30.6$\pm$9.5 & -- & 0.7 & 0 & NIs & NI3\\[0.5ex]
131 & 679.4$\pm$277.9 & 262.6$\pm$96.9 & 286.0$\pm$99.5 & -- & -- & -- & 3.0 & 1 & Hs & H1\\[0.5ex]
132 & 237.7$\pm$48.0 & 126.0$\pm$18.6 & 286.0$\pm$33.8 & 101.8$\pm$8.0 & 63.8$\pm$7.7 & -- & 1.4 & 0 & NSQA & NSQA\\[0.5ex]
133 & 159.7$\pm$18.5 & 177.0$\pm$15.6 & 286.0$\pm$21.3 & 89.4$\pm$3.6 & 40.2$\pm$3.4 & 19.2$\pm$3.5 & 1.0 & 0 & NIs & NI2\\[0.5ex]
134 & 202.5$\pm$72.4 & 53.4$\pm$20.9 & 286.0$\pm$59.0 & 111.6$\pm$19.3 & 69.1$\pm$18.2 & -- & 1.3 & 0 & NIs & NI3\\[0.5ex]
135 & 135.9$\pm$55.2 & -- & 286.0$\pm$84.6 & 126.0$\pm$21.7 & 86.6$\pm$20.4 & -- & 0.3 & 0 & NSQA & NSQA\\[0.5ex]
136 & 258.3$\pm$79.2 & 123.7$\pm$27.6 & 286.0$\pm$50.4 & 108.8$\pm$11.6 & 62.0$\pm$10.9 & -- & 1.6 & 0 & NSQA & NSQA\\[0.5ex]
137 & 180.8$\pm$49.4 & 87.9$\pm$17.0 & 286.0$\pm$39.1 & 116.2$\pm$10.9 & 45.8$\pm$9.7 & -- & 1.4 & 0 & NSQA & NSQA\\[0.5ex]
138 & 189.6$\pm$44.7 & 85.9$\pm$20.0 & 286.0$\pm$46.5 & 112.1$\pm$8.6 & 62.5$\pm$8.0 & 36.0$\pm$8.0 & 1.1 & 0 & NIs & NI2\\[0.5ex]
139 & 100.5$\pm$21.0 & 124.6$\pm$14.2 & 286.0$\pm$26.4 & 80.3$\pm$6.8 & 52.9$\pm$6.8 & -- & 0.3 & 0 & NSQA & NSQA\\[0.5ex]
140 & 222.9$\pm$68.9 & 90.8$\pm$19.7 & 286.0$\pm$46.2 & 109.4$\pm$15.9 & 44.2$\pm$14.3 & -- & 1.5 & 0 & NSQA & NSQA\\[0.5ex]
141 & 192.7$\pm$36.7 & 120.9$\pm$19.6 & 286.0$\pm$36.0 & 94.6$\pm$6.7 & 53.7$\pm$6.4 & -- & 1.0 & 0 & NSQA & NSQA\\[0.5ex]
142 & 179.4$\pm$42.9 & 104.9$\pm$21.8 & 286.0$\pm$44.8 & 118.0$\pm$10.1 & 75.1$\pm$9.5 & -- & 1.1 & 0 & NSQA & NSQA\\[0.5ex]
143 & 155.9$\pm$45.0 & 81.4$\pm$16.6 & 286.0$\pm$39.6 & 119.9$\pm$11.7 & 76.9$\pm$11.0 & -- & 1.0 & 0 & NSQA & NSQA\\[0.5ex]
144$^\dagger$ & -- & 146.3$\pm$61.3 & 286.0$\pm$116.3 & 102.9$\pm$51.0 & -- & -- &0.4&0& SDR & SDR\\[0.5ex]
145 & -- & 176.5$\pm$60.1 & 254.5$\pm$82.9 & -- & -- & -- & 0.0 & 0 & NSQA & NSQA\\[0.5ex]
146 & 161.5$\pm$38.8 & 82.3$\pm$14.6 & 286.0$\pm$35.2 & 111.2$\pm$7.3 & 59.4$\pm$6.8 & -- & 1.0 & 0 & NSQA & NSQA\\[0.5ex]
147 & 814.9$\pm$336.9 & 127.8$\pm$48.3 & 286.0$\pm$91.8 & 105.3$\pm$28.4 & -- & -- & 2.7 & 0 & NSQA & NSQA\\[0.5ex]
148 & 270.5$\pm$109.7 & 65.1$\pm$25.2 & 286.0$\pm$71.5 & 112.9$\pm$19.9 & 66.7$\pm$19.1 & -- & 1.5 & 0 & Ls & L4\\[0.5ex]
149 & 300.8$\pm$115.5 & 87.1$\pm$29.8 & 286.0$\pm$72.9 & 105.2$\pm$26.6 & 85.2$\pm$26.8 & -- & 0.8 & 0 & NIs & NI4\\[0.5ex]
150 & 254.5$\pm$76.3 & 64.4$\pm$25.2 & 286.0$\pm$68.8 & 94.5$\pm$20.4 & 70.4$\pm$20.4 & -- & 1.2 & 0 & NIs & NI4\\[0.5ex]
151 & -- & 94.5$\pm$40.6 & 286.0$\pm$90.0 & 147.8$\pm$25.8 & 89.5$\pm$24.8 & -- & 0.6 & 0 & NIs & NI4\\[0.5ex]
152 & 175.9$\pm$54.9 & 81.9$\pm$28.9 & 286.0$\pm$69.6 & 64.3$\pm$15.5 & 79.0$\pm$17.2 & -- & 1.5 & 0 & NIs & NI4\\[0.5ex]
153 & 193.7$\pm$48.2 & 84.3$\pm$24.7 & 286.0$\pm$58.6 & 86.6$\pm$11.0 & 50.2$\pm$10.9 & -- & 1.2 & 0 & NIs & NI4\\[0.5ex]
154 & 138.0$\pm$37.8 & 114.7$\pm$22.0 & 286.0$\pm$42.5 & 98.2$\pm$9.6 & 41.1$\pm$9.0 & -- & 0.7 & 0 & Ls & L4\\[0.5ex]
155 & 199.1$\pm$53.5 & 215.2$\pm$48.7 & 286.0$\pm$57.8 & 70.6$\pm$11.0 & -- & -- & 1.3 & 0 & Ls & L4\\[0.5ex]
156 & 197.6$\pm$53.1 & 121.6$\pm$23.9 & 286.0$\pm$49.6 & 80.3$\pm$18.0 & -- & -- & 0.2 & 0 & SDR & SDR\\[0.5ex]
157 & 194.9$\pm$55.9 & 124.3$\pm$29.1 & 286.0$\pm$53.7 & 89.8$\pm$12.9 & 49.3$\pm$12.5 & -- & 1.3 & 0 & NIs & NI4\\[0.5ex]
158 & -- & 81.5$\pm$27.1 & 212.1$\pm$65.0 & -- & 134.1$\pm$46.7 & -- & 0.0 & 0 & SDR & SDR\\[0.5ex]
159 & -- & 75.2$\pm$26.0 & 286.0$\pm$67.7 & 104.8$\pm$16.8 & 68.1$\pm$16.1 & -- & 0.6 & 0 & NIs & NI4\\[0.5ex]
160 & -- & 185.9$\pm$61.3 & 286.0$\pm$87.7 & 86.8$\pm$23.2 & -- & -- & 2.2 & 0 & SDR & SDR\\[0.5ex]
161$^{a}$ & -- & -- & 286.0$\pm$75.7 & 48.6$\pm$14.0 & -- & -- & 2.0 & 0 & Ls & L3\\[0.5ex]
161$^{b}$ & 353.9$\pm$152.1 & -- & 286.0$\pm$112.0 & 71.8$\pm$28.4 & -- & -- & 0.5 & 1 & Ls & L3\\[0.5ex]
\hline\\[-2ex]
\end{tabular}
\tablefoot{Continued.}
\end{table*}

\newpage
\clearpage

\begin{table*}
\setlength{\tabcolsep}{5pt}
\tiny
\begin{tabular}{c | c c c c c c c c c c }
\hline\hline  \\[-2ex]
(1) & (2) & (3) & (4) & (5) & (6) & (7) & (8) & (9) & (10) & (11) \\[0.5ex] 
Region & [\ion{O}{ii}]$\lambda$3727 & [\ion{O}{iii}]$\lambda$5007 & H$\alpha$ & [\ion{N}{ii}]$\lambda$6584 & [\ion{S}{ii}]$\lambda$6716 & [\ion{S}{ii}]$\lambda$6731 & A$_v$ & Vel & Zone & Subzone\\[0.5ex]
ID &  &  &  &  &  &  &  & (mag) &  &  \\[0.5ex]
\hline\\[-2ex]
162 & -- & -- & 221.4$\pm$59.1 & 103.4$\pm$53.1 & -- & -- & 0.0 & 0 & Ls & L3\\ [0.5ex]
163 & -- & 154.3$\pm$56.5 & 286.0$\pm$94.6 & -- & -- & -- & 0.4 & 0 & SDR & SDR\\[0.5ex]
164 & 192.5$\pm$26.8 & 167.8$\pm$19.3 & 286.0$\pm$27.4 & 80.0$\pm$4.3 & 56.5$\pm$4.3 & -- & 0.8 & 0 & NIs & NI5\\[0.5ex]
165 & 173.3$\pm$24.5 & 158.9$\pm$19.5 & 286.0$\pm$29.1 & 77.9$\pm$4.7 & 44.2$\pm$4.6 & -- & 0.7 & 0 & NIs & NI5\\[0.5ex]
166 & 205.5$\pm$37.4 & 194.9$\pm$26.1 & 286.0$\pm$33.3 & 76.8$\pm$6.0 & 47.8$\pm$6.0 & -- & 1.2 & 0 & NIs & NI5\\[0.5ex]
167 & 230.3$\pm$47.9 & 202.9$\pm$31.4 & 286.0$\pm$38.9 & 56.8$\pm$6.8 & 46.4$\pm$7.1 & -- & 1.1 & 0 & NIs & NI5\\[0.5ex]
168 & 157.4$\pm$37.0 & 207.4$\pm$32.2 & 286.0$\pm$39.4 & 66.9$\pm$7.7 & 39.1$\pm$7.9 & 39.0$\pm$11.3 & 0.0 & 0 & SDR & SDR\\[0.5ex]
169 & 218.5$\pm$41.8 & 168.6$\pm$25.8 & 286.0$\pm$38.7 & 78.7$\pm$10.8 & 75.7$\pm$11.5 & -- & 0.7 & 0 & SDR & SDR\\[0.5ex]
170 & 100.6$\pm$26.3 & 164.3$\pm$22.6 & 269.2$\pm$33.4 & 62.3$\pm$10.6 & 40.9$\pm$11.0 & -- & 0.0 & 0 & SDR & SDR\\[0.5ex]
171 & 175.2$\pm$37.2 & 213.9$\pm$27.4 & 286.0$\pm$34.0 & 67.5$\pm$9.0 & 54.9$\pm$9.5 & -- & 0.2 & 0 & SDR & SDR\\[0.5ex]
172 & 124.8$\pm$21.9 & 341.0$\pm$32.3 & 286.0$\pm$27.0 & 36.8$\pm$6.7 & 21.9$\pm$7.2 & -- & 0.6 & 0 & SDR & SDR\\[0.5ex]
173 & 252.7$\pm$54.6 & 239.3$\pm$39.9 & 286.0$\pm$43.0 & 52.7$\pm$6.7 & 41.6$\pm$7.1 & -- & 1.0 & 0 & SDR & SDR\\[0.5ex]
174 & 336.0$\pm$73.4 & 189.2$\pm$35.7 & 286.0$\pm$46.1 & 71.0$\pm$7.7 & 46.5$\pm$8.2 & -- & 1.3 & 0 & SDR & SDR\\[0.5ex]
175 & 443.3$\pm$164.6 & 307.9$\pm$101.0 & 286.0$\pm$89.5 & -- & -- & -- & 2.0 & 0 & SDR & SDR\\[0.5ex]
176 & 600.4$\pm$241.0 & 80.0$\pm$33.9 & 286.0$\pm$81.0 & 57.4$\pm$20.2 & -- & -- & 3.6 & 1 & NG & NG\\[0.5ex]
\hline\\[-2ex]
\end{tabular}
\tablefoot{The columns correspond to: 
(1) Identifier of the SQ H$\alpha$ emission regions; 
(2) Integrated [\ion{O}{ii}]$\lambda$3727 flux, relative to H$\beta$=100, and its uncertainty;
(3) Integrated [\ion{O}{iii}]$\lambda$5007 flux, relative to H$\beta$=100, and its uncertainty;
(4) Integrated H$\alpha$ flux, relative to H$\beta$=100, and its uncertainty;
(5) Integrated [\ion{N}{ii}]$\lambda$6584 flux, relative to H$\beta$=100, and its uncertainty;
(6) Integrated [\ion{S}{ii}]$\lambda$6716 flux, relative to H$\beta$=100, and its uncertainty;
(7) Integrated [\ion{S}{ii}]$\lambda$6731 flux, relative to H$\beta$=100, and its uncertainty;
(8) A$_v$ extinction (mag); 
(9) Velocity of the region belongs to the LV (0) or HV (1) sub-sample;
(10) Global localisation zone;
(11) Local zone.
``*'' ID 1 was assigned to the galaxy NGC7320c in the kinematical study done in paper I.
``$^\dagger$'' Line ratios of the regions where H$\beta$ is not measured; these line ratios are computed by estimating an H$\beta$ flux (3$\sigma$, see paper I) value and are not used in our analysis. ``$^\triangledown$'' These line ratios may be affected by the kinematics and the conspicuous complex line profile of the region. 
For regions with two velocity components we use `$^{a}$' and `$^{b}$' for low- and high-velocity components, respectively.
}
\end{table*}

\begin{table*}
\setlength{\tabcolsep}{3pt}
\begin{minipage}[t]{.48\linewidth}
\tiny
\caption{Star formation rate, H$\alpha$ luminosity, BPT classification, and the radial velocity sub-sample.}
\label{table:table4} 
\centering
\begin{tabular}{c | c c c c}
\hline\hline  \\[-2ex]
(1) & (2) & (3) & (4) & (5)\\[0.5ex] 
Region & SFR & L$_{H\alpha}$ & BPT & Vel\\[0.5ex]
ID & (x10$^{-3}\,M_\odot\,yr^{-1}$) & (x10$^{38}erg\,s^{-1}$) &  & \\[0.5ex]
\hline\\[-2ex]
2 & 12.81$\pm$0.69 & 16.21$\pm$0.87 & SF & 1 \\[0.5ex]
3 & 11.08$\pm$0.78 & 14.02$\pm$0.99 & SF & 1 \\[0.5ex]
4 & 27.53$\pm$0.70 & 34.84$\pm$0.89 & SF & 1 \\[0.5ex]
5 & -- & 0.63$\pm$0.09 & C & 1 \\[0.5ex]
6 & 17.70$\pm$1.35 & 22.40$\pm$1.71 & SF & 1 \\[0.5ex]
8 & 31.37$\pm$1.80 & 39.71$\pm$2.27 & SF & 1 \\[0.5ex]
9 & 10.26$\pm$0.46 & 12.99$\pm$0.58 & SF & 1 \\[0.5ex]
10 & 39.85$\pm$0.87 & 50.44$\pm$1.10 & SF & 1 \\[0.5ex]
11 & 3.33$\pm$0.24 & 4.21$\pm$0.30 & SF & 1 \\[0.5ex]
12 & 1.66$\pm$0.16 & 2.10$\pm$0.21 & SF & 1 \\[0.5ex]
13 & 76.48$\pm$1.61 & 96.82$\pm$2.03 & SF & 1 \\[0.5ex]
14 & 94.95$\pm$4.06 & 120.19$\pm$5.14 & SF & 1 \\[0.5ex]
15 & 196.79$\pm$5.31 & 249.10$\pm$6.72 & SF & 1 \\[0.5ex]
24 & 29.67$\pm$2.38 & 37.55$\pm$3.01 & SF & 1 \\[0.5ex]
25 & 17.79$\pm$1.51 & 22.52$\pm$1.91 & SF & 1 \\[0.5ex]
26 & -- & 2690.21$\pm$192.90 & AGN & 1 \\[0.5ex]
27 & -- & 552.11$\pm$55.75 & AGN & 1 \\[0.5ex]
29 & -- & 3.45$\pm$0.53 & AGN & 1 \\[0.5ex]
30 & -- & 23.05$\pm$3.23 & C & 1 \\[0.5ex]
33 & 2.31$\pm$0.27 & 2.93$\pm$0.34 & SF & 0 \\[0.5ex]
35 & 23.79$\pm$2.45 & 30.12$\pm$3.10 & SF & 1 \\[0.5ex]
39 & 6.37$\pm$0.58 & 8.06$\pm$0.73 & SF & 0 \\[0.5ex]
40 & -- & 11.44$\pm$1.22 & C & 1 \\[0.5ex]
42 & 13.18$\pm$0.81 & 16.68$\pm$1.03 & SF & 0 \\[0.5ex]
43 & -- & 3.63$\pm$0.48 & C & 0 \\[0.5ex]
45 & 12.63$\pm$1.32 & 15.99$\pm$1.67 & SF & 0 \\[0.5ex]
46 & 12.23$\pm$1.01 & 15.48$\pm$1.28 & SF & 0 \\[0.5ex]
47 & 0.67$\pm$0.10 & 0.85$\pm$0.12 & SF & 0 \\[0.5ex]
48 & 6.19$\pm$0.96 & 7.83$\pm$1.22 & SF & 0 \\[0.5ex]
50 & 3.00$\pm$0.42 & 3.79$\pm$0.53 & SF & 0 \\[0.5ex]
51$^{b}$ & 15.59$\pm$2.35 & 19.73$\pm$2.97 & SF & 1 \\[0.5ex]
52 & 41.39$\pm$1.85 & 52.39$\pm$2.34 & SF & 0 \\[0.5ex]
53 & -- & 0.82$\pm$0.15 & C & 0 \\[0.5ex]
54 & -- & 2.28$\pm$0.33 & C & 0 \\[0.5ex]
58$^{a}$ & -- & 7.57$\pm$1.26 & AGN & 0 \\[0.5ex]
59$^{a}$ & -- & 10.63$\pm$1.54 & C & 0 \\[0.5ex]
59$^{b}$ & 36.96$\pm$7.87 & 46.78$\pm$9.96 & SF & 1 \\[0.5ex]
60$^{a}$ & -- & 1.26$\pm$0.19 & C & 0 \\[0.5ex]
60$^{b}$ & 11.77$\pm$2.09 & 14.90$\pm$2.64 & SF & 1 \\[0.5ex]
64 & 6.14$\pm$0.63 & 7.77$\pm$0.80 & SF & 1 \\[0.5ex]
65 & 36.82$\pm$2.38 & 46.60$\pm$3.01 & SF & 0 \\[0.5ex]
66$^{a}$ & -- & 17.69$\pm$2.30 & AGN & 0 \\[0.5ex]
67 & -- & 8.74$\pm$1.23 & C & 0 \\[0.5ex]
68$^{a}$ & -- & 16.09$\pm$1.70 & C & 0 \\[0.5ex]
69$^{a}$ & -- & 1.52$\pm$0.23 & C & 0 \\[0.5ex]
70 & 14.75$\pm$1.00 & 18.68$\pm$1.27 & SF & 0 \\[0.5ex]
71 & -- & 2.83$\pm$0.32 & C & 0 \\[0.5ex]
74 & 14.51$\pm$1.27 & 18.37$\pm$1.61 & SF & 1 \\[0.5ex]
75 & 17.09$\pm$1.11 & 21.63$\pm$1.41 & SF & 0 \\[0.5ex]
76$^{a}$ & 7.37$\pm$0.99 & 9.33$\pm$1.26 & SF & 0 \\[0.5ex]
76$^{b}$ & 3.56$\pm$0.67 & 4.51$\pm$0.85 & SF & 1 \\[0.5ex]
77 & 8.81$\pm$0.92 & 11.15$\pm$1.16 & SF & 1 \\[0.5ex]
78 & 2.20$\pm$0.31 & 2.78$\pm$0.39 & SF & 0 \\[0.5ex]
81$^{a}$ & -- & 2.45$\pm$0.39 & C & 0 \\[0.5ex]
82 & 7.07$\pm$0.39 & 8.95$\pm$0.50 & SF & 0 \\[0.5ex]
83$^{b}$ & 22.20$\pm$3.06 & 28.10$\pm$3.87 & SF & 1 \\[0.5ex]
86$^{a}$ & -- & 15.92$\pm$3.06 & AGN & 0 \\[0.5ex]
90 & -- & 19.96$\pm$1.45 & C & 0 \\[0.5ex]
\hline\\[-2ex]
\end{tabular}
\end{minipage}
\begin{minipage}[t]{.48\linewidth}\vspace{1.05cm}
\tiny
\centering
\begin{tabular}{c | c c c c}
\hline\hline  \\[-2ex]
(1) & (2) & (3) & (4) & (5)\\[0.5ex] 
Region & SFR & L$_{H\alpha}$ & BPT & Vel\\[0.5ex]
ID & (x10$^{-3}\,M_\odot\,yr^{-1}$) & (x10$^{38}erg\,s^{-1}$) &  & \\[0.5ex]
\hline\\[-2ex]
91 & 4.29$\pm$0.50 & 5.43$\pm$0.64 & SF & 0 \\[0.5ex]
92 & -- & 6.18$\pm$0.48 & AGN & 0 \\[0.5ex]
96 & 19.53$\pm$2.10 & 24.72$\pm$2.66 & SF & 1 \\[0.5ex]
97 & 3.88$\pm$0.24 & 4.92$\pm$0.31 & SF & 0 \\[0.5ex]
98 & -- & 15.75$\pm$1.40 & C & 0 \\[0.5ex]
102$^{a}$ & 3.40$\pm$0.63 & 4.30$\pm$0.80 & SF & 0 \\[0.5ex]
104$^{b}$ & 11.35$\pm$1.59 & 14.36$\pm$2.01 & SF & 1 \\[0.5ex]
109 & 0.54$\pm$0.13 & 0.69$\pm$0.17 & SF & 1 \\[0.5ex]
112 & 23.20$\pm$1.29 & 29.37$\pm$1.63 & SF & 0 \\[0.5ex]
113 & 3.69$\pm$0.55 & 4.68$\pm$0.70 & SF & 1 \\[0.5ex]
117 & 119.16$\pm$2.67 & 150.84$\pm$3.38 & SF & 0 \\[0.5ex]
120 & -- & 20.64$\pm$1.11 & C & 0 \\[0.5ex]
121 & 139.98$\pm$6.46 & 177.19$\pm$8.18 & SF & 1 \\[0.5ex]
122 & 733.22$\pm$13.18 & 928.12$\pm$16.69 & SF & 1 \\[0.5ex]
123 & 0.57$\pm$0.08 & 0.72$\pm$0.10 & SF & 1 \\[0.5ex]
125 & 7.57$\pm$0.98 & 9.58$\pm$1.24 & SF & 0 \\[0.5ex]
127 & 14.95$\pm$0.80 & 18.93$\pm$1.01 & SF & 0 \\[0.5ex]
128 & 18.60$\pm$1.03 & 23.54$\pm$1.30 & SF & 0 \\[0.5ex]
129 & 59.93$\pm$2.10 & 75.86$\pm$2.65 & SF & 0 \\[0.5ex]
130 & 26.72$\pm$1.25 & 33.82$\pm$1.59 & SF & 0 \\[0.5ex]
132 & 38.31$\pm$1.36 & 48.49$\pm$1.72 & SF & 0 \\[0.5ex]
133 & 94.48$\pm$1.55 & 119.59$\pm$1.97 & SF & 0 \\[0.5ex]
134 & 13.96$\pm$1.16 & 17.66$\pm$1.46 & SF & 0 \\[0.5ex]
136 & 33.43$\pm$1.69 & 42.32$\pm$2.14 & SF & 0 \\[0.5ex]
137 & 43.84$\pm$2.03 & 55.50$\pm$2.57 & SF & 0 \\[0.5ex]
138 & 32.01$\pm$1.18 & 40.52$\pm$1.49 & SF & 0 \\[0.5ex]
139 & 19.86$\pm$0.64 & 25.14$\pm$0.81 & SF & 0 \\[0.5ex]
140 & 21.83$\pm$1.51 & 27.63$\pm$1.91 & SF & 0 \\[0.5ex]
141 & 42.05$\pm$1.28 & 53.22$\pm$1.62 & SF & 0 \\[0.5ex]
142 & 31.38$\pm$1.33 & 39.73$\pm$1.68 & SF & 0 \\[0.5ex]
143 & 29.21$\pm$1.43 & 36.98$\pm$1.81 & SF & 0 \\[0.5ex]
146 & 44.23$\pm$1.39 & 55.99$\pm$1.76 & SF & 0 \\[0.5ex]
147 & 8.64$\pm$1.08 & 10.94$\pm$1.37 & SF & 0 \\[0.5ex]
148 & 8.24$\pm$0.74 & 10.43$\pm$0.94 & SF & 0 \\[0.5ex]
149 & 2.93$\pm$0.34 & 3.70$\pm$0.43 & SF & 0 \\[0.5ex]
150 & 5.24$\pm$0.49 & 6.63$\pm$0.61 & SF & 0 \\[0.5ex]
151 & -- & 2.43$\pm$0.22 & C & 0 \\[0.5ex]
152 & 10.59$\pm$0.81 & 13.41$\pm$1.02 & SF & 0 \\[0.5ex]
153 & 25.37$\pm$1.31 & 32.12$\pm$1.65 & SF & 0 \\[0.5ex]
154 & 19.76$\pm$0.86 & 25.01$\pm$1.09 & SF & 0 \\[0.5ex]
155 & 15.12$\pm$0.82 & 19.14$\pm$1.04 & SF & 0 \\[0.5ex]
156 & 2.69$\pm$0.23 & 3.40$\pm$0.29 & SF & 0 \\[0.5ex]
157 & 19.74$\pm$1.18 & 24.98$\pm$1.49 & SF & 0 \\[0.5ex]
159 & 14.09$\pm$1.04 & 17.84$\pm$1.32 & SF & 0 \\[0.5ex]
160 & 6.16$\pm$0.66 & 7.80$\pm$0.84 & SF & 0 \\[0.5ex]
164 & 85.55$\pm$1.73 & 108.29$\pm$2.19 & SF & 0 \\[0.5ex]
165 & 93.67$\pm$2.09 & 118.57$\pm$2.65 & SF & 0 \\[0.5ex]
166 & 53.28$\pm$1.53 & 67.44$\pm$1.93 & SF & 0 \\[0.5ex]
167 & 34.23$\pm$1.17 & 43.33$\pm$1.48 & SF & 0 \\[0.5ex]
168 & 8.00$\pm$0.30 & 10.12$\pm$0.39 & SF & 0 \\[0.5ex]
169 & 7.31$\pm$0.38 & 9.26$\pm$0.48 & SF & 0 \\[0.5ex]
170 & 5.95$\pm$0.32 & 7.54$\pm$0.40 & SF & 0 \\[0.5ex]
171 & 10.65$\pm$0.47 & 13.48$\pm$0.60 & SF & 0 \\[0.5ex]
172 & 23.95$\pm$0.84 & 30.32$\pm$1.06 & SF & 0 \\[0.5ex]
173 & 21.09$\pm$0.73 & 26.69$\pm$0.92 & SF & 0 \\[0.5ex]
174 & 24.79$\pm$0.76 & 31.38$\pm$0.96 & SF & 0 \\[0.5ex]
176 & 90.27$\pm$9.22 & 114.26$\pm$11.66 & SF & 1 \\[0.5ex]
\hline\\[-2ex]
\end{tabular}
\end{minipage}
\tablefoot{The columns correspond to: 
(1) Identifier of the HII region; 
(2) Star formation rate (x10$^{-3}\,M_\odot\,yr^{-1}$);
(3) H$\alpha$ luminosity (x10$^{38}erg\,s^{-1}$);
(4) BPT classification;
(5) Velocity of the region belongs to the LV (0) or the HV (1) sub-samples. For regions with two velocity components we use `$^{a}$' and `$^{b}$' for low- and high-velocity components, respectively.
}
\end{table*}

\begin{table*}
\setlength{\tabcolsep}{1pt}
\begin{minipage}[t]{.48\linewidth}
\tiny
\caption{Oxygen abundances for the SQ HII regions using N2 and O3N2 \citep{2009MNRAS.398..949P}, and R calibrators \citep{2016MNRAS.457.3678P}, respectively, nitrogen to oxygen abundances ratio using \cite{2016MNRAS.457.3678P}, and the radial velocity sub-sample.}
\label{table:table5} 
\begin{tabular}{c | c c c c c }
\hline\hline  \\[-2ex]
(1) & (2) & (3) & (4) & (5) & (6) \\[0.5ex] 
Region & 12+log(O/H)$_{N2}$ & 12+log(O/H)$_{O3N2}$ & 12+log(O/H)$_{R}$ & log(N/O)$_{Pi16}$ & Vel\\[0.5ex]
ID &  &  &  &  & \\[0.5ex]
\hline\\[-2ex]
2 & 8.30$\pm$0.11 & 8.35$\pm$0.05 & 8.22$\pm$0.11 & -1.29$\pm$0.13 & 1 \\[0.5ex]
3 & 8.61$\pm$0.09 & 8.48$\pm$0.06 & 8.46$\pm$0.10 & -1.17$\pm$0.14 & 1 \\[0.5ex]
4 & 8.55$\pm$0.03 & 8.47$\pm$0.02 & 8.49$\pm$0.04 & -0.98$\pm$0.09 & 1 \\[0.5ex]
6 & 8.48$\pm$0.10 & 8.36$\pm$0.06 & -- & -- & 1 \\[0.5ex]
8 & 8.59$\pm$0.07 & 8.59$\pm$0.05 & -- & -- & 1 \\[0.5ex]
9 & 8.61$\pm$0.04 & 8.50$\pm$0.03 & 8.55$\pm$0.04 & -0.86$\pm$0.09 & 1 \\[0.5ex]
10 & 8.46$\pm$0.03 & 8.38$\pm$0.02 & 8.47$\pm$0.03 & -0.95$\pm$0.06 & 1 \\[0.5ex]
11 & 8.66$\pm$0.06 & 8.54$\pm$0.04 & 8.60$\pm$0.04 & -0.74$\pm$0.11 & 1 \\[0.5ex]
12 & 8.62$\pm$0.09 & 8.63$\pm$0.06 & -- & -- & 1 \\[0.5ex]
13 & 8.55$\pm$0.02 & 8.42$\pm$0.02 & 8.54$\pm$0.03 & -0.84$\pm$0.09 & 1 \\[0.5ex]
14 & 8.72$\pm$0.03 & 8.66$\pm$0.03 & 8.60$\pm$0.03 & -0.79$\pm$0.10 & 1 \\[0.5ex]
15 & 8.59$\pm$0.03 & 8.49$\pm$0.02 & 8.58$\pm$0.02 & -0.74$\pm$0.07 & 1 \\[0.5ex]
24 & 8.76$\pm$0.05 & 8.70$\pm$0.05 & 8.65$\pm$0.04 & -0.67$\pm$0.10 & 1 \\[0.5ex]
25 & 8.81$\pm$0.05 & 8.74$\pm$0.05 & -- & -- & 1 \\[0.5ex]
33 & 8.70$\pm$0.09 & 8.59$\pm$0.05 & 8.62$\pm$0.06 & -0.74$\pm$0.11 & 0 \\[0.5ex]
35 & 8.84$\pm$0.08 & 8.70$\pm$0.05 & 8.62$\pm$0.08 & -1.01$\pm$0.10 & 1 \\[0.5ex]
39 & 8.60$\pm$0.09 & 8.47$\pm$0.04 & 8.55$\pm$0.07 & -0.87$\pm$0.11 & 0 \\[0.5ex]
42 & 8.64$\pm$0.07 & 8.56$\pm$0.04 & 8.53$\pm$0.06 & -0.95$\pm$0.10 & 0 \\[0.5ex]
45 & 8.79$\pm$0.08 & 8.64$\pm$0.06 & 8.61$\pm$0.08 & -1.00$\pm$0.12 & 0 \\[0.5ex]
46 & 8.58$\pm$0.08 & 8.62$\pm$0.06 & -- & -- & 0 \\[0.5ex]
47 & 8.81$\pm$0.09 & 8.72$\pm$0.06 & 8.68$\pm$0.05 & -0.58$\pm$0.13 & 0 \\[0.5ex]
48 & 8.82$\pm$0.11 & 8.65$\pm$0.07 & 8.66$\pm$0.09 & -0.85$\pm$0.15 & 0 \\[0.5ex]
50 & 8.83$\pm$0.09 & 8.66$\pm$0.06 & 8.62$\pm$0.10 & -1.06$\pm$0.12 & 0 \\[0.5ex]
51$^{b}$ & 8.71$\pm$0.12 & 8.63$\pm$0.07 & 8.50$\pm$0.13 & -1.11$\pm$0.13 & 1 \\[0.5ex]
52 & 8.71$\pm$0.04 & 8.66$\pm$0.04 & 8.59$\pm$0.04 & -0.82$\pm$0.09 & 0 \\[0.5ex]
59$^{b}$ & 8.39$\pm$0.13 & 8.52$\pm$0.07 & 7.98$\pm$0.17 & -1.58$\pm$0.13 & 1 \\[0.5ex]
60$^{b}$ & 8.71$\pm$0.09 & 8.55$\pm$0.06 & 8.44$\pm$0.12 & -1.32$\pm$0.13 & 1 \\[0.5ex]
64 & 8.72$\pm$0.12 & 8.74$\pm$0.07 & 8.60$\pm$0.08 & -0.75$\pm$0.11 & 1 \\[0.5ex]
65 & 8.65$\pm$0.06 & 8.60$\pm$0.06 & 8.50$\pm$0.07 & -1.01$\pm$0.14 & 0 \\[0.5ex]
70 & 8.79$\pm$0.05 & 8.71$\pm$0.05 & 8.61$\pm$0.05 & -0.92$\pm$0.10 & 0 \\[0.5ex]
74 & 8.57$\pm$0.11 & 8.69$\pm$0.06 & 8.42$\pm$0.10 & -1.02$\pm$0.11 & 1 \\[0.5ex]
75 & 8.69$\pm$0.09 & 8.65$\pm$0.05 & 8.60$\pm$0.06 & -0.75$\pm$0.11 & 0 \\[0.5ex]
76$^{a}$ & 8.77$\pm$0.09 & 8.70$\pm$0.05 & 8.58$\pm$0.09 & -0.97$\pm$0.10 & 0 \\[0.5ex]
76$^{b}$ & 8.76$\pm$0.12 & 8.67$\pm$0.07 & 8.56$\pm$0.12 & -1.01$\pm$0.13 & 1 \\[0.5ex]
77 & 8.66$\pm$0.11 & 8.64$\pm$0.06 & 8.56$\pm$0.09 & -0.85$\pm$0.13 & 1 \\[0.5ex]
78 & 8.42$\pm$0.12 & 8.52$\pm$0.06 & 8.27$\pm$0.13 & -1.23$\pm$0.14 & 0 \\[0.5ex]
82 & 8.52$\pm$0.07 & 8.46$\pm$0.03 & 8.46$\pm$0.06 & -1.01$\pm$0.09 & 0 \\[0.5ex]
83$^{b}$ & 8.52$\pm$0.10 & 8.55$\pm$0.05 & 8.36$\pm$0.10 & -1.16$\pm$0.11 & 1 \\[0.5ex]
91 & 8.71$\pm$0.09 & 8.66$\pm$0.06 & 8.55$\pm$0.08 & -0.95$\pm$0.12 & 0 \\[0.5ex]
96 & 8.60$\pm$0.09 & 8.62$\pm$0.06 & 8.44$\pm$0.09 & -1.05$\pm$0.12 & 1 \\[0.5ex]
97 & 8.35$\pm$0.12 & 8.28$\pm$0.05 & 8.45$\pm$0.07 & -0.92$\pm$0.12 & 0 \\[0.5ex]
102$^{a}$ & 8.77$\pm$0.11 & 8.61$\pm$0.06 & 8.60$\pm$0.12 & -1.01$\pm$0.15 & 0 \\[0.5ex]
104$^{b}$ & 8.67$\pm$0.09 & 8.67$\pm$0.06 & 8.49$\pm$0.09 & -1.02$\pm$0.11 & 1 \\[0.5ex]
109 & 8.75$\pm$0.14 & 8.66$\pm$0.07 & -- & -- & 1 \\[0.5ex]
112 & 8.72$\pm$0.04 & 8.56$\pm$0.03 & 8.62$\pm$0.03 & -0.85$\pm$0.08 & 0 \\[0.5ex]
\hline\\[-2ex]
\end{tabular}
\end{minipage}
\begin{minipage}[t]{.48\linewidth}\vspace{1.75cm}
\tiny
\centering
\begin{tabular}{c | c c c c c }
\hline\hline  \\[-2ex]
(1) & (2) & (3) & (4) & (5) & (6) \\[0.5ex] 
Region & 12+log(O/H)$_{N2}$ & 12+log(O/H)$_{O3N2}$ & 12+log(O/H)$_{R}$ & log(N/O)$_{Pi16}$ & Vel\\[0.5ex]
ID &  &  &  &  & \\[0.5ex]
\hline\\[-2ex]
113 & 8.72$\pm$0.11 & 8.60$\pm$0.07 & -- & -- & 1 \\[0.5ex]
117 & 8.57$\pm$0.02 & 8.49$\pm$0.02 & 8.54$\pm$0.02 & -0.83$\pm$0.05 & 0 \\[0.5ex]
121 & 8.63$\pm$0.04 & 8.48$\pm$0.03 & 8.51$\pm$0.05 & -1.05$\pm$0.09 & 1 \\[0.5ex]
122 & 8.46$\pm$0.02 & 8.41$\pm$0.02 & 8.45$\pm$0.03 & -0.99$\pm$0.06 & 1 \\[0.5ex]
123 & 8.79$\pm$0.11 & 8.60$\pm$0.06 & -- & -- & 1 \\[0.5ex]
125 & 8.74$\pm$0.09 & 8.65$\pm$0.07 & 8.58$\pm$0.09 & -0.95$\pm$0.14 & 0 \\[0.5ex]
127 & 8.77$\pm$0.04 & 8.64$\pm$0.03 & 8.66$\pm$0.02 & -0.65$\pm$0.08 & 0 \\[0.5ex]
128 & 8.75$\pm$0.04 & 8.59$\pm$0.05 & 8.60$\pm$0.04 & -0.97$\pm$0.12 & 0 \\[0.5ex]
129 & 8.71$\pm$0.03 & 8.60$\pm$0.03 & 8.59$\pm$0.03 & -0.88$\pm$0.07 & 0 \\[0.5ex]
130 & 8.65$\pm$0.04 & 8.64$\pm$0.05 & 8.60$\pm$0.03 & -0.63$\pm$0.10 & 0 \\[0.5ex]
132 & 8.72$\pm$0.03 & 8.57$\pm$0.02 & 8.61$\pm$0.02 & -0.86$\pm$0.07 & 0 \\[0.5ex]
133 & 8.67$\pm$0.01 & 8.51$\pm$0.01 & 8.61$\pm$0.01 & -0.77$\pm$0.04 & 0 \\[0.5ex]
134 & 8.75$\pm$0.06 & 8.70$\pm$0.06 & 8.61$\pm$0.05 & -0.79$\pm$0.12 & 0 \\[0.5ex]
136 & 8.74$\pm$0.04 & 8.58$\pm$0.03 & 8.62$\pm$0.04 & -0.87$\pm$0.10 & 0 \\[0.5ex]
137 & 8.76$\pm$0.03 & 8.64$\pm$0.03 & 8.65$\pm$0.03 & -0.74$\pm$0.09 & 0 \\[0.5ex]
138 & 8.75$\pm$0.03 & 8.63$\pm$0.03 & 8.63$\pm$0.02 & -0.77$\pm$0.08 & 0 \\[0.5ex]
139 & 8.63$\pm$0.03 & 8.54$\pm$0.02 & 8.61$\pm$0.02 & -0.64$\pm$0.07 & 0 \\[0.5ex]
140 & 8.74$\pm$0.05 & 8.62$\pm$0.04 & 8.62$\pm$0.04 & -0.82$\pm$0.10 & 0 \\[0.5ex]
141 & 8.69$\pm$0.02 & 8.57$\pm$0.02 & 8.60$\pm$0.02 & -0.81$\pm$0.06 & 0 \\[0.5ex]
142 & 8.77$\pm$0.03 & 8.61$\pm$0.03 & 8.66$\pm$0.02 & -0.74$\pm$0.08 & 0 \\[0.5ex]
143 & 8.77$\pm$0.03 & 8.65$\pm$0.03 & 8.66$\pm$0.02 & -0.69$\pm$0.09 & 0 \\[0.5ex]
146 & 8.75$\pm$0.02 & 8.64$\pm$0.03 & 8.64$\pm$0.02 & -0.72$\pm$0.08 & 0 \\[0.5ex]
147 & 8.73$\pm$0.09 & 8.57$\pm$0.06 & 8.49$\pm$0.12 & -1.25$\pm$0.15 & 0 \\[0.5ex]
148 & 8.75$\pm$0.06 & 8.67$\pm$0.06 & 8.60$\pm$0.06 & -0.88$\pm$0.13 & 0 \\[0.5ex]
149 & 8.73$\pm$0.09 & 8.62$\pm$0.06 & 8.58$\pm$0.08 & -0.93$\pm$0.14 & 0 \\[0.5ex]
150 & 8.69$\pm$0.07 & 8.65$\pm$0.06 & 8.56$\pm$0.07 & -0.90$\pm$0.11 & 0 \\[0.5ex]
152 & 8.56$\pm$0.08 & 8.57$\pm$0.06 & 8.51$\pm$0.07 & -0.88$\pm$0.12 & 0 \\[0.5ex]
153 & 8.66$\pm$0.04 & 8.60$\pm$0.04 & 8.57$\pm$0.04 & -0.84$\pm$0.09 & 0 \\[0.5ex]
154 & 8.70$\pm$0.03 & 8.58$\pm$0.03 & 8.63$\pm$0.02 & -0.70$\pm$0.09 & 0 \\[0.5ex]
155 & 8.59$\pm$0.05 & 8.45$\pm$0.04 & 8.54$\pm$0.04 & -0.89$\pm$0.10 & 0 \\[0.5ex]
156 & 8.63$\pm$0.08 & 8.54$\pm$0.04 & 8.56$\pm$0.06 & -0.86$\pm$0.10 & 0 \\[0.5ex]
157 & 8.67$\pm$0.05 & 8.55$\pm$0.04 & 8.59$\pm$0.04 & -0.83$\pm$0.10 & 0 \\[0.5ex]
159 & 8.73$\pm$0.05 & 8.64$\pm$0.05 & -- & -- & 0 \\[0.5ex]
160 & 8.66$\pm$0.09 & 8.50$\pm$0.06 & -- & -- & 0 \\[0.5ex]
164 & 8.63$\pm$0.02 & 8.50$\pm$0.02 & 8.57$\pm$0.02 & -0.85$\pm$0.05 & 0 \\[0.5ex]
165 & 8.62$\pm$0.02 & 8.50$\pm$0.02 & 8.57$\pm$0.02 & -0.83$\pm$0.05 & 0 \\[0.5ex]
166 & 8.62$\pm$0.03 & 8.47$\pm$0.02 & 8.56$\pm$0.02 & -0.88$\pm$0.06 & 0 \\[0.5ex]
167 & 8.52$\pm$0.04 & 8.43$\pm$0.03 & 8.47$\pm$0.04 & -1.00$\pm$0.07 & 0 \\[0.5ex]
168 & 8.57$\pm$0.04 & 8.45$\pm$0.03 & 8.55$\pm$0.03 & -0.83$\pm$0.08 & 0 \\[0.5ex]
169 & 8.63$\pm$0.05 & 8.50$\pm$0.03 & 8.56$\pm$0.04 & -0.90$\pm$0.07 & 0 \\[0.5ex]
170 & 8.55$\pm$0.06 & 8.47$\pm$0.03 & 8.56$\pm$0.03 & -0.70$\pm$0.09 & 0 \\[0.5ex]
171 & 8.57$\pm$0.05 & 8.44$\pm$0.02 & 8.54$\pm$0.04 & -0.86$\pm$0.08 & 0 \\[0.5ex]
172 & 8.37$\pm$0.06 & 8.30$\pm$0.03 & 8.46$\pm$0.04 & -0.91$\pm$0.07 & 0 \\[0.5ex]
173 & 8.49$\pm$0.04 & 8.39$\pm$0.03 & 8.45$\pm$0.04 & -1.04$\pm$0.08 & 0 \\[0.5ex]
174 & 8.59$\pm$0.04 & 8.47$\pm$0.03 & 8.49$\pm$0.04 & -1.06$\pm$0.08 & 0 \\[0.5ex]
176 & 8.52$\pm$0.12 & 8.55$\pm$0.07 & 8.27$\pm$0.15 & -1.31$\pm$0.16 & 1 \\[0.5ex]
\hline\\[-2ex]
\end{tabular}
\end{minipage}
\tablefoot{The columns correspond to: 
(1) Identifier of the HII region;
(2) Oxygen abundances using N2 calibrator \citep{2009MNRAS.398..949P};
(3) Oxygen abundances using O3N2 calibrator \citep{2009MNRAS.398..949P};
(4) Oxygen abundances using R calibrator \citep{2016MNRAS.457.3678P};
(5) Nitrogen to oxygen abundances ratio using \cite{2016MNRAS.457.3678P} calibrator;
(6) Velocity of the region belongs to the LV (0) or the HV (1) sub-samples. For regions with two velocity components we use `$^{a}$' and `$^{b}$' for low- and high-velocity components, respectively.
}
\end{table*}

\end{appendix}

\end{document}